\documentclass[10pt]{article}

\usepackage{amsmath}
\usepackage{graphicx}
\usepackage{amssymb}
\usepackage{changepage}
\usepackage{amsthm}
\usepackage{enumerate}
\usepackage{enumitem}
\usepackage{multicol}
\usepackage{multirow}
\usepackage{makecell}
\usepackage{bm}
\usepackage{physics}
\usepackage{hyperref}
\usepackage{float}
\usepackage{subcaption}
\usepackage{listings}
\usepackage{natbib}
\usepackage{bbm}
\usepackage{longtable}
\usepackage{geometry}
\usepackage{blkarray}
\usepackage{setspace}
\usepackage{pdfpages}
\usepackage{algorithm2e} 
\usepackage{placeins}
\usepackage{lscape}
\usepackage{color,soul}
\usepackage{pdfpages}
\RestyleAlgo{ruled}
\newcommand{\reffig}[2]{\hyperref[#1]{figure~\ref*{#1}#2}}
\newcommand{\refeq}[1]{\hyperref[#1]{equation~(\ref*{#1})}}
\newcommand{\refeqsystem}[1]{\hyperref[#1]{equations~(\ref*{#1})}}
\newcommand{\refeqshort}[1]{\hyperref[#1]{Eq~(\ref*{#1})}}
\newcommand{\refeqs}[2]{\hyperref[#1]{equations~(\ref*{#1})} and \hyperref[#2]{(\ref*{#2})}}
\newcommand{\RN}[1]{\textup{\uppercase\expandafter{\romannumeral#1}}}
\title{Bayesian spatiotemporal modelling of political violence and conflict events using discrete-time Hawkes processes}
\author{Raiha Browning$^{1,2}$, Hamish Patten$^3$, Judith Rousseau$^{4}$, Kerrie Mengersen$^{1}$}
\date{% 
	\small
    $^1$Centre for Data Science, Queensland University of Technology, Australia\\%
    $^2$The Walter and Eliza Hall Institute of Medical Research, Melbourne, Australia\\%
    $^3$Information Management, International Federation of the Red Cross Red Crescent Societies (IFRC)\\%
    $^4$Ceremade, CNRS, UMR 7534, Université Paris-Dauphine, PSL University, France}
    
\begin{document}

\maketitle
\vfill

{\bf Keywords:} Bayesian inference, conflict risk monitoring, spatiotemporal Hawkes processes, self-exciting processes 

\clearpage
\pagenumbering{arabic} 

\section*{Abstract}

The monitoring of conflict risk in the humanitarian sector is largely based on simple historic averages. The overarching goal of this work is to assess the potential for using a more statistically rigorous approach to monitor the risk of political violence and conflict events in practice, and thereby improve our understanding of their temporal and spatial patterns, to inform preventative measures.

In particular, a Bayesian, spatiotemporal variant of the Hawkes process is fitted to data gathered by the Armed Conflict Location and Event Data (ACLED) project to obtain sub-national estimates of conflict risk in South Asia over time and space. Our model can effectively estimate the risk level of these events within a statistically sound framework, with a more precise understanding of uncertainty than was previously possible. The model also provides insights into differences in behaviours between countries and conflict types. We also show how our model can be used to monitor short and long term trends, and that it is more stable and robust to outliers compared to current practices that rely on historical averages. 

\newpage 

\section{Introduction} \label{sec:intro}

A key indicator of disorder in society is the occurrence of political violence and conflict events, such as protests and riots. To promote sustainable development and prevent future violence and displacement, the underlying causes of conflict must first be understood and addressed. From a humanitarian perspective, understanding both the real-time and longer term trends of these events is crucial as it allows organisations to protect civilians, provide aid, and foster peace among communities. This knowledge is also important for other key parties such as governments, journalists and researchers, especially during times of instability and unrest. A recent report by the United Nations Office for the Coordination of Humanitarian Affairs \citep{OCHAprediction} also highlighted this fact, emphasising the need to focus on understanding the risk of conflict events, in favour of other activities such as forecasting and event classification, to drive anticipatory action.

Improved modelling of conflict and political violence events has the potential to significantly advance several UN Sustainable Development Goals (SDGs). For SDG 16 (Peace, Justice and Strong Institutions), better modelling can help to identify drivers of instability, provide early warning signs, and highlight high-risk regions or populations that may need particular attention, supporting SDG 16’s targets on reducing violence, ensuring equal access to justice and building accountable institutions. It also contributes to SDG 1 (No Poverty) and SDG 2 (Zero Hunger), as conflict is a major cause of poverty and food insecurity. Furthermore, it has implications for SDG 3 (Good Health and Wellbeing) and SDG 4 (Quality Education), since conflict and violence can inhibit the effective functioning of health systems and educational institutions. 

The Armed Conflict Location and Event Data (ACLED) project \citep{Raleigh:2010} provides a comprehensive, geo-referenced database containing details of political violence and protests worldwide. Currently, ACLED uses a combination of historical averages and machine learning models to assess and predict conflict risk respectively, and these results are generally aggregated to the national level. A key use case for these data is in determining appropriate measures to take in anticipation of a future hazard to reduce the humanitarian impact of the event before it fully unfolds, referred to as anticipatory action by the United Nations Office for the Coordination of Humanitarian Affairs. In this article, we introduce a new statistical approach to conflict monitoring, that is more sophisticated than historical averages, but remains easily interpretable with an enhanced understanding of model uncertainty compared to traditional machine learning approaches. 

Hawkes processes \citep{Hawkes:1971vr} are self-exciting stochastic processes whereby previous events in the process increase the probability of future events occurring. They have previously been used to study phenomena such as earthquakes \citep{Ogata:1988jd}, crime and terrorism \citep{White:2013,Mohler:vf}, and social media interactions \citep{Chen:2018fw,Dutta:2020}. In this article, we extend the standard temporal model to include a spatially varying baseline rate of events, in addition to a spatial kernel that quantifies the spatial impact of events to the intensity of the process. A discrete-time variant of the spatiotemporal Hawkes process is fit to the ACLED data at a fine spatial scale to obtain sub-national estimates of conflict risk. This approach aids in characterising the temporal and spatial patterns that emerge from these types of conflicts, and provides an estimate of risk over space and time that can be used both for real-time monitoring and for making short-term predictions. We use data from the region of South Asia to showcase our approach; however, the methodology can easily be extended to other countries. For the purposes of this example, the four countries considered are Bangladesh, Sri Lanka, Nepal and Pakistan. In addition to modelling each of the countries separately, the various types of conflict events in each country are also modelled individually. By doing so, we can compare the spatial and temporal dynamics of events between both countries and conflict types.

 The overarching goal of this article is to determine whether spatiotemporal discrete-time Hawkes processes can provide new spatial and temporal insights that can help practitioners and key actors in the humanitarian sector to better understand conflict events. Specifically, we address three main questions throughout the article: (i) are spatiotemporal discrete-time Hawkes processes useful for explaining conflict events, (ii) what can we learn from them, that we cannot learn from other approaches, and (iii) how can these models be used in practice. 

ACLED provides two main tools for tracking conflict risk. The first tool is the ACLED Trendfinder Dashboard, that monitors previous and current trends and identifies periods in which an unusually high level of activity occurred. These thresholds are based on historical moving averages. The second tool is the Conflict Alert System (CAST), a forecasting tool used to predict future conflict activity for battles, explosions/remote violence, and violence against civilians. CAST is trained using a series of machine learning models. These two tools are intended to be used in conjunction with one another. While the approach taken in this article could also be extended to provide forecasts, we focus instead on gaining a better understanding of conflict risk as discussed in \cite{OCHAprediction}.

Several authors have proposed HPs as a method for modelling conflict events using the ACLED project data. \cite{Zhu:2020ti} models the temporal triggering kernel for the HP through a mixture of neural networks that encode a latent function to capture underlying dynamics of community states and apply this model to the ACLED data. \cite{Campedelli:2021st} use HPs to cluster events and explain the temporal behaviour patterns of disorder events during the COVID-19 pandemic. To account for spatial dependency among events, the authors used K-means clustering to group spatial locations with similar characteristics. None of these approaches directly incorporate a spatial triggering kernel into the Hawkes model. By doing so in this article, we are able to better understand the extent of spatial correlation inherent within the process.

\cite{Hengel:2020st} does model an explicit spatial triggering kernel within the context of a marked variant of the Hawkes process known as the Epidemic Type Aftershock Sequence (ETAS), which assumes a specific form for the kernel. \cite{Okawa:2022} introduce the convolutional Hawkes process, incorporating spatial heterogeneity through a convolutional neural network that takes images as an input. To our knowledge, there have been no Bayesian approaches to modelling the ACLED data with spatiotemporal Hawkes processes. 

Other classes of spatiotemporal models have also been fitted to the ACLED data. \cite{Xie:2023} and \cite{Weidmann:2010} use boosted regression trees and autoregressive regression techniques respectively. Recent work by \cite{Egbon:2024} introduces an autoregressive model with spatial effects modelled through a stochastic partial differential equation. Estimation was carried out using the Integrated Nested Laplace Approximation (INLA). There are also other examples of Bayesian spatiotemporal models used to model other conflict datasets \citep{Zammit:2012,Tench:2018st,Python:2019go} that employ, respectively, log-Gaussian Cox processes, Dirichlet process mixture models and stochastic partial differential equations.

While the focus of this paper is on the ACLED database, other geo-referenced conflict databases also exist. The Global Terrorism Database (GTD) \citep{GTD} contains records of terrorist activity that occurred worldwide since 1970. Naturally, the GTD has a smaller range of conflict types compared to ACLED, since events in the former database must be outside the context of legitimate warfare activities. Several authors have proposed Hawkes processes to model these data \citep{White:2013,Porter:2012fh}; however, these are limited to studying the temporal dimension of the data. The model proposed by \cite{Jun:2024} considers the spatial dimension through a flexible representation of the spatiotemporal triggering kernel. For parameter inference, these authors rely on maximum likelihood estimation, whereas our analysis is focused on utilising the Bayesian framework analysing conflict to reap the benefits discussed in Section \ref{sec:uv_st_dthp}. \cite{Clark:2018} use a spatiotemporal Hawkes process within the Bayesian framework, however using a simplified construction of the temporal and spatial components where the temporal component is represented as a point mass and there is no explicit spatial kernel.

The Uppsala Conflict Data Program (UCDP) \citep{UCDP} is another rich source of data for events involving armed conflict. While the ACLED data has a larger scope of conflict types recorded, the UCDP records have a longer history, in some cases going back to 1946. Using these data, \citet{Mueller:2022} propose a framework to generate predictions of conflict using a combination of unsupervised and supervised machine learning techniques. This work differs substantially from ours as it is prediction-focused, with the authors integrating a web-scraping machine learning algorithm to characterise conflict risk. Spatiotemporal models considered for these data include spatiotemporal regression techniques \citep{Racek:2024} and spatial Poisson processes \citep{Schutte:2017}.

To our knowledge, we propose the first Bayesian, spatiotemporal Hawkes model applied to the ACLED data. We study countries in South Asia as an exemplar of how this model can be applied to these types of data worldwide. The models are fitted on a univariate level, so given sufficient computational resources the number of countries included in the study can be extended to include a much larger sample. Different types of conflict events are also considered individually, allowing us to compare the patterns of both temporal and spatial excitation for different conflict types. The proposed model enables a data-driven approach to monitoring and understanding the current risk of conflict events throughout time and space. Combined with input and expertise from key actors in the humanitarian and political sciences sectors, such as ACLED, this work can have important implications for the understanding and management of these events.

Section \ref{sec:methods} introduces the data, the spatiotemporal models, and the inference methods used in this article. Section \ref{sec:results} follows by presenting the results of our analysis. We first present our Bayesian, spatiotemporal model, assess the model fit and interpret the outputs. In Section \ref{sec:practicals}, we demonstrate how our models can be used to assess conflict risk in practice and benchmark against the types of risk models used by organisations, such as ACLED, that rely on historical averages. Finally, Section \ref{sec:discussion} concludes with a discussion of the outcomes of this analysis, limitations of the model and avenues for future work.

\section{Materials and methods} \label{sec:methods}

\subsection{Data} \label{sec:data}

The ACLED project \citep{Raleigh:2010} provides a comprehensive geo-referenced database containing details of political violence and protests worldwide and has recorded in total more than one million individual events. The earliest recorded events in the database are from 1997 for many countries in Africa, and the coverage has been progressively expanded over recent years to now having achieved global coverage with data being updated weekly. In addition to the date and location of events, the database also records the number of fatalities, type of violence and the actors involved, which allows the user to further analyse and understand conflict events. Summaries and reports on the ACLED website are often segmented by conflict type, indicating that this type of segmentation is of interest to key stakeholders in the humanitarian sector. The types of conflicts recorded in the ACLED database and their respective definitions \citep{ACLED} are,

\begin{itemize}
	\item Battles: violent interactions between two organised armed groups;
	\item Explosions/Remote violence: one-sided violence events in which the tool for engaging in conflict creates asymmetry by taking away the ability of the target to respond;
	\item Violence against civilians: violent events where an organised armed group deliberately inflicts violence upon unarmed non-combatants;
	\item Protests: a public demonstration against a political entity, government institution, policy or group in which the participants are not violent;
	\item Riots: violent events where demonstrators or mobs engage in disruptive acts or disorganised acts of violence against property or people;
	\item Strategic development: accounts for often non-violent activity by conflict and other agents within the context of the war/dispute. Recruitment, looting and arrests are included.
\end{itemize}

As mentioned in Section \ref{sec:intro}, for the purposes of illustrating our approach, we focus on South Asia as an exemplar region. We consider weekly counts of conflict events in South Asia over a 5 year period from 2010--2014. During the selected 5 year period, there were four countries with conflict data recorded in the ACLED database, namely Bangladesh, Sri Lanka, Nepal and Pakistan. Data for South and Southeast Asia start from 2010 in the ACLED dataset, and the 5 year period was chosen to reduce computational expense and because it avoids the period in 2016 when ACLED significantly increased capacity and updated their reporting methodology. Several other countries comprise the South Asian region, but did not have data available during the observation window for this study.

\begin{figure}[H]
	\centering
	\includegraphics[width=\textwidth]{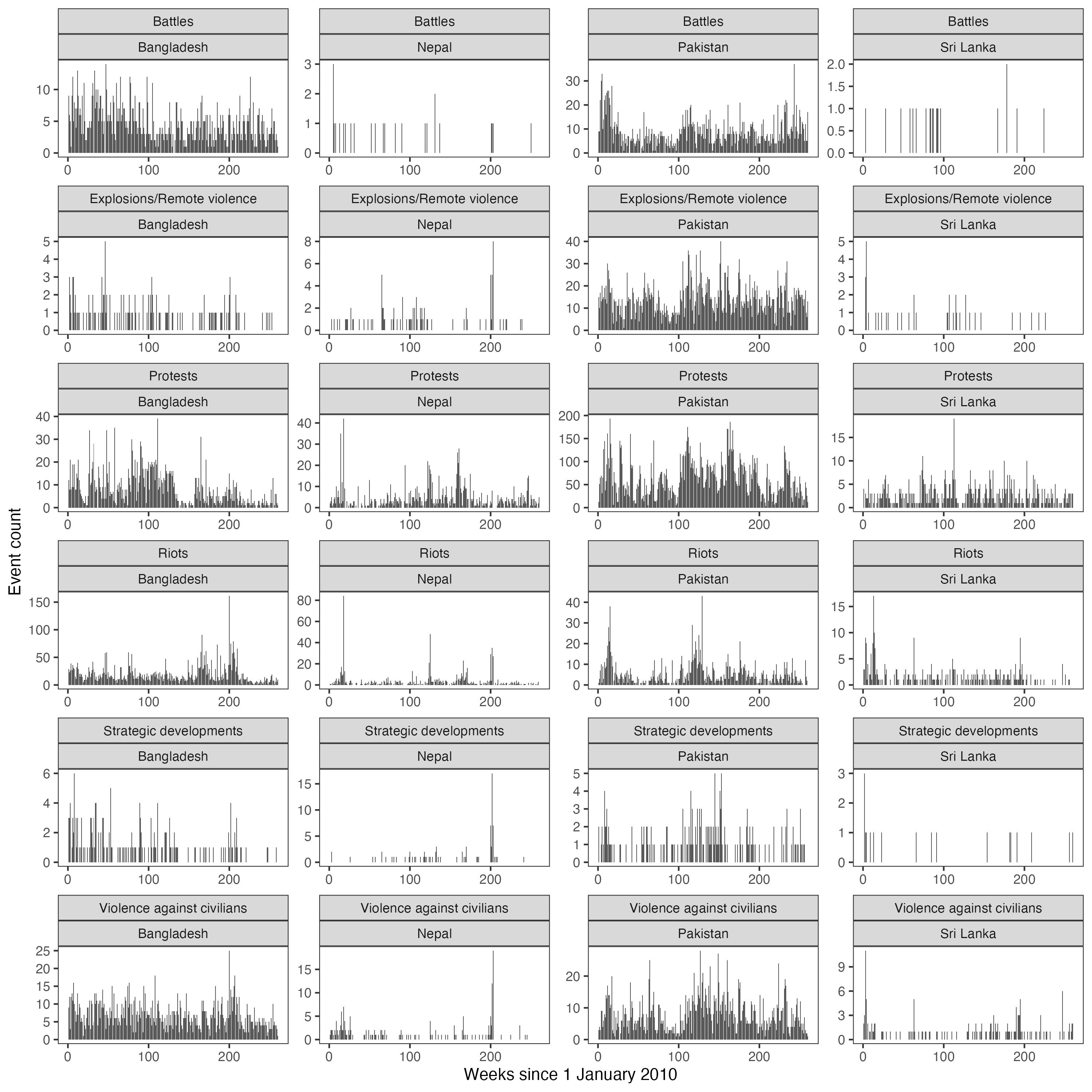}
	\caption{Weekly events in each country for each conflict type.}
\label{fig:events_weekly}
\end{figure}

\begin{figure}[H]
	\centering
	\includegraphics[width=0.5\textwidth]{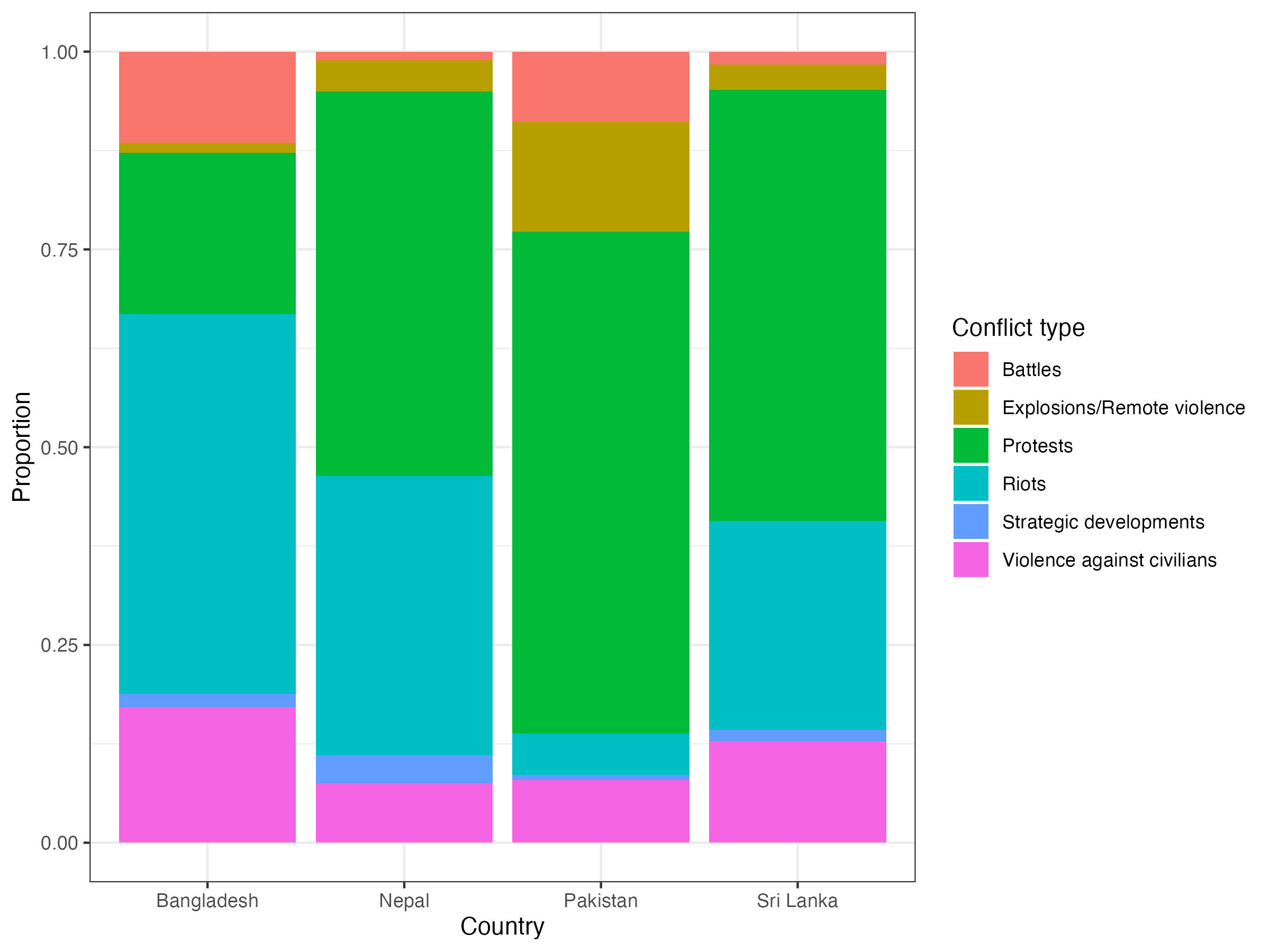}
	\caption{Proportion of total events of each conflict type by country.}
\label{fig:prop_event_type}
\end{figure}

Figure \ref{fig:events_weekly} shows the number of conflict events recorded each week for each conflict type and country over the observation window. There is a wide range of dynamics inherent in these data. We can see that Pakistan experienced substantially higher event counts over this period and Sri Lanka experienced the lowest number of conflict events. We also observe that some event types are more prevalent across all countries, such as protests and riots. Figure \ref{fig:prop_event_type} displays the observed proportion of each conflict type for each country throughout the observation window. While protests and riots account for the majority of conflict events recorded, we see that Pakistan also has a high proportion of Explosions/remote violence compared to the other countries. We also see that Pakistan and Bangladesh experienced higher proportions of events classed as Battles, and that Sri Lanka and Bangladesh have higher proportions of events classed as Violence against civilians.

We are also interested in modelling spatial correlation in these data to understand how conflict intensity is influenced by events in neighbouring regions and to compare spatial dynamics between countries and conflict types. To determine the spatial aggregation we discretised according to administrative boundaries, which define the borders of a geographic area under a given political or social structure, rather than considering the individual location of each event. These boundaries were obtained from the Database of Global Administrative Areas (GADM) \citep{ADM}. For Nepal and Pakistan we used the Level 2 county/district level administrative boundaries. For Bangladesh and Sri Lanka, we used the coarser aggregation of Level 1 state/province level administrative boundaries since the Level 2 boundaries for these countries are very finely defined. The coordinates for each administrative boundary was then defined as the centroid of each region.

\begin{figure}[H]
	\centering
	\includegraphics[width=\textwidth]{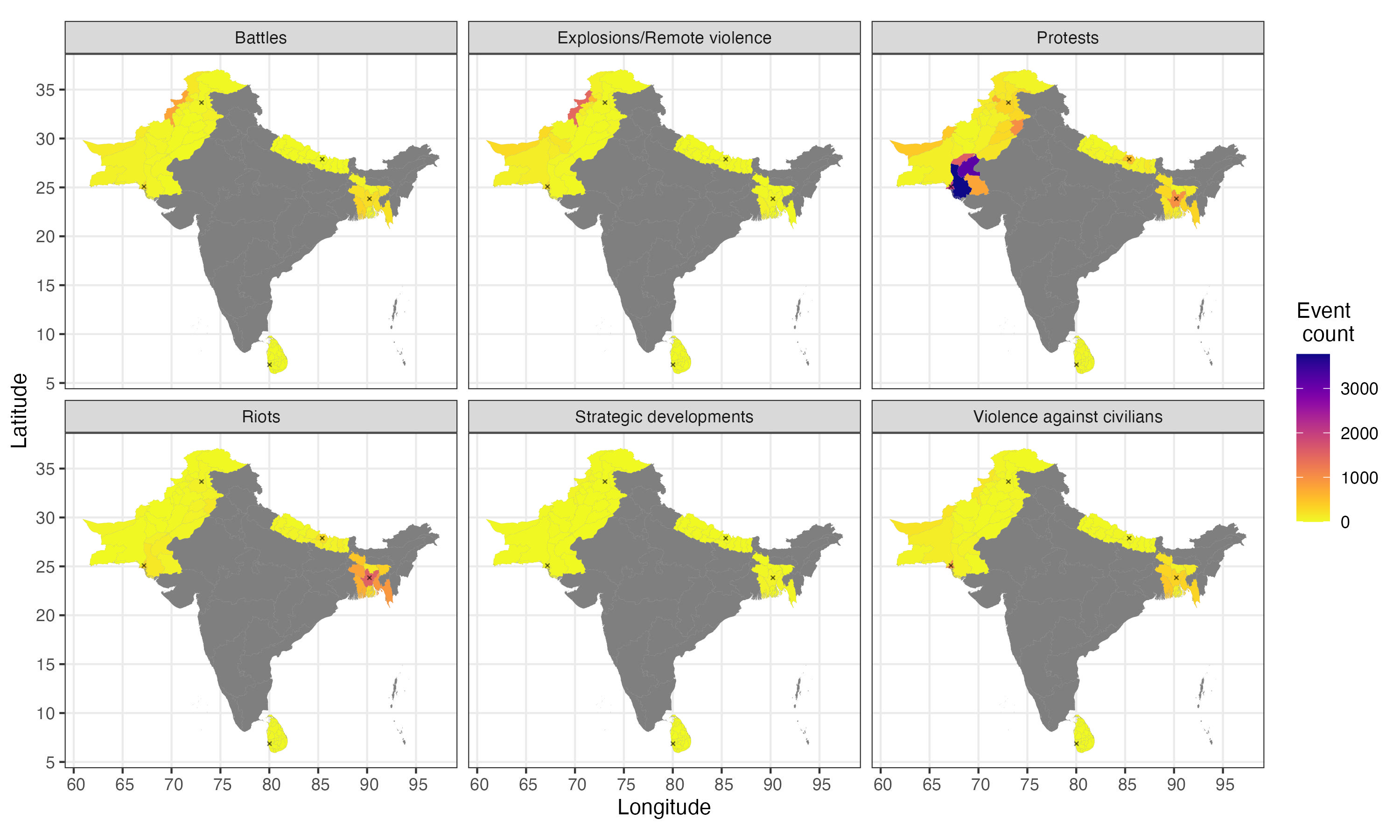}
	\caption{Total number of events over 5 year observation window. Black crosses indicate capital cities (including Karachi in Pakistan as the biggest city and economic capital). }
\label{fig:total_events}
\end{figure}

\begin{figure}[H]
	\centering
	\includegraphics[width=\textwidth]{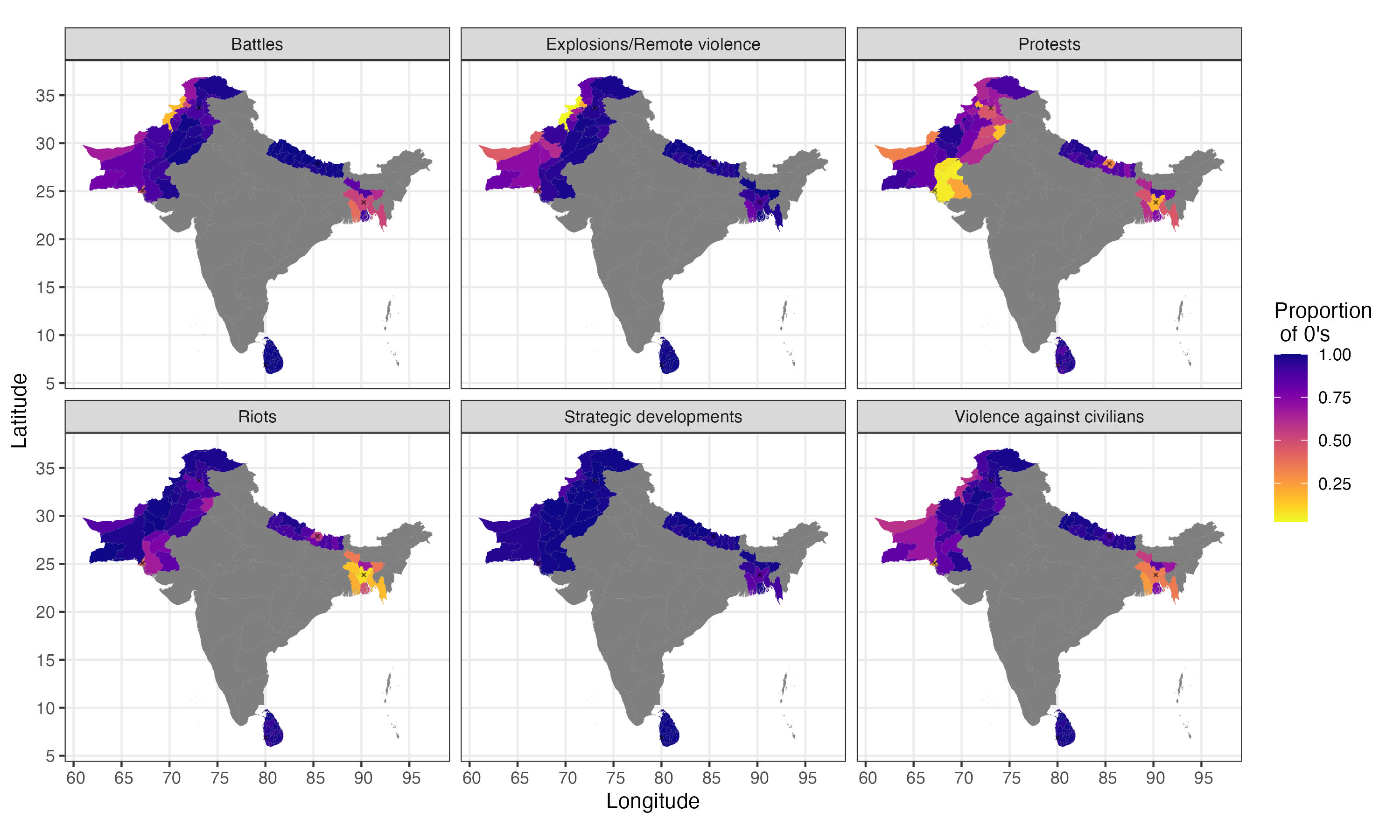}
	\caption{Proportion of weeks with zero events. Black crosses indicate capital cities (including Karachi in Pakistan as the biggest city and economic capital). }
\label{fig:zero_inf}
\end{figure}

Figure \ref{fig:total_events} presents the total number of events of each conflict type to occur over the 5 year observation window at the administrative regions specified above for each country. The maps show that most activity generally occurs in or near the capital cities. Pakistan also has a high concentration of protests on the southeastern border near the biggest city and economic capital of Karachi. By modelling the spatial correlation in these data we can characterise how events from these regions influence neighbouring areas. There is also significant zero-inflation present in these weekly event counts. Figure \ref{fig:zero_inf} presents the proportion of weeks for each administrative region where the number of events for each conflict type was 0. In general, Sri Lanka and to a lesser extent Nepal have a higher proportion of zero event counts compared to Bangladesh and Pakistan.

\subsection{Discrete-time Hawkes processes} \label{sec:dthp}

Hawkes processes (HPs) \citep{Hawkes:1971vr} were introduced to model phenomena that exhibit self-exciting behaviour whereby events occur in groups or clusters. HPs were originally introduced as a continuous-time process. However, since the data collected by the ACLED project is in the form of daily counts we adopt a discrete-time version of the HP \citep{Browning:2022,Browning:2021,Linderman:2015tk,Mohler:2013hy}, referred to in this article as the discrete-time Hawkes process (DTHP). 

For a given DTHP, the number of events that occur at a particular time can be distributed according to any appropriate distribution defined on $\mathbb{Z}_{\ge 0}$ \citep{Porter:2012fh, Laub:2025}. In our analysis, we assume that events are distributed according to a Poisson distribution to approximate the continuous-time counterpart \citep{Kirchner:2016}. The DTHP is characterised by this distribution and its conditional intensity function which governs the rate at which events occur. The intensity function is comprised of two components: the baseline process and the self-exciting process. The baseline process describes the arrival rate of events that occur independently, and not as a result of other events. The self-exciting component on the other hand, describes the clustering behaviour of the process. The aspect of the self-exciting component that describes the impact that past events have on the current intensity of the process is referred to as the triggering kernel, and is a function of the time that has elapsed since an event occurred. 

To state this formally, consider a linear DTHP $N$, where $N_t$ represents the number of events up to and including the $t$-th time interval and let the random variable $Y_t= N_t - N_{t-1}$ represent the number of event occurrences in $t$-th time interval. $N_t$ is then dependent on the set of events to occur up to but not including the $t$-th time interval, denoted by $\mathcal{H}_{t-1}= \{y_s: s < t \}$, where $y_s$ represents the observed number of events during the $s$-th time interval. We denote the corresponding conditional intensity function as $\lambda_t$. This quantity represents the number of events expected to occur in the counting process $N_t$ during the $t$-th time interval, conditionally on the past, and is given by, 
\begin{align}
	\lambda_t &=  \mathbb{E}\{N_t - N_{t-1} | \mathcal{H}_{t-1}\} \nonumber  \\ 
		&= \mu + \alpha \sum_{s<t} y_{s} g(t-s) \label{eq:cond_int_dis} 
\end{align}
where $\mu$ represents the baseline mean of the process and the second term represents the self-exciting component of the DTHP. In particular, $\alpha$ is a magnitude parameter that describes the expected number of offspring events that a single event is expected to produce. If $\alpha>1$, then the dynamics of the process will be explosive and tend towards infinity, otherwise if $\alpha<1$ then the process will be stationary. The triggering kernel $g(t-s)$ is a probability mass function that describes the influence of these past events on the intensity of the process, given the time elapsed since the $s$-th time interval, where $s<t$.

The number of events observed during the $t$-th time interval, namely $y_t$, is distributed according to the random variable, $Y_t$, which has conditional mean $\mathbb{E}\{Y_t|\mathcal{H}_{t-1}\} = \lambda_t$ as defined in \eqref{eq:cond_int_dis}. In this analysis, $Y_t$ is assumed to be Poisson distributed such that $Y_t|\mathcal{H}_{t-1} \sim \text{Pois}(\lambda_t)$. The corresponding likelihood function is then,

\begin{align}
	L &= \prod_{t=1}^{T} \frac{\lambda_t ^{y_{t}}}{y_t!} \exp(- \lambda_t ).
\end{align}

This model can be seen as a special case of a Poisson autoregressive model \citep{Fokianos:2009,Armillotta:2024}, and the endemic-epidemic model \citep{Held:2005,Bracher:2022,Lu:2023} which has been used widely in epidemiological research. 

\subsection{Spatiotemporal discrete-time Hawkes process}\label{sec:uv_st_dthp}

In this article, we consider several ways to extend the temporal DTHP defined in Section \ref{sec:dthp} to account for spatial patterns in the data. The first approach we consider is to include a spatially varying baseline rate $\mu_{x,y}$ for each region with coordinates $(x,y)$. Since we are aggregating to the level of administrative areas, this spatially varying parameter is also expected to indirectly capture underlying characteristics inherent in each region such as population size and other demographic factors. The second approach we consider is, in addition to the spatially varying baseline rate $\mu_{x,y}$, we also incorporate a spatial triggering kernel $h(\cdot)$ to capture any spatial clustering patterns that may be present. This means the level of excitation is then governed by both the time since an event elapsed and distance from the location where the event occurred. In both of these models we have also incorporated a term for zero-inflation to account for excess zeros in the data, as observed in Figure \ref{fig:zero_inf}.

The first approach that includes a spatially varying baseline rate and no spatial self-excitation, referred to in this article as ``nospatialSE", results in the following rate parameter $\lambda_{t,x,y}$ for the $t$-th time interval and location with coordinates $(x,y)$,

\begin{equation}
	\lambda_{t,x,y} = \exp(\mu_0 + \mu_{x,y}) + \alpha \sum_{s<t}  y_{s,x,y} g(t-s)  \label{eq:cond_int_dis_nospatialSE} 
\end{equation}
where $\mu_0 \in \mathbb{R}$ is a global baseline rate parameter, $\mu_{x,y} \in \mathbb{R}$ is a region-specific baseline rate, and their sum is exponentiated to ensure positivity. Furthermore, $y_{s,x,y}$ denotes the number of events to occur in the $s$-th time interval in location with coordinates $(x,y)$.

The second approach which incorporates a spatial triggering kernel in addition to the spatially varying baseline rate, referred to as ``ST\_SE", gives the following rate parameter $\lambda_{t,x,y}$ for the $t$-th time interval and location with coordinates $(x,y)$,

\begin{equation}
	\lambda_{t,x,y} = \exp(\mu_0 + \mu_{x,y}) + \alpha \sum_{s<t} \sum_{(x_l,y_l) \in \mathcal{S}} y_{s,x_l,y_l} g(t-s) h(x-x_l,y-y_l) \label{eq:cond_int_dis_st} 
\end{equation}
where $\mathcal{S}$ is the set of centroids for all spatial regions, and $h(\cdot)$ is a spatial triggering kernel that determines how the distance between events affects the intensity of the process. Note also the subtle differences between \eqref{eq:cond_int_dis_nospatialSE} and \eqref{eq:cond_int_dis_st}, whereby the excitation term in the latter includes a summation over all other regions, and the former is only affected by events that have occurred in its own region.

The likelihood function for both of these models is constructed in the same manner. We assume the number of events observed during $t$-th time interval in location with coordinates $(x,y)$, $Y_{t,x,y}$ follows a zero-inflated Poisson distributed such that $Y_{t,x,y}|\mathcal{H}_{t-1} \sim \text{Pois-ZeroInf}(\lambda_{t,x,y}, \pi)$, where $\pi$ is the probability of a structural zero. The likelihood function is then, 

\begin{align}
	L & = \prod_{t=1}^{T} \prod_{(x,y) \in \mathcal{S}}    \Big[\pi + (1 - \pi) \exp(-\lambda_{t,x,y})\Big]^{\mathbb{I}\{y_{t,x,y}=0\}}
  \Big[(1 - \pi) \dfrac{\lambda_{t,x,y}^{\,y_{t,x,y}}}{y_{t,x,y}!} \exp(-\lambda_{t,x,y})\Big]^{\mathbb{I}\{y_{t,x,y}\ge1\}}
\end{align}

An important distinction between temporal and spatiotemporal DTHPs as defined above is in the interpretation of $\alpha$. In the spatiotemporal setting under both models we propose, the incorporation of a zero-inflation factor means that the non-zero component of the likelihood function must be scaled by the probability of structural zeroes $\pi$, and as such $\alpha>1$ no longer necessarily corresponds to explosive dynamics. Additionally, when considering the model that includes a spatial triggering kernel as proposed in \eqref{eq:cond_int_dis_st}, the level of excitation from a single event also depends on the location of each event. 

As it is presented in \eqref{eq:cond_int_dis_st}, the temporal triggering kernel $g(\cdot)$ is defined on $\mathbb{Z}^+$. However, in many contexts it is reasonable to assume that the maximum length of excitation is not infinite, and the impact of past events will eventually dissipate. Thus, the triggering kernel could be truncated to some finite value $t_{\max}$ with little impact when there is negligible mass at higher time intervals. Truncating the triggering kernel also reduces computational expense since the summations involved in evaluating $\lambda_{t,x,y}$ comprise less terms.

We performed comparisons to assess three modelling assumptions, namely the maximum excitation time $t_{\max}$ (where we considered 12, 26 and 52 weeks), the form of $\lambda_{t,x,y}$, and the choice of triggering kernels $g(\cdot)$ and $h(\cdot)$. Across all event types and countries, we found that a longer $t_{\max}$ of 52 weeks provided the best fit to the data, and we therefore proceed with this value. Regarding the choice of model for $\lambda_{t,x,y}$, we considered three choices: no self-excitation (only a spatially varying baseline rate), a spatially varying baseline rate and only temporal self-excitation as in \eqref{eq:cond_int_dis_nospatialSE}, and a spatially varying baseline rate with self-excitation both temporally and spatially as in \eqref{eq:cond_int_dis_st}. The first model with only a spatially varying baseline rate underperformed compared to the two alternatives. From these two models that include some form of self-excitation mechanism, neither emerged as a clear winner so both are considered in further analysis. Fitting both of these models for all country and conflict type scenarios also allows us to characterise more fully the spatial clustering dynamics by determining whether they are important in specific scenarios. In terms of triggering kernels, we compared a geometric temporal triggering kernel combined with a radial basis function (RBF) spatial triggering kernel against power law forms for both temporal and spatial kernels. There was no strong evidence supporting the inclusion of the more complex power law triggering kernels, as the additional parameters did not substantially improve model fit. We therefore retain the more parsimonious kernel specifications, which are described in detail below. Additional details on these comparisons are provided in the supplementary material.

In this study, we therefore specify the temporal triggering kernel in the form of a truncated, re-normalised geometric distribution such that,
 
\begin{equation}
    g(u)=\frac{\beta (1-\beta)^{u-1}}{\sum_{j=1}^{t_{\max}} \beta (1-\beta)^{j-1}}
\end{equation}
where $\beta \in [0,1]$ represents the success probability, or equivalently the inverse mean, for the geometric distribution. In the context of our study, a higher value of $\beta$ indicates that recent events have a stronger influence on the conditional intensity function $\lambda_{t,x,y}$. In other words, the impact of past events decays more rapidly over time, giving greater weight to events that occur closer to the current $t$-th time interval.

Furthermore, we represent the spatial triggering kernel by a radial basis function (RBF) kernel, which for vectors of coordinates $z=(x,y)$ and $z_l=(x_l,y_l)$ is given by,

\begin{equation}
	h(z,z_l) = \exp(-\frac{d(z,z_l)}{2 \sigma^2})
\end{equation}
where $\sigma$ is the bandwidth parameter and $d$ is assumed to be the Euclidean distance function such that,
\begin{equation}
	d(z,z_l) = \sqrt{(x-x_l)^2+(y-y_l)^2}.
\end{equation}
In the context of our study, $\sigma$ determines how spatial proximity influences the conditional intensity function $\lambda_{t,x,y}$. A smaller value of $\sigma$ indicates that events occurring close to the region with coordinates $(x, y)$ have a stronger effect on $\lambda_{t,x,y}$, resulting in more localised spatial interactions. In contrast, a larger $\sigma$ implies a smoother spatial influence, whereby events occurring farther away can contribute more to the intensity of events at $(x, y)$.

We also note that distance in the latitude-longitude coordinate space can be poorly defined, as the latitude coordinate grid is equally spaced but the longitude has a strong dependence on the associated latitude value. This distorts the comparison of distances, thus if countries further away from the equator were being considered, then an alternative distance such as the Haversine distance should be used. 

In this work we adopt a Bayesian framework due to its advantages, such as the ability to directly quantify uncertainty and introduce prior knowledge from experts or previous studies directly in the inference. This model was fitted using the probabilistic programming language \texttt{Stan} \citep{Stan}, which implements MCMC sampling through the No U-Turn Sampler (NUTS). 

We modelled the spatially varying baseline rate parameters $\mu_{x,y}$ using the Intrinsic Conditional Auto-Regressive (ICAR) model \citep{Besag:1974}. To simplify notation, we will simply write $\mu_r$ to refer to the region-specific baseline rate parameter for region $r$ with centroid $(x,y) \in \mathcal{S}$. We can then define the ICAR model based on \cite{Morris:2019} such that each $\mu_r$ is distributed according to a normal distribution with mean equal to the average of its neighbours and such that,

\begin{equation}
	p(\mu_r | \mu_{j \in N(l)}) = \mathcal{N}\bigg(\frac{\sum_{j \in N(l)} \mu_j}{d_r}, \frac{s_r^2}{d_r}\bigg)
\end{equation}
where $d_r$ is the number of neighbours of region $r$, $N(r)$ is the set of its neighbours, and $s_r^2$ is an unknown variance. The joint distribution of $M = (\mu_r, \ r \in \{1,\dots,|\mathcal{S}|\})$ is then,

\begin{equation}
	M \sim \mathcal{N}(0, Q^{-1})
\end{equation}
where $Q$ is the precision matrix. \cite{Morris:2019} defines $Q$ in terms of a diagonal matrix $D$ containing the number of neighbours in each region, and an adjacency matrix $A$ that identifies neighbouring regions. The precision matrix can then be written,

\begin{equation}
	Q = D-A.
\end{equation} 

For the non-spatially varying parameters, we considered two prior specifications: weakly informative priors and priors centered around the maximum likelihood estimates (MLEs). We found that the algorithm often did not converge when placing weakly informative priors on the non-spatial model parameters due to difficulties in estimating the correlation structure of the spatially varying baseline parameters. We therefore specify the following priors centered around the MLEs,

\begin{align}
	\mu_0 &\sim \mathcal{N}(\hat{\mu_0},0.1) \nonumber \\
	\alpha &\sim \log \mathcal{N}(m_{\hat{\alpha}},0.1) \nonumber \\
	\beta &\sim \text{Beta}(a_{\hat{\beta}},b_{\hat{\beta}}), \ 0<\beta <1 \nonumber \\
	\sigma &\sim \log \mathcal{N}(m_{\hat{\sigma}},0.1) \nonumber \\
	\pi &\sim \text{Beta}(a_{\hat{\pi_0}},b_{\hat{\pi_0}}), \ 0<\pi <1 
\end{align}
where \^{} indicates the MLEs which were calculated in \texttt{R} using the \texttt{optim()} function and $m$, $a$ and $b$ are constants calculated to ensure the mean of the prior is equal to the MLE. Further details of this prior sensitivity analysis are provided in the supplementary material.\\

\section{Results} \label{sec:results}

In this section we present the results obtained from fitting the Bayesian spatiotemporal models described in Section \ref{sec:uv_st_dthp} to the weekly counts of conflict events from the 2010 -- 2014 ACLED data for each country and conflict type combination detailed in Section \ref{sec:data}. We first describe the fitting process, in which we compare two specifications for the spatiotemporal DTHP and, for each scenario, select the best model to use in downstream analysis. We then compare the observed event counts with the expected number of events estimated from our model. The posterior distributions for the model parameters and triggering kernels are also visualised. Lastly, out-of-sample posterior predictive checks are shown. The code to implement these models is available at \url{https://github.com/RaihaTuiTaura/st-hawkes-thesis}.

As discussed in Section \ref{sec:uv_st_dthp}, we considered two specifications for the form of the spatiotemporal DTHP. We refer to the model defined in \eqref{eq:cond_int_dis_nospatialSE} with only temporal self-excitation as ``nospatialSE" and the model in \ref{eq:cond_int_dis_st} with self-excitation both temporally and spatially as ``ST\_SE". To choose an appropriate model form, we consider the following criteria. As a first pass, the algorithm must not show signs of non-convergence. This was assessed based on diagnostic criteria such as the $\hat{\mathcal{R}}$ \citep{Vehtari2021}, diagnostic plots including autocorrelation functions and trace plots, and by confirming that there were no divergent transitions that would indicate that \texttt{Stan} could not find algorithm settings that allowed the posterior to be explored without failure. Only two scenarios failed this criteria, namely Battles and Protests in Nepal under the ``nospatialSE" model, so for these two scenarios we use the ``ST\_SE" model. These diagnostics can be found in the supplementary materials.

If satisfied that the algorithm has likely converged, we consider the following goodness-of-fit criteria for model comparison to consider various aspects of the data:

\begin{enumerate}
	\item Out-of-sample coverage: proportion of observations where the observed event counts fall within the 95\% posterior interval based on 100 out-of-sample predictions.
	\item In-sample coverage: proportion of observations where the observed event counts fall within the median 95\% quantile interval (2.5\% - 97.5\% quantiles) calculated based on 100 posterior samples.
	\item In-sample residuals (non-zero; median): sum of the absolute value of the difference between the observed event count and the median of the 50\% quantile calculated based on 100 posterior samples.
	\item In-sample residuals (non-zero; upper quantile): sum of the absolute value of the difference between the observed event count and the median of the 97.5\% quantile calculated based on 100 posterior samples.
	\item In-sample residuals (zeros): sum of the absolute value of the difference between the observed number of zero event counts and the proportion in which the median of the 50\% quantile is equal to 0 calculated based on 100 posterior samples.
	\item BIC: median BIC with likelihood calculated based on 100 posterior samples.
\end{enumerate}
If the temporally and spatially self-exciting model (``ST\_SE") had better performance in greater than 50\% of these goodness-of-fit criteria, we select this model; otherwise, we resort to the more parsimonious model that only includes a temporal self-excitation mechanism (``nospatialSE"). Table \ref{tab:model_choice} presents the chosen model for each scenario. 

\begin{table}[ht]
\centering
\begin{tabular}{rllll}
  \hline
\textbf{Conflict type} & \textbf{Bangladesh} & \textbf{Nepal} & \textbf{Pakistan} & \textbf{Sri Lanka} \\ 
  \hline
 Battles & ST\_SE & ST\_SE & nospatialSE & ST\_SE \\ 
 Explosions/Remote violence & nospatialSE & ST\_SE & nospatialSE & ST\_SE \\ 
  Protests & ST\_SE & ST\_SE & nospatialSE & nospatialSE \\ 
  Riots & nospatialSE & ST\_SE & nospatialSE & nospatialSE \\ 
  Strategic developments & nospatialSE & ST\_SE & nospatialSE & nospatialSE \\ 
   Violence against civilians & nospatialSE & ST\_SE & nospatialSE & nospatialSE \\ 
   \hline
\end{tabular}
\caption{Choice of spatiotemporal DTHP for each scenario}
\label{tab:model_choice}
\end{table}

The maximum excitation time $t_{\max}$ was assumed to be 12 months. The impact of this assumption in our modelling is that past events cannot affect the intensity of the process after 12 months have elapsed. As such, the model was trained on data starting from the beginning of 2011 to mitigate edge effects and ensure that the model is not biased due to past events not being included in the intensity function calculations. 

\begin{figure}[H]
    \centering
    \includegraphics[width=\textwidth]{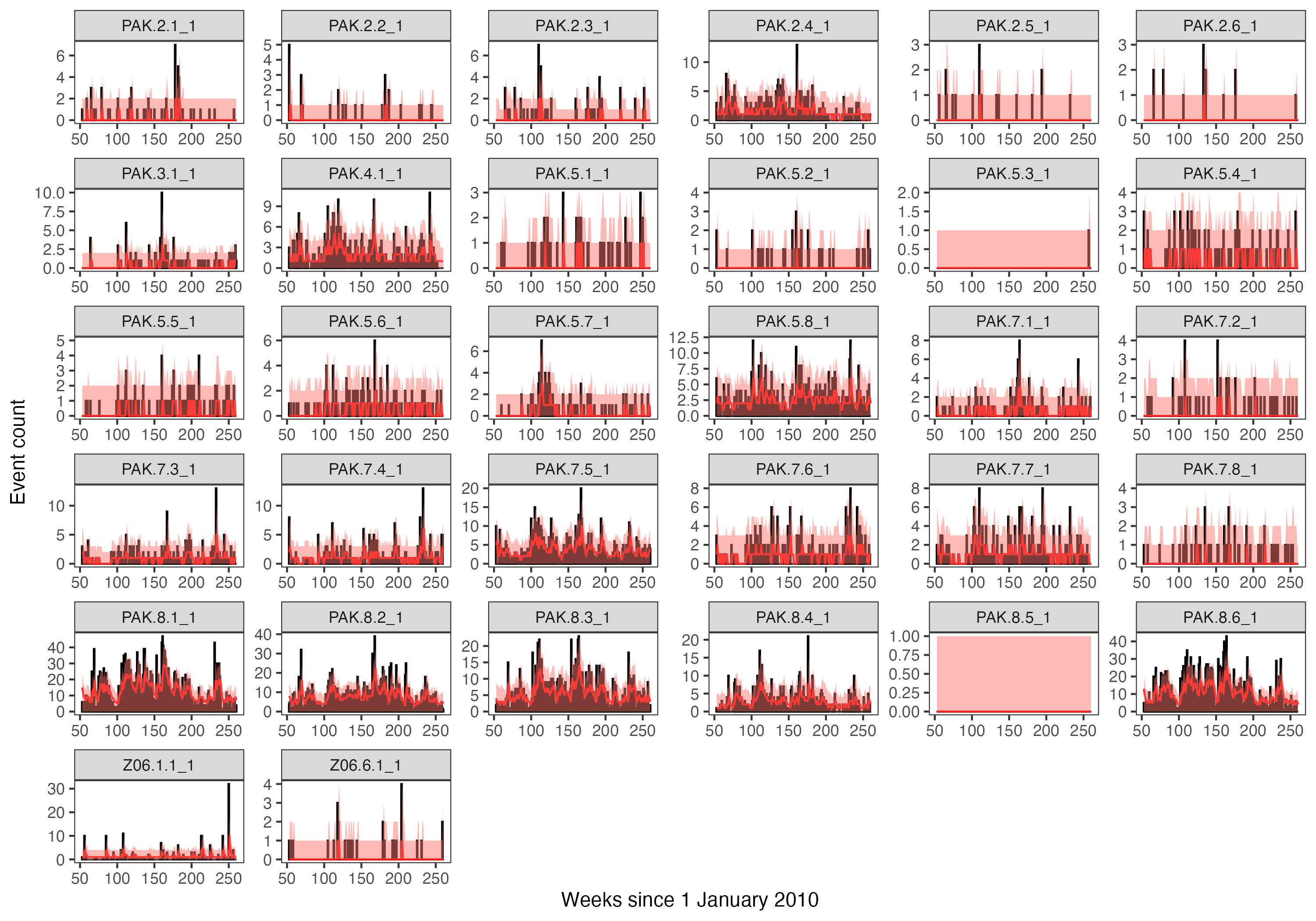}
    \caption{Protests in Pakistan (unaggregated). Observed data on week $t$ (black bars) versus median of the estimated 50\% quantile (red line) and median 95\% interval of event counts calculated based on 100 posterior samples (red ribbon).}
	\label{fig:st_est_lambda_PakProtest}
\end{figure}

To learn about short term trends, Figure \ref{fig:st_est_lambda_PakProtest} shows the modelled event counts for a single scenario, namely Protests in Pakistan and compares these to the observed event counts. Our Bayesian approach also allows us to characterise uncertainty around our estimates. We obtained the median 50\% quantile and 95\% uncertainty intervals for the estimated event counts by calculating, for 100 posterior samples, the 2.5\%, 50\% and 97.5\% quantiles of event counts arising from the estimated Poisson distributions. 

We see in our example for Protests in Pakistan that the median 50\% quantile and the corresponding uncertainty intervals around the counts are closely aligned to the observed counts within the individual regions and the model is able to react quickly to recent events. This is generally consistent for all countries and conflict type scenarios that we considered, as can be seen in the supplementary material. We note that, for scenarios with high zero-inflation, the median of the estimated 50\% quantiles is often also 0, however the uncertainty intervals generally encompass the observed event counts. 

\begin{figure}[H]
    \centering
    \includegraphics[width=0.85\textwidth]{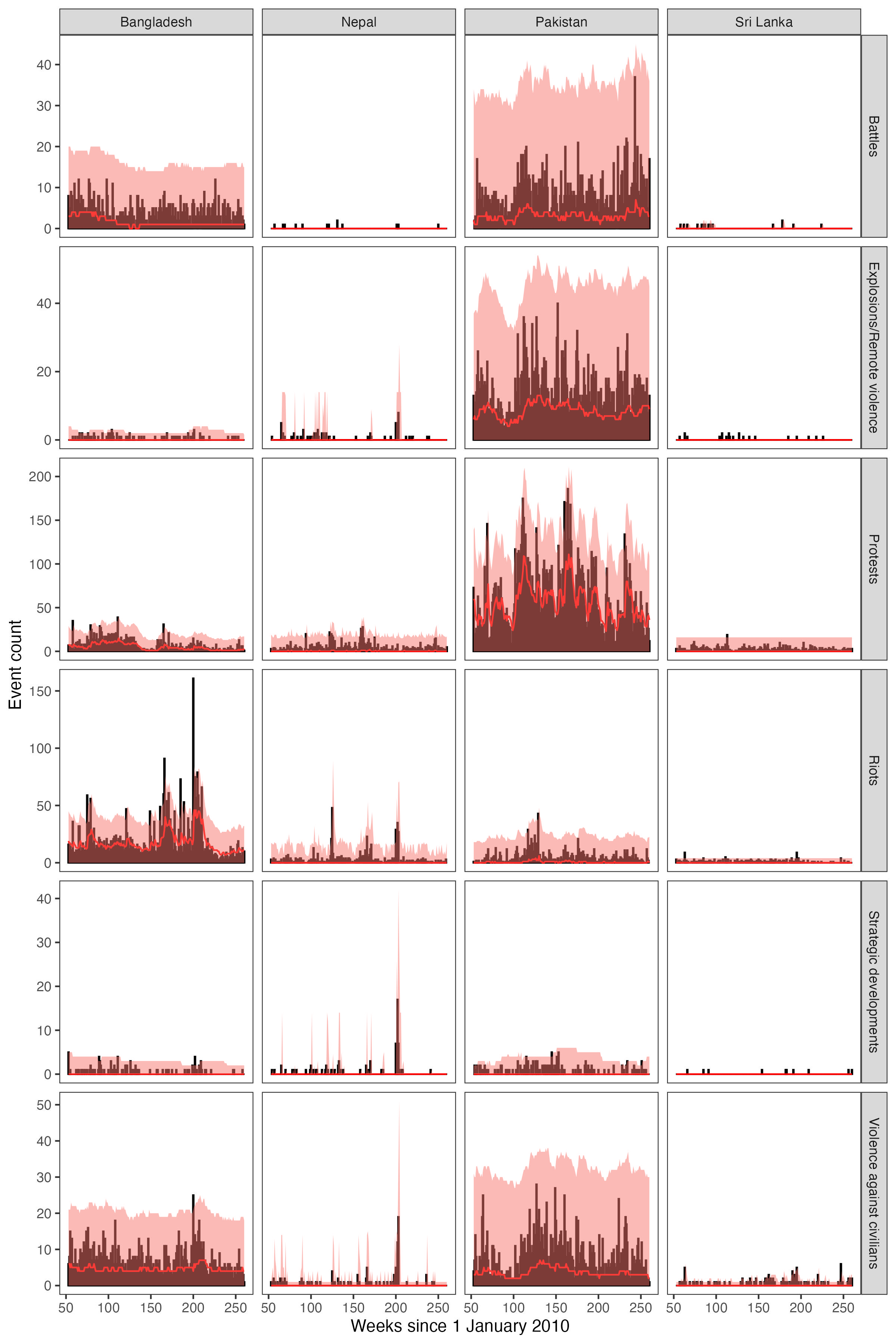}
    \caption{All scenarios (aggregated). Observed data on week $t$ (black bars) versus median of the estimated 50\% quantile (red line) and median 95\% interval of event counts calculated based on 100 posterior samples  (red ribbon).}
	\label{fig:st_est_lambda}
\end{figure}

We now consider aggregating over administrative regions such that, for each week and region, we calculate and sum the 5\%, 50\% and 95\% quantiles from 100 posterior samples over all regions to obtain weekly totals, from which we then calculate the median for each quantile. Figure \ref{fig:st_est_lambda} presents the median 90\% quantile intervals around these aggregated event counts and the corresponding median of the 50\% quantile, compared to the observed aggregated event counts for all scenarios. The estimated median number of events in each week is in line with the observed data. We considered tighter median quantile intervals of 90\% rather than 95\% since aggregating across administrative regions is expected to reduce uncertainty. Although these median quantile intervals of event counts appear to overestimate the upper bound in some scenarios, this is largely an artefact of the aggregation process. In cases where many regions experience no events, the upper bound is often a small, non-zero integer, and the accumulation of such low-count regions naturally inflates the resulting uncertainty bounds.

\begin{figure}[H]
	\centering
    \includegraphics[width=\textwidth]{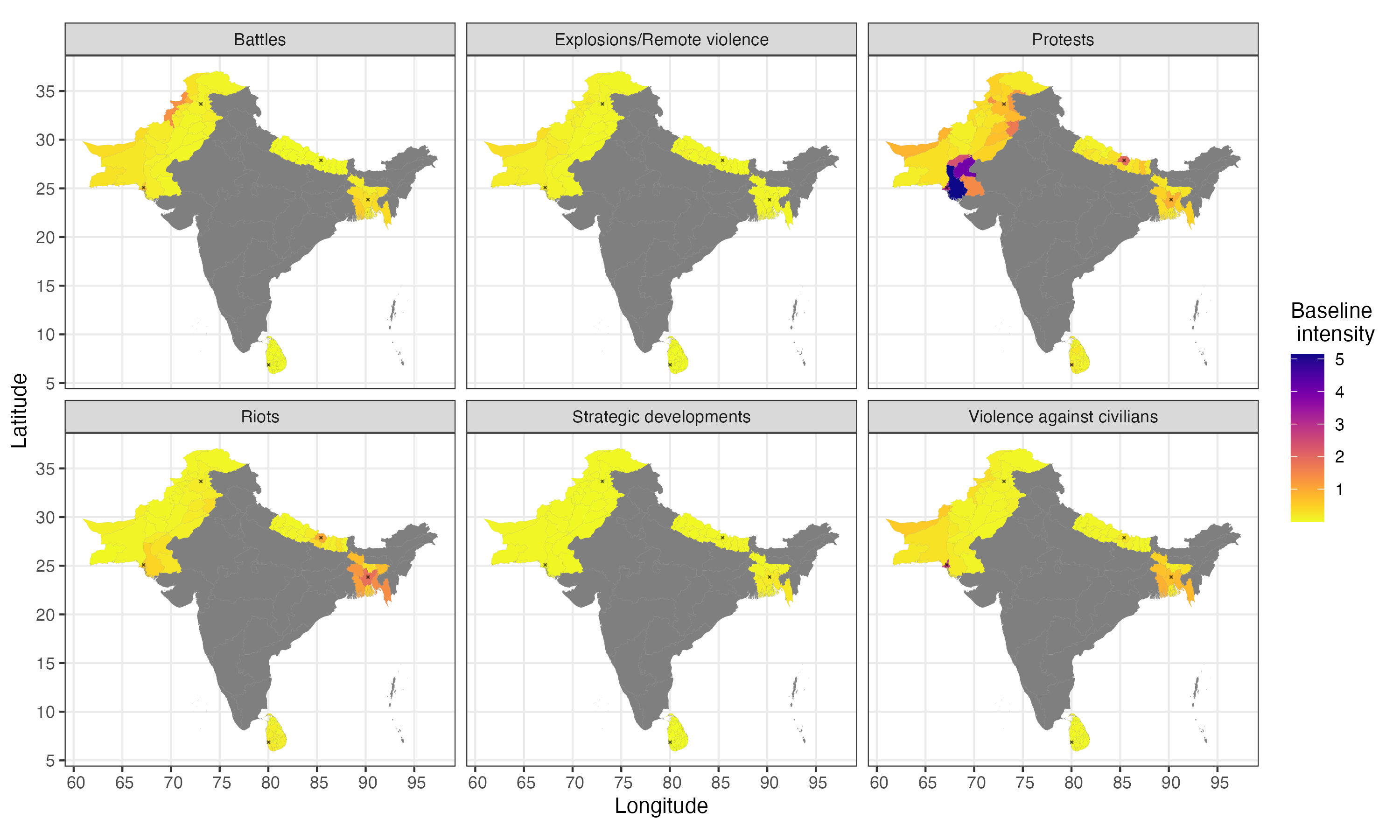}
	\caption{Median estimated baseline rates. Note that while displayed on a single map models for each country were estimated individually. Black crosses indicate capital cities (including Karachi in Pakistan as the biggest city and economic capital). }
	\label{fig:baseline_median}
\end{figure}

Figure \ref{fig:baseline_median} shows $\exp(\mu_0+\mu_l)$, namely the median estimated spatially varying baseline rate, for each region $l$ across conflict types. This rate reflects the level of conflict activity expected to arise independently of other events. Regions in Sri Lanka generally have a low level of baseline risk of conflict events across all conflict types. Bangladesh has higher baseline risk of Protests, Riots and Violence against civilians compared to other event types. Nepal has higher risk of Protests and Riots, and Pakistan has higher risk of Battles in the northwest region and Protests and Riots in the southeastern regions. In practice these estimates can provide actors in the humanitarian sector with a useful tool to discern the level of activity to be expected independently of any other events and the corresponding uncertainty in these estimates. 

\begin{figure}[H]
    \centering
    \includegraphics[width=0.85\textwidth]{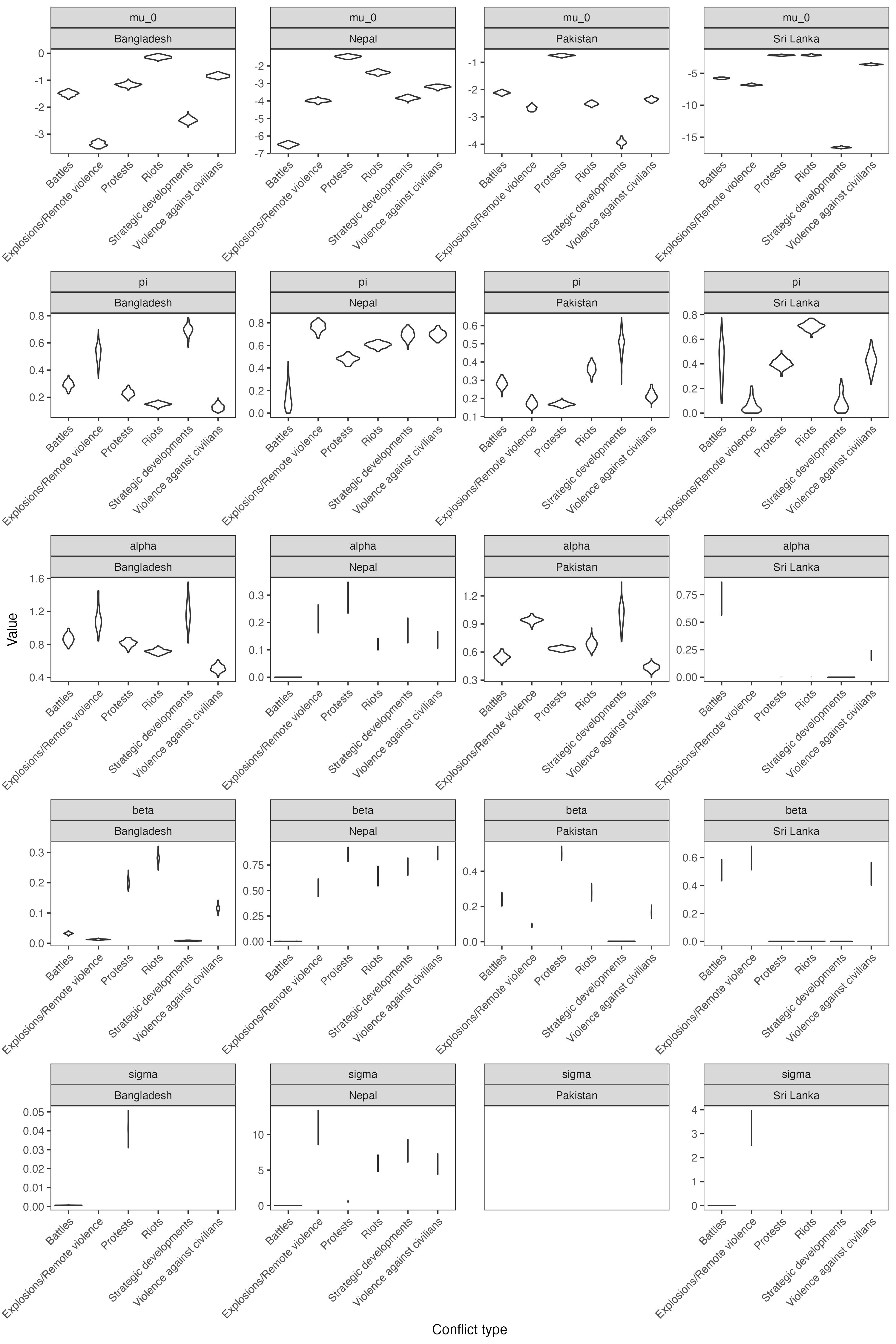}
    \caption{Distribution of estimated parameters}
	\label{fig:st_parameters}
\end{figure}

Figure \ref{fig:st_parameters} shows the estimated distribution of non-spatially varying model parameters. As mentioned in Section \ref{sec:uv_st_dthp}, the zero-inflation factor in our model means that the magnitude parameter $\alpha$ does not directly correspond to the expected number of offspring overall; rather, it indicates the level of self-excitation within the non-structural zero portion of the likelihood and should be interpreted in conjunction with the probability of structural zeros $\pi$. There are a number of scenarios for which the estimated $\alpha$ is essentially zero, particularly in Sri Lanka, indicating that a baseline only model may be sufficient for these scenarios. The scenarios for which this was the case fall into two categories: very few events occurring, or events occurring reasonably frequently but for which the baseline rate alone fits the data well. Large values of $\alpha$ typically correspond to high $\pi$, as seen often with Explosions/remote violence and Strategic developments, indicating that there are many zero counts but events that do occur are highly excitatory. There is a range in values of the inverse mean of the geometric temporal triggering kernel $\beta$, ranging from very low, which indicates that the influence of past events on the future intensity is more sustained throughout the 12 month excitation time, to higher values indicating that the influence drops off more sharply as time passes. The global baseline rate $\mu_0$ is generally higher for protests and riots indicating that, compared to other conflict types, more protests and riots arise independently rather than as a result of self-excitation mechanisms. For scenarios in which the ``ST\_SE" model was selected, the bandwidth for the spatial triggering kernel $\sigma$ was also estimated. In several cases $\sigma$ is very small, indicating that these conflict types also tend to be more localised; however, particularly in Nepal, we see larger $\sigma$ values which indicates a broader reach.

\begin{figure}[H]
	\centering
	\begin{subfigure}{\textwidth}
	\centering
	\includegraphics[width=\textwidth]{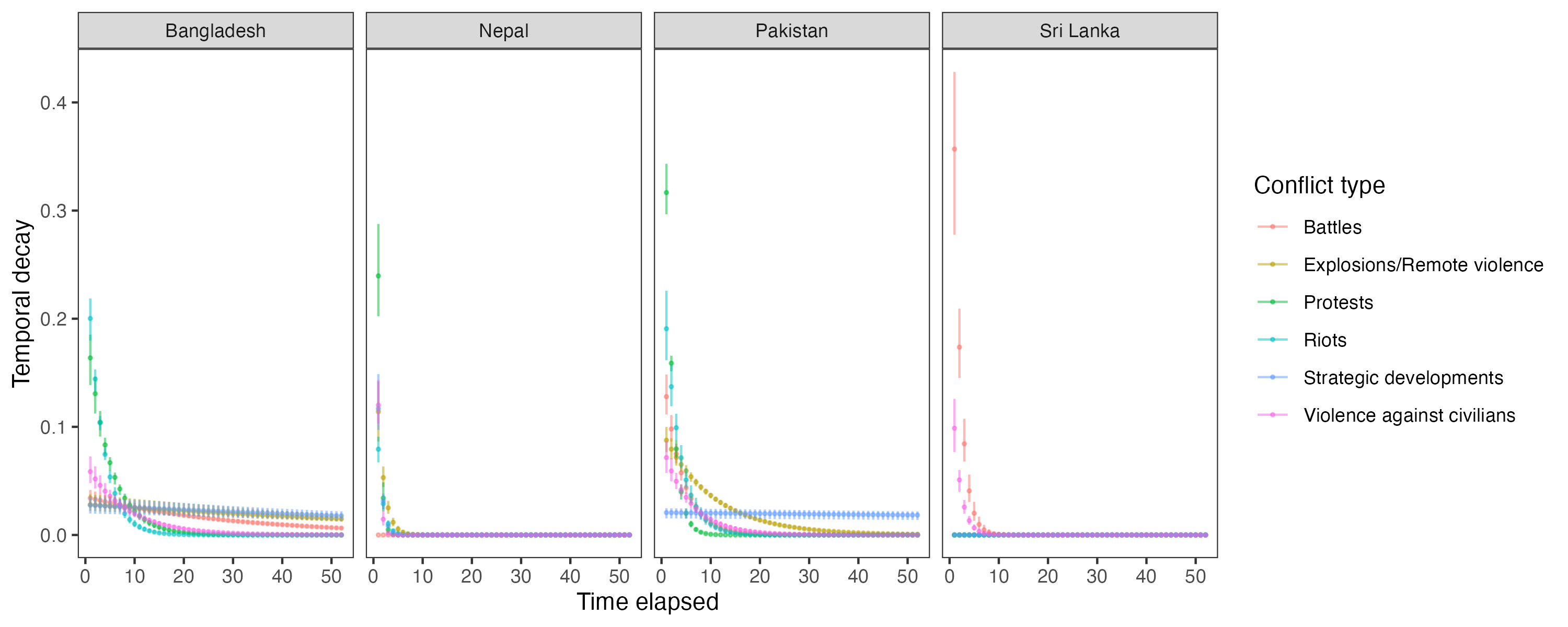}
	\caption{Estimated posterior distribution of temporal excitation kernels multiplied by $\alpha$.}
    \label{fig:temporal_kernels}
	\end{subfigure}
	\hfill
	\begin{subfigure}{\textwidth}
	\centering
	\includegraphics[width=\textwidth]{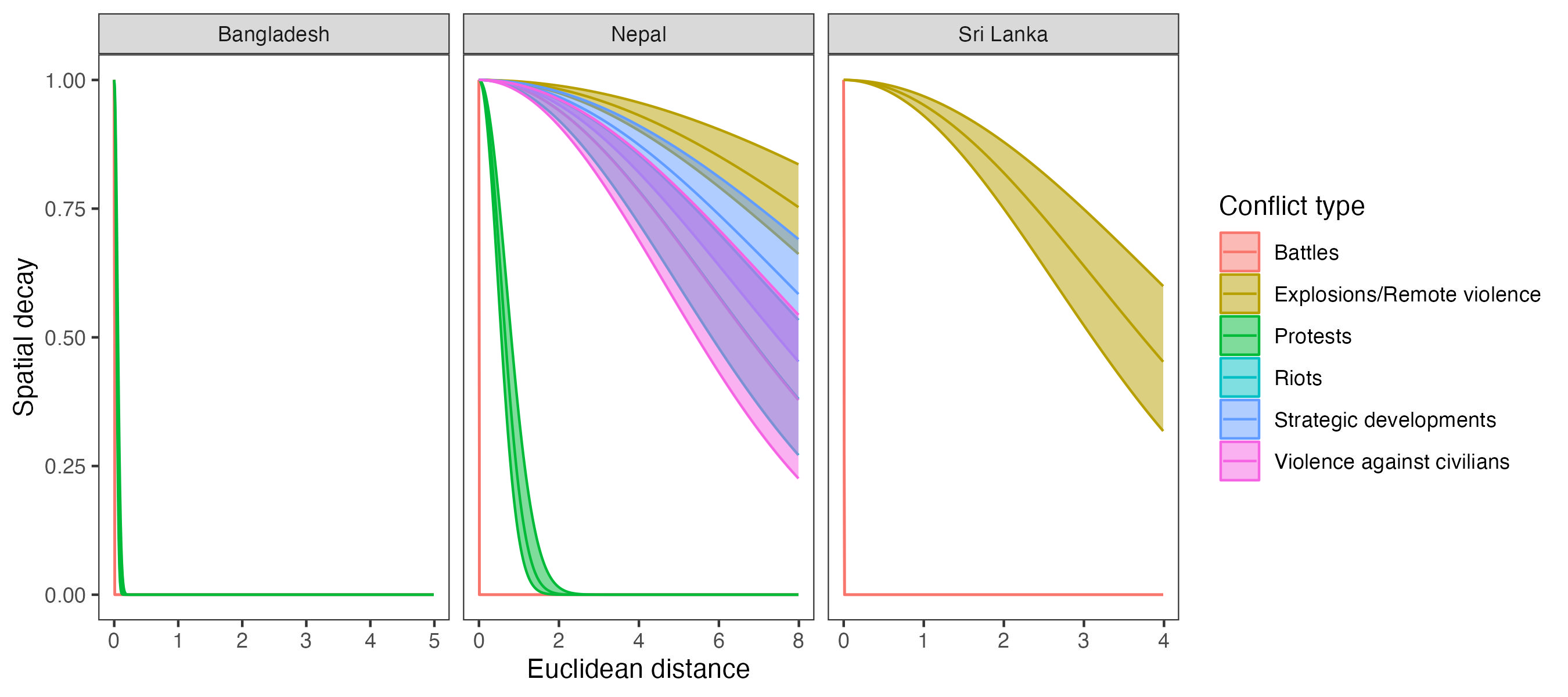}
	\caption{Estimated posterior distribution of spatial excitation kernels for scenarios using the ``ST\_SE" model. Units for horizontal axis are distance between two sets of coordinates, measured by Euclidean distance.}
	\end{subfigure}
	\caption{Posterior distribution of estimated excitation kernels for all countries and corresponding 95\% posterior interval.}
\label{fig:kernels_weekly_allcountries}
\end{figure}

The posterior median excitation kernels and corresponding 95\% intervals estimated from 100 posterior samples for each country and conflict type are presented in Figure \ref{fig:kernels_weekly_allcountries}. In Figure \ref{fig:temporal_kernels}, we multiplied the temporal excitation kernels by $\alpha$ to reflect the effective level of excitation from a single past event. The temporal excitation kernels for Nepal and Sri Lanka generally decay very quickly for all conflict types. On the contrary, conflicts in Bangladesh and Pakistan generally have longer excitation memories, meaning the influence of past events can persist for longer as time elapses. This is particularly true for the conflict types Battles, Explosions/remote violence and Strategic developments. 

For scenarios in which the spatial triggering kernel was estimated, we notice that spatial clustering of conflict events is more localised in Bangladesh, characterised by quickly decaying spatial triggering kernels. We also observe this pattern for Battles across all countries. On the other hand, for most conflict types in Nepal and for Explosions/remote violence in Sri Lanka we see more slowly decaying kernels, suggesting that the influence on the intensity function for events occurring in nearby locations and those further away is more similar compared to other conflict types.

\begin{figure}[H]
	\centering
    \includegraphics[width=0.85\textwidth]{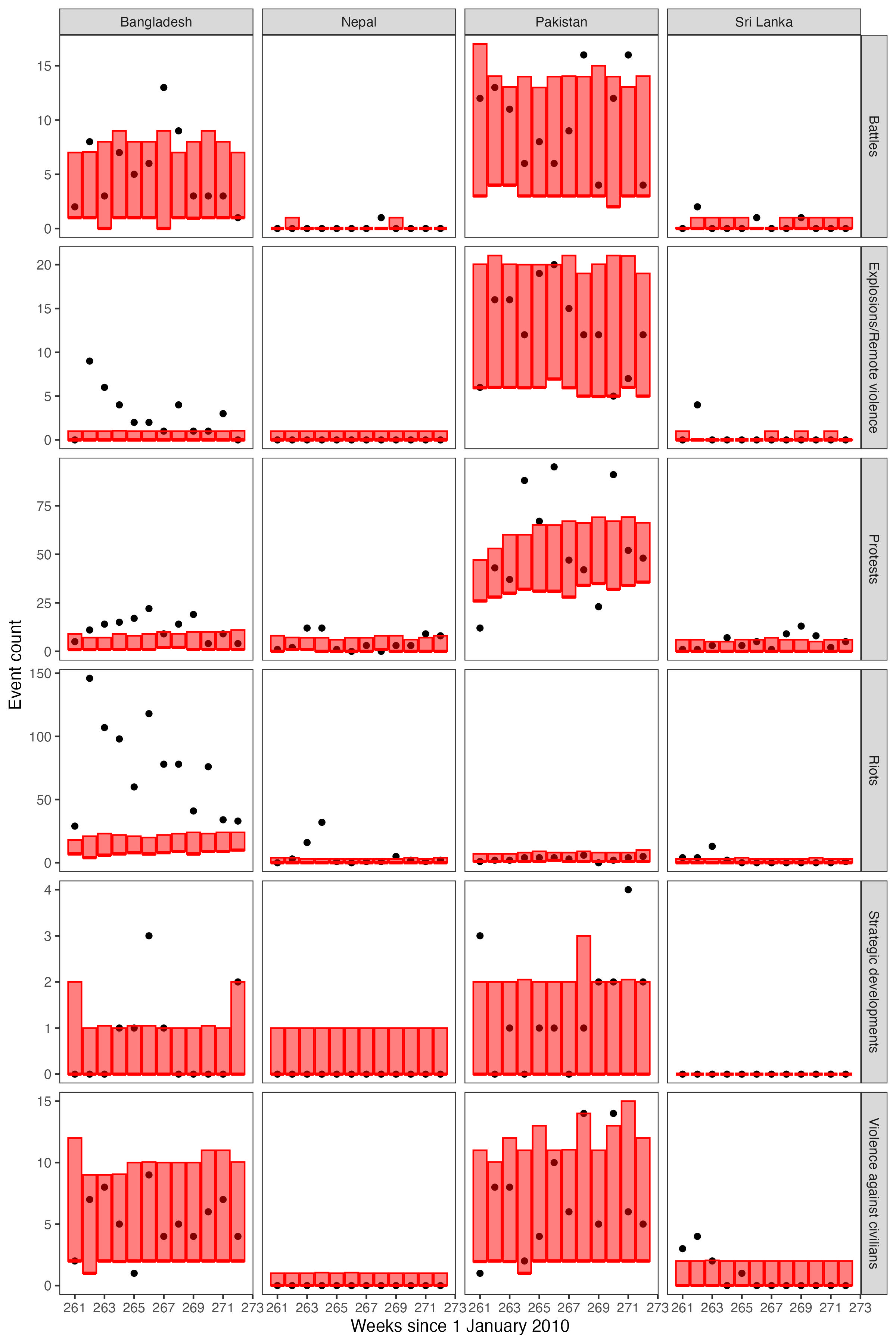}
	\caption{Out of sample predictive checks (aggregated over 3 month prediction interval for each location and conflict type). The black dots show the observed number of events. The red bars show the 95\% interval of simulated events using parameter values from 100 posterior samples. }
\label{fig:val_st_weekly}
\end{figure}

To assess the predictive ability of our model, Figure \ref{fig:val_st_weekly} shows the results from out-of-sample posterior predictive checks. Using 100 posterior samples, we simulated three months of data following from the end of the five year observation window, which corresponds to predicting the first three months of 2015. The results were then summed over locations to obtain event counts for each week for each country and conflict type pair. Overall, the 95\% interval for the predicted number of events usually encapsulates the observed number of events. However there are some larger than expected event counts that are not captured in our prediction intervals, particularly in Bangladesh for Riots, Protests and Explosions/Remote Violence. This is due to the testing period coinciding with a substantial surge in events associated with the 2015 Bangladeshi political crisis \citep{ICGBangladesh}. The 12 months preceding the test period exhibited low levels of conflict activity, meaning that the contribution from the self-exciting component of the intensity function was based on a period of low event counts. Consequently, the model was unable to reproduce the excess events observed during the crisis, as out-of-sample data were not used in generating the forecasts.

\FloatBarrier

\section{Assessing conflict risk in practice} \label{sec:practicals}

Compared to relying on historical moving averages, a key advantage of using our model is the ability to mitigate the impact of noisy data and obtain stable estimates of risk. Our approach is also more interpretable than other statistical methods based on machine learning techniques, and our Bayesian approach provides a more precise understanding of uncertainty. By incorporating the time elapsed since past events and when appropriate considering the influence of distance between events in a sophisticated, yet interpretable way, we can provide a more robust foundation for monitoring risk and developing early warning tools, thus informing the allocation of resources for anticipatory actions.

In this section, we demonstrate through practical examples how our statistical model provides valuable insights that enable stakeholders to make more informed decisions about how to address conflict risks. In particular, we illustrate three types of insights from our modelling approach that may be useful for decision makers: an early warning tool for monitoring trends, visualisations of the spatial risk profile of conflict events, and short term predictions about the future risk of conflict events.

Early warning tools are essential for funding bodies to be able to make decisions on how to allocate resources. While defining the thresholds for these early warnings is beyond the scope of this work, we aim to provide a more nuanced perspective by statistically modelling the trend of events, rather than using simple historical averages. To capture longer-term trends of increasing threat, we adopt a methodology closely based on that used by ACLED in their Trendfinder dashboard, with the necessary modifications made to the time scale and spatial units to directly compare our model. 

The Trendfinder dashboard calculates a rolling 12-month historical average of event counts, and flags time periods where the event count is more than two standard deviations above the rolling average. To benchmark this approach against our model, we instead used the quantile function of a Poisson distribution with mean given by the estimated intensity function from our model to determine time periods with unusually high activity. In particular, the estimated intensity function was calculated using 100 posterior samples of the model parameters. Then, for each posterior sample and corresponding estimate of the intensity function, we obtained the 95\% interval of event counts at each week using the quantile function. Rolling 12-month averages were then taken of the median of these quantiles at each week. Time periods were then flagged if the observed number of events exceeded the rolling average of the 97.5\% quantile.

\begin{figure}[H]
	\centering
	\includegraphics[width=\textwidth]{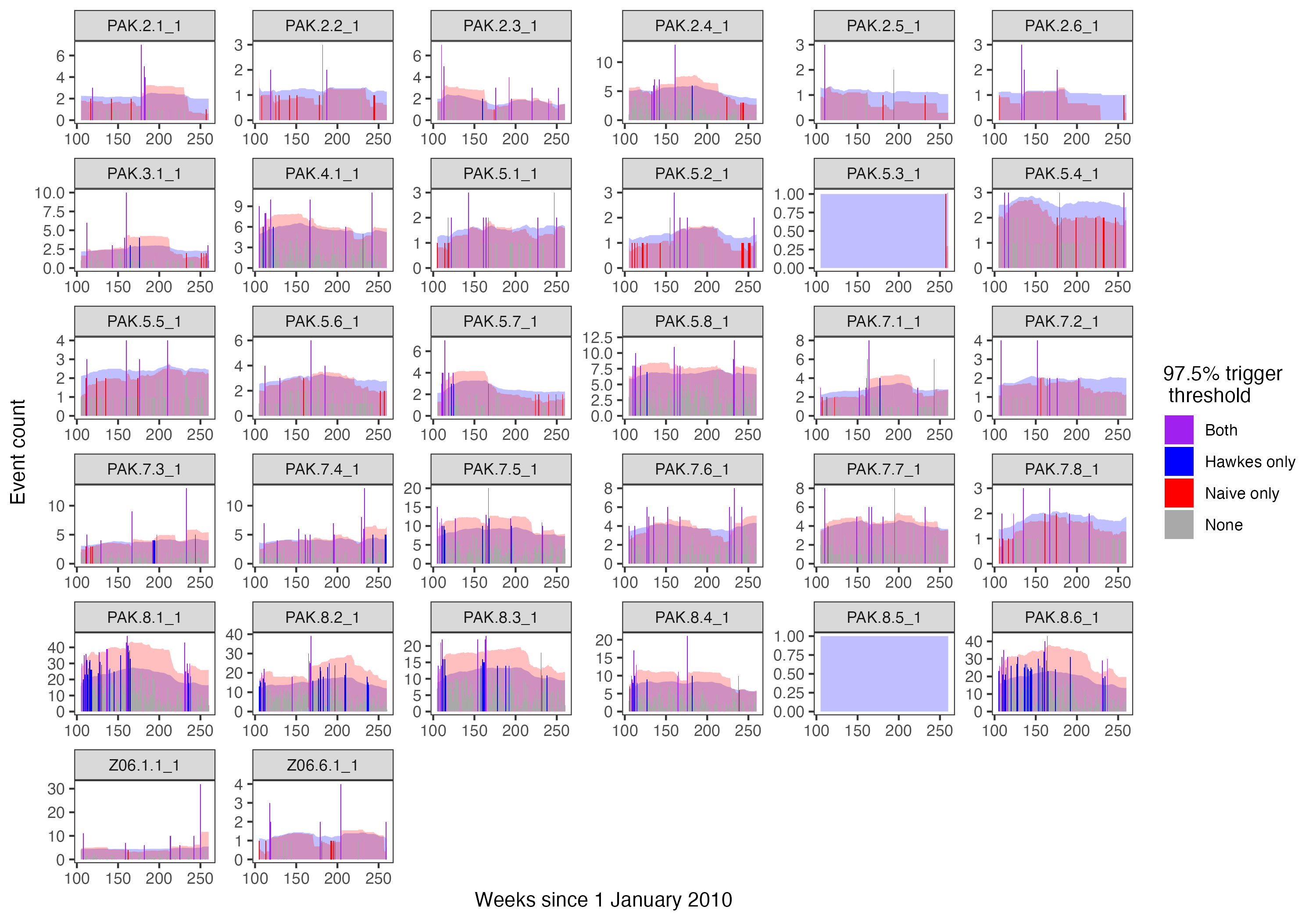}
	\caption{Comparison of early warning triggers for Protests in Pakistan. \\
	\textbf{Grey bars:} no flag. \\\textbf{Blue: flagged via naive model.} Observed counts greater than two standard deviations above 12-month rolling average. \\ \textbf{Red: flagged via proposed Hawkes model.} Observed counts greater than 12-month rolling average of the median 97.5\% quantiles from the Poisson distribution, calculated from 100 posterior samples. \\\textbf{Purple: flagged in both approaches.}}
\label{fig:comp_weekly_hawkes}
\end{figure}

Figure \ref{fig:comp_weekly_hawkes} presents an example of this early warning tool, comparing our modelled risk estimates and a naive approach using simple historical moving averages over the same time period, inspired by the ACLED Trendfinder dashboard. We present here the result for Protests in Pakistan as an example, and the figures for the remaining scenarios are available in the supplementary materials. In general, we find that the upper bounds used as the threshold in the naive approach are highly volatile and sensitive to outliers, often resulting in large, abrupt increases and reductions in the threshold. Our approach has much more consistent intervals and is also more robust to outliers, since these events do not directly contribute to the calculation of the rolling average. For regions with lower event counts, the naive model often identifies weeks with just a single event as having unusually high activity, whereas our model often considers this to fall within the reasonable range of values expected based on the estimated intensity function. For regions with higher event counts, under the naive model, outliers can strongly influence the 12-month rolling average and inflate the threshold for a trigger, whereas our model may still identify such events as anomalous. The 97.5\% trigger threshold was chosen arbitrarily in this example, but this threshold percentage can also be adapted to suit the risk appetite of the user. 

\begin{figure}[H]
    \centering
    \includegraphics[width=\textwidth]{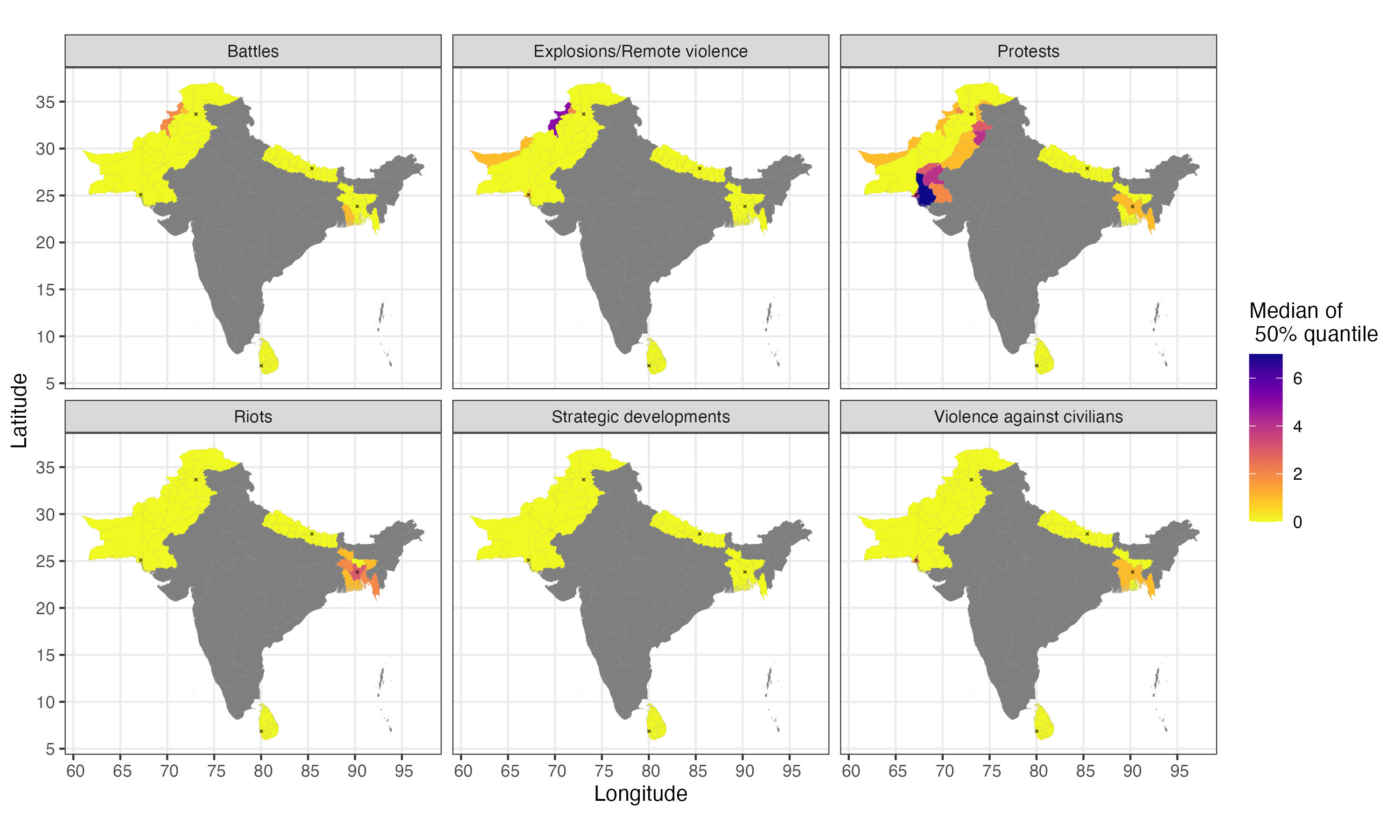}
    \caption{Median of the 50\% quantile, based on 100 posterior samples, at December 2014 (i.e. the end of the observation window). Black crosses indicate capital cities (including Karachi in Pakistan as the biggest city and economic capital). }
\label{fig:risk_st_weekly}
\end{figure}

The visualisation of the spatial risk profile for conflict events is also an important aspect of conflict risk monitoring. As an example, Figure \ref{fig:risk_st_weekly} presents the median of the 50\% quantile, calculated from 100 posterior samples, in the last week of December in 2014, namely the last week of the observation window considered in these analyses. Our modelling can operate at a finer spatial scale than current methods produced by organisations such as ACLED. This granularity can then take into account the unique attributes of each administrative region. Furthermore, the estimates can easily be aggregated up as required to obtain more accurate estimates of a coarser spatial resolution, with corresponding estimates of uncertainty.

\begin{figure}[H]
    \centering
    \includegraphics[width=\textwidth]{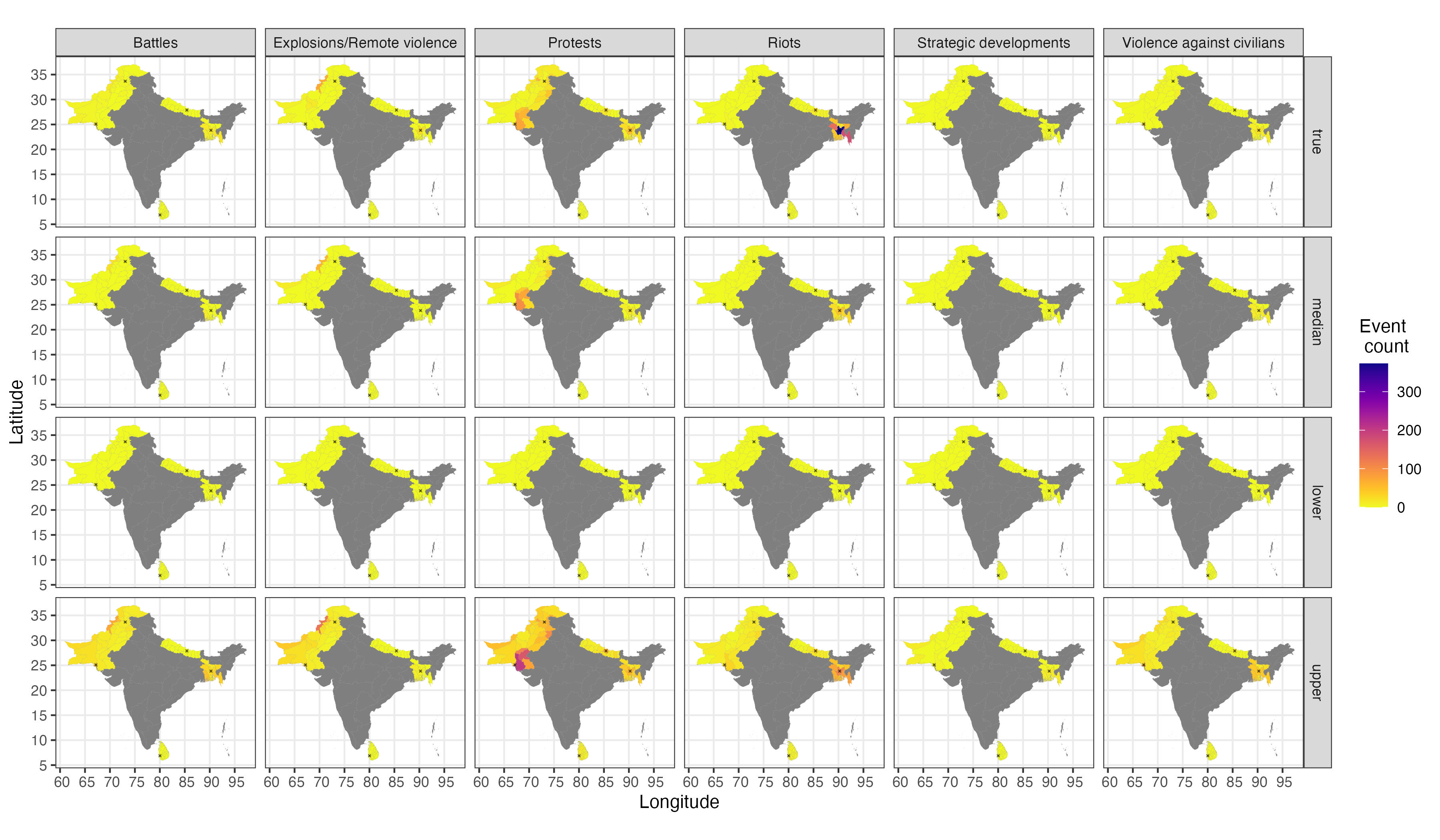}
    \caption{\textbf{Out-of-sample 3-month-ahead predictions.} \\
    First row: observed cumulative event counts. \\
    Second row: median of cumulative event counts.\\
    Third row: 2.5\% quantile of cumulative event counts. \\
    Fourth row: 97.5\% quantile of cumulative event counts. \\
    Black crosses indicate capital cities (including Karachi in Pakistan as the biggest city and economic capital). }
\label{fig:spat_pred_st_weekly}
\end{figure}

Another notable advantage of adopting a statistical model for analysing conflict risk is the ability to generate short-term predictions of future events. In addition to estimating the level of future risk, our Bayesian approach also produces estimates of uncertainty surrounding the predicted number of events. In Section \ref{sec:results}, we used 100 posterior samples to predict the expected number of events in the first 3 months of 2015, namely 3 months from the end of our modelling observation window. Figure \ref{fig:spat_pred_st_weekly} presents the 2.5\%, 50\% and 97.5\% quantiles of the cumulative number of events predicted over these 3 months. Across most regions, the projected risk level for these three months is relatively low. However, there are numerous hotspots, particularly in Pakistan and Bangladesh. These spatial predictions, along with their uncertainty intervals, provide valuable insights into the distribution and intensity of anticipated conflict events, enabling policymakers and stakeholders to prioritise resources and interventions accordingly.

\FloatBarrier

\section{Discussion} \label{sec:discussion}

This work addressed three main questions about spatiotemporal DTHPs: (i) are they useful for modelling conflict risk, (ii) what can we learn about conflict events using these models, and (iii) how can these models be used to improve current practices for conflict monitoring. Our main contributions and the results, and how they relate to these questions, are first summarised. This is followed by a discussion of the limitations and future work, and finally concluding remarks.

\subsection{Summary}

In this analysis we presented a unified framework using spatiotemporal DTHPs to analyse both the short and long term trends of conflict risk. Through the Bayesian framework, we can gain a more precise understanding of the uncertainty inherent in our estimates of conflict risk, compared to other approaches based on machine learning techniques and optimisation methods such as maximum likelihood estimation. To our knowledge this is the first analysis of its kind applied to the ACLED database, and the first to characterise and compare the spatial and temporal dynamics of the various conflict types between each other and also among neighbouring countries. We also demonstrated from the practitioners' perspective the advantages of our approach and examples of how to interact with our model to gain valuable insights.

In response to our first research question, we find that spatiotemporal DTHPs are indeed a useful tool for policy makers to model and monitor conflict events. Our model was fitted to each of the individual countries within South Asia, and broken down further by the conflict type. The resulting models performed well allowing us to capture and understand the self-exciting nature of conflict in the short term. We also performed out-of-sample predictive checks, which showed that our model was generally capable of producing a reasonable range of estimates for the predicted number of events over a 3 month period, with the exception of some extreme cases. 

There are several key learnings from using spatiotemporal DTHPs for modelling conflict events that address our second question. While other modelling approaches can also provide a measure of conflict risk, interpretation of the parameters in our model offers unique insights. The level of independent events and those that are caused by past events can be quantified through the baseline and magnitude parameters respectively. This can help to inform preventative measures, for example where the baseline rate of events is high, there should be a focus on preventing new events from occurring. Whereas, if the magnitude parameter is high, there should be focus on monitoring ongoing events. We also fitted several types of spatiotemporal DTHP and investigated whether a spatially varying baseline rate alone was sufficient to the characterise spatial dynamics in the data, or if spatial excitation was a factor in driving conflict events. In doing so, we quantified the effect that time and distance have on the risk of future conflict events by explicitly parameterising the temporal and spatial patterns of the process through their respective triggering kernels. We then used these kernels to compare the self-exciting characteristics of the various conflict types and countries to further deepen our understanding of these processes. 

To address our final research question, we provided specific examples of how to use and adopt our models in practice and benchmark against current practices. We demonstrated how an early warning tool, used to monitor longer-term trends in the data and thus inform anticipatory actions, can be developed based on our statistical model. Specifically, this tool identifies when the observed level of events exceeds a threshold percentile determined by the posterior distribution, and we showed that our model is more stable and robust than the alternative based on historical averages. Visualisations for the current level of conflict risk estimated from our model were also presented. Our modelling can be performed at a finer spatial resolution compared to similar analyses, allowing for more accurate estimates of conflict risk, even when aggregating into coarser spatial regions. Lastly, our statistical modelling approach allows us to make short-term predictions, and calculate the corresponding uncertainty in these estimates. 

Our model provides a more sophisticated outlook on conflict risk compared to relying on historical averages, while remaining interpretable. Compared to ACLED's new prediction tool for conflict activity, CAST, the outputs of our models can be easily interpreted to quantify specific characteristics of conflict. Currently ACLED uses a combination of techniques to understand conflict, namely historical moving averages to analyse both short and long term trends and machine learning models for prediction. However, our model can fulfil both roles in one cohesive framework, meaning it could be a useful addition to the toolkit for conflict monitoring. 

\subsection{Limitations and future work}   \label{sec:st-diss-fw}

A major direction for future work is connecting the modelling of conflict risk with the human impact of conflict. To achieve this, covariates for the number of fatalities or other economic or demographic variables could be included in the model to inform and refine parameter estimates. An alternative approach is to jointly model the number of events and the number of people impacted by each of these events. The financial impact of conflict events, and how this corresponds with the type of anticipatory actions taken could also be explored. 

Given sufficient computational resources or a more computationally efficient inference method, a global, multivariate model \citep{Hawkes:1971vr, Li:2024} for conflict events could be estimated to characterise the interactions between countries and regions. Alternatively, if interactions were less important, and instead similarities between regions were of interest, univariate studies could be conducted in parallel and the estimated parameters grouped via a clustering algorithm to determine countries or regions with similar risk profiles. Furthermore, one could extend the spatiotemporal DTHP to a marked process whereby the type of conflict event is a mark which also influences the excitation dynamics \citep{Ogata:1988jd,DiLoro:2025}. A more complete option which allows for different triggering kernels for each pair of conflict types is a multivariate model between conflict types to capture and quantify dependencies between each of the conflict types. The level of risk identified in our analysis could also be compared to other trends, for example economic or demographic time series to determine any interactions.

In the present work, we incorporated a spatially varying baseline rate to better capture spatial heterogeneity in event dynamics. This extension allows for more realistic modelling of the underlying spatial structure while maintaining computational feasibility. Several further developments could be explored in future work. First, it may be possible to extend branching structure–based approaches \citep{DiLoro:2025,Dangelo:2024,Fox:2016,Zhuang:2018} to the discrete-time Hawkes framework, though this remains technically challenging due to the absence of a clear branching interpretation in discrete time. Additionally, the baseline rate could be refined through regression-based adjustments using covariates such as population density or other socio-demographic factors of interest. Alternative modelling strategies include adopting a multivariate (regional) formulation as discussed above or employing Gaussian Process-based methods, such as the Log Cox Gaussian Process \citep{Miscouridou:2023}, for encoding spatial variation.

In this study, Euclidean centroid-to-centroid distances based on longitude and latitude were used as a measure of spatial separation. Given that the study countries lie relatively close to the equator and the administrative units are generally small, this approximation is reasonable for our purposes. However, we acknowledge that this approach becomes less appropriate in settings involving larger spatial domains or regions at latitudes further from the equator. In such contexts, projections to suitable coordinate reference systems, for example UTM, or the use of other distances, such as Haversine distances, would provide more accurate distance representations, although care is needed when countries span multiple UTM zones. Furthermore, centroid-based distances may oversimplify interactions between neighbouring areas, and alternatives such as adjacency-based structures or shared border lengths could offer more meaningful representations of spatial connectivity. Finally, extensions of continuous-space methodologies, such as those used in the discrete approximations of LGCPs \citep{Li:2012,Johnson:2019}, may offer a route toward more flexible spatial interaction modelling within the Hawkes process framework. 

One phenomenon observed throughout this analysis was that the proposed model does not always capture more extreme event counts. However, if it were reasonable to assume that these excess counts have some inherent structure, some modifications could be made to the model to better represent this range in values. For example, one could incorporate an over-dispersion term by adopting a negative binomial likelihood as opposed to the Poisson likelihood. Another approach could use a hurdle model \citep{Porter:2012fh}, that models a binary response for whether there will be a non-zero number of events, and the magnitude of events given a non-zero count. Further extensions could be considered to accommodate excess counts under this framework, such as a hidden Markov model or a simple mixture model to capture varying levels of intensity.

Due to the model formulation there is a systematic delay in the modelled risk estimates whereby events that occur in the current time period can only influence the intensity of the process in the subsequent time period. To offset this delay, a potential future direction could consider covariates derived from topic modelling of newspaper articles \citep{Mueller:2022} or other exogenous, real-time covariates. This information could mitigate this delayed eﬀect by allowing the model to respond to external drivers that correlate with imminent changes in event frequency.
    
The level of discretisation, both spatial and temporal, is also an important consideration in the modelling. The inference is more stable given a coarser discretisation as it avoids complications arising from excessive zero-inflated event counts, and it also reduces the computational expense of the inference. However, there is an approximation error introduced in the discrete-time model that assumes events that occur in the same interval are independent, and thus cannot have been produced by one another. Therefore, if the discretisation is too coarse, there is the potential to artificially inflate the baseline rate. The discretisation error is explored for temporal Hawkes processes in \cite{Kirchner:2016gz}. Further work could be undertaken to explore the impact of this approximation error, and to determine the optimal discretisation level for a given dataset. The same arguments also apply to the spatial aggregation, however in this case study we believe the administrative boundaries used are sensible given the implicit modelling assumption of homogeneity between all events that occur in the same region. \cite{Siviero2024} also investigate this discretisation error for spatiotemporal Hawkes processes. 

In this work we adopted general forms of the modelling assumptions across all scenarios, such as the choice of triggering kernels and aggregation levels, to assess the suitability of spatiotemporal DTHPs for analysing conflict activity. This analysis can then inform specific scenarios which may require further refinement of the model assumptions; for example, a different temporal aggregation may be more appropriate depending on the research question. We expect that it will be useful for users to be able to tailor these assumptions to their specific scenarios. As such, a more automated and streamlined workflow for determining the key modelling decisions, such as the form of $\lambda$, maximum time lag, choice of kernels, temporal and spatial aggregation level and choice of likelihood distribution, should be developed. It would also be helpful for assessing goodness-of-fit to extend the existing, formal residual analyses available for continuous-time Hawkes processes \citep{Ogata:1988jd,Daley:2006} to the discrete-time setting.

\subsection{Conclusion}

Overall, a better statistical model of both shorter and longer term trends in the risk of conflict can lead to more accurate predictions, more effective policy interventions and enhanced early warning systems. These benefits can ultimately help to prevent conflicts and minimise their impact on individuals and communities. To our knowledge, we propose the first spatiotemporal Hawkes model for conflict data in the Bayesian setting, with full characterisation of the spatial and temporal triggering kernels. While this article explores its use among countries in South Asia, the methodology could easily be extended to other countries or continents to provide an estimate of the spatiotemporal risk of conflict events worldwide. We propose South Asia as an intrinsically meritorious study and as an important exemplar of how these models can be used. 

Our proposed model provides a means for up to date estimates of the risk of conflict events. Our model can provide reasonable estimates of both short and long term trends in the occurrence of conflict events and also provide short-term predictions, determined using the posterior predictive distribution. This is also the first study we are aware of that captures and compares the characteristics of conflict events across different countries and types of events, and that demonstrates the advantage of using a statistical model to monitor conflict risk, compared to relying on historical averages. These features make this work a useful tool for actors in the humanitarian and social sciences sectors to better understand these risks and make more informed decisions. 

This work paves the way for new explorations in spatiotemporal modelling for conflict events. Our desire is that this work advances the current state of conflict risk modelling by providing a nuanced foundation for which anticipatory action and financing decisions can be based. We hope that it enables further discussions with key stakeholders to enable more sophisticated and data driven risk estimates of conflict events worldwide.

\FloatBarrier

\newpage 

% References and appendices

\newpage



\section*{Supplementary material}

# A  Analysis of spatiotemporal DTHPs

Here we aim to quantify the effect of incorporating a spatial triggering kernel into the DTHP by comparing a spatiotemporal model to one with only a temporal element. First, the maximum likelihood estimate for each model for all countries and conflict types in the data was estimated. We then compared the respective root mean squared error (RMSE) and Bayesian information criterion (BIC) for the two models. The RMSE was calculated via the square of the difference between $y$, the observed number of events, and $\lambda$, the estimated expected number of events given by the intensity function evaluated at the MLE, averaged over all event locations and times for the duration of the 5-year observation window considered in this study. Let $k$ be an index denoting each unique time and location pairing $(t, x, y)$. Then the RMSE for a particular country and conflict type has the form,

$$\text{RMSE} = \sqrt{\frac{\sum_{k=1}^{n}(y_k - \lambda_k)^2}{n}} \qquad (7)$$

where $n$ is equal to the product of the total number of locations and time points in the sample. This RMSE is a measure of predictive risk and represents the average residual value for each country and conflict type pair over all spatial regions and times.

To construct a DTHP that does not account for spatial dependencies, while enabling direct comparison to the spatiotemporal alternative defined by the intensity function (3), we fix the spatial kernel $h(\cdot) = 1$ and restrict events that contribute to the intensity function to those occurring in the same region. The temporal comparison model then has a conditional mean function given by,

$$\lambda(t, x, y) = \mu + \alpha \sum_{i:t_i<t} y_(t_i, x, y) g(t - t_i). \qquad (8)$$

Table 1 compares the respective RMSE and BIC, defined in (7), that were obtained from the temporal and spatiotemporal models.

| Country | Conflict Type | RMSE | | BIC | |
|---|---|---|---|---|---|
| | | Temporal | Spatiotemporal | Temporal | Spatiotemporal |
| Bangladesh | Battles | 0.71 | 0.71 | 5071.95 | 5068.44 |
| | Explosions/ remote violence | 0.31 | 0.30 | 827.48 | 827.18 |
| | Protests | 1.35 | 1.35 | 6097.99 | 6102.04 |
| | Riots | 2.17 | 2.17 | 11002.97 | 11014.16 |
| | Strategic developments | 0.28 | 0.28 | 1108.90 | 1102.70 |
| | Violence against civilians | 0.87 | 0.87 | 6379.67 | 6383.71 |
| Sri Lanka | Battles | 0.15 | 0.13 | 195.51 | 190.77 |
| | Explosions/ remote violence | 0.14 | 0.13 | 310.33 | 308.29 |
| | Protests | 0.38 | 0.38 | 3942.13 | 3954.76 |
| | Riots | 0.20 | 0.20 | 2307.79 | 2290.09 |
| | Strategic developments | 0.11 | 0.11 | 132.65 | 136.70 |
| | Violence against civilians | 0.17 | 0.17 | 1222.23 | 1224.40 |
| Nepal | Battles | 0.21 | 0.21 | 170.64 | 173.66 |
| | Explosions/ remote violence | 0.44 | 0.42 | 576.15 | 579.24 |
| | Protests | 2.69 | 2.69 | 2884.76 | 2889.93 |
| | Riots | 2.16 | 2.17 | 2606.76 | 2605.53 |
| | Strategic developments | 0.47 | 0.46 | 554.72 | 525.04 |
| | Violence against civilians | 0.62 | 0.61 | 870.54 | 845.14 |
| Pakistan | Battles | 1.84 | 1.84 | 4426.67 | 4430.71 |
| | Explosions/ remote violence | 2.57 | 2.57 | 4934.38 | 4938.43 |
| | Protests | 12.19 | 12.20 | 14655.04 | 14670.93 |
| | Riots | 1.47 | 1.47 | 3732.09 | 3723.38 |
| | Strategic developments | 0.41 | 0.41 | 1046.76 | 1050.82 |
| | Violence against civilians | 2.13 | 2.13 | 4092.02 | 4096.07 |

Table 1: BIC and RMSE between the observed event counts and the estimated number of events from a temporal model versus a spatiotemporal model.

# B  Daily temporal aggregation

We initially considered a daily temporal aggregation of the spatiotemporal DTHP in Bangladesh from the period 2010 – 2011, for which the intensity function is given by (3). In this section, to obtain the set of locations $\mathcal{S}$, a naive discretisation was performed whereby the spatial boundaries in the data for each country were divided evenly into a $20 \times 20$ grid. The model parameters were then estimated via maximum likelihood estimation. The parameter estimates are presented in Table 2.

| Conflict type | $\mu$ | $\alpha$ | $\beta$ | $\sigma$ |
|---|---|---|---|---|
| Battles | 0.0054 | 0.1862 | 0.092 | 0.1784 |
| Protests | 0.0069 | 0.6664 | 0.0513 | 0.0993 |
| Riots | 0.0086 | 0.5989 | 0.0583 | 0.0901 |
| Violence against civilians | 0.0057 | 0.2092 | 0.1108 | 0.1624 |
| Strategic developments | 0 | 0 | 0.2665 | 0 |
| Explosions/Remote violence | 0.0058 | 0.0344 | 0.989 | 0.0269 |

Table 2: MLEs for spatiotemporal DTHP with daily temporal aggregation in Bangladesh

Figure 14 presents the observed events on day $t$ versus the estimated mean number of events on day $t$, grouping together events on the same day at different locations. We find that the estimated number of events on each day closely follows the trend of the observed data, but some larger residuals are observed due to excess zero counts and also larger than expected daily counts.

Figure 15 assesses the predictive capability of the proposed model using out of sample predictive checks, using the same approach as in Section 3. It is difficult to compare these predictions to the actual number of events due to an excess number of zero counts, since most locations over the three-week testing window experienced few or no events. In most cases, the 80% prediction interval spans from 0 to 1 events, whereas in reality we know that most of these locations indeed experienced no events.

A Bayesian version of this spatiotemporal model was also considered at the daily temporal aggregation level using `Stan`. A large proportion of the model runs for each of the conflict types did not converge and had divergent transitions, indicating that Stan could not find algorithm settings that allowed the posterior to be explored, while not failing. This was likely due to model misspecification. There are many regions at each time increment that experienced no events, and the likelihood function used does not account for the level of excess number of zero counts present in these data.

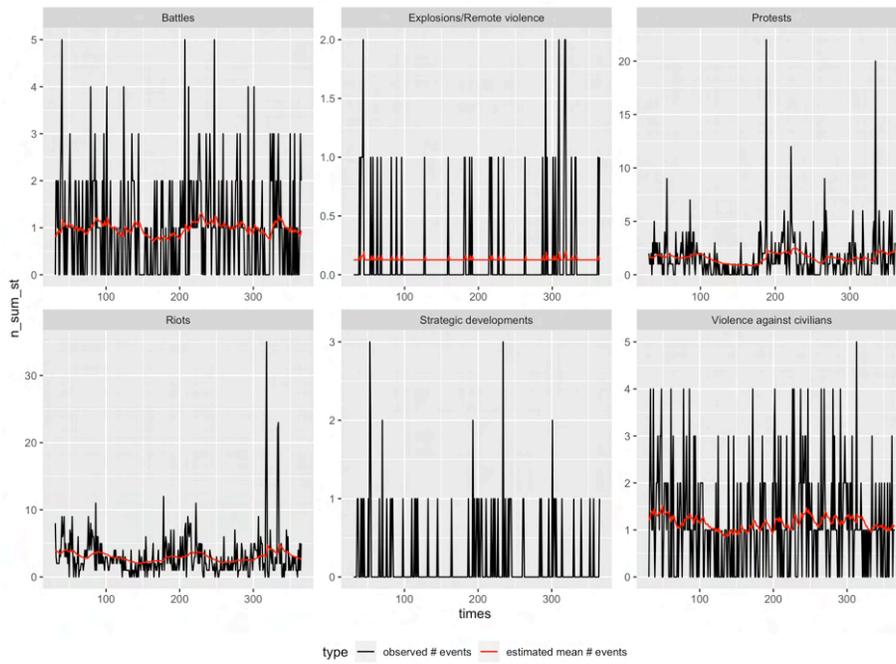

Figure 14: Spatiotemporal DTHP by conflict type for Bangladesh. Observed data on day $t$ (black line) versus estimated $\lambda(t)$ (red line)

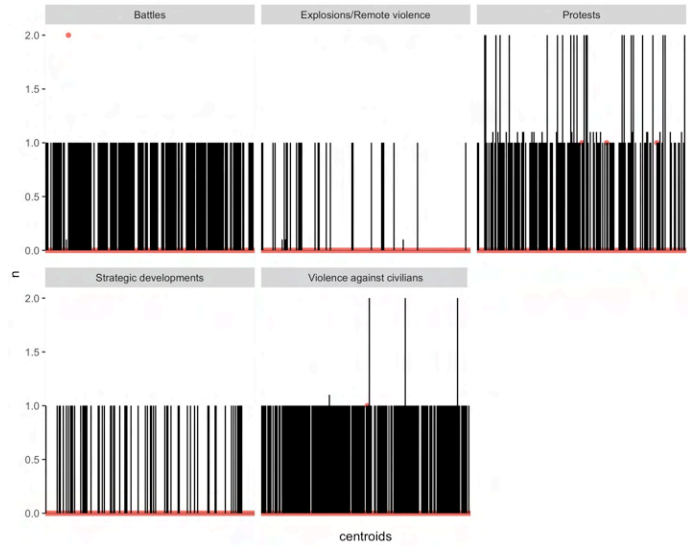

(a) Grouped by location and conflict type

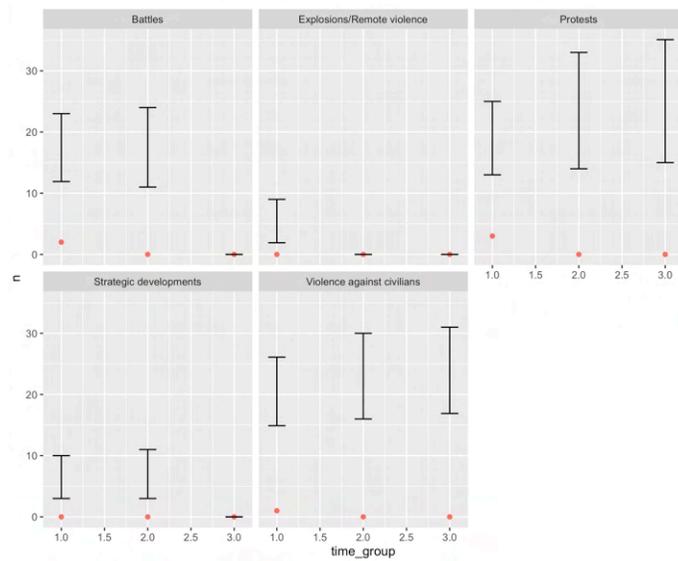

(b) Grouped by months and conflict type

Figure 15: Out of sample predictive checks. The red dots show the observed number of events. The bars show the 80% interval of simulated events using the MLEs (from 100 simulations).

37

# C Population model

Population data by administrative boundaries was obtained from WorldPop [2022]. We incorporated population into the model by introducing a population indicator defined by a binary variable indicating areas with "Low" and "High" population density. The population density was calculated by taking the total population in a given region relative to the area of the administrative boundary. The indicator was then determined by whether the population density for a particular region with centroid $(x, y)$ fell below or above a given quantile of the population distribution for each country. To determine an appropriate threshold, each octile from $[0.5, 1)$ was considered. The root mean squared error (RMSE) for each of these thresholds was calculated using the approach outlined in Section A. While the difference in the various thresholds was not significant, the highest threshold of 87.5% had the lowest RMSE in the most cases. However, under this scenario there were several country and conflict type pairs that did not converge, indicating this more extreme threshold is less robust for all scenarios. Thus, the next lowest threshold of 75% was selected.

We assume that the population of a region could impact both the baseline and self-exciting components of the process, and thus we included population based parameters for the baseline rate $\mu$ and the magnitude parameter $\alpha$. The model then has the following rate parameter $\lambda(t, x, y)$ for a particular time interval $t$ and location with coordinates $(x, y)$,

$$\lambda(t, x, y) = (\mu_{\text{low}}\mathbb{I}_{\text{low}}(t, x, y) + \mu_{\text{high}}\mathbb{I}_{\text{high}}(t, x, y)) \\ + \sum_{(t_i, x_i, y_i): t_i < i} (\alpha_{\text{low}}\mathbb{I}_{\text{low}}(t, x, y) + \alpha_{\text{high}}\mathbb{I}_{\text{high}}(t, x, y))g(t - t_i)h(x - x_i, y - y_i) \quad (9)$$

where $\mathbb{I}_{\text{low}}$ and $\mathbb{I}_{\text{high}}$ indicate whether the location $(x, y)$ is a region of low or high population density, respectively, at time interval $t$.

Here we compare the spatiotemporal models, namely the models that include and exclude population density information. For both of these models, RMSE is used to compare the observed event counts with the expected number of events, estimated via maximum likelihood estimation.

Table 3 presents the RMSE for the population and non-population models. There are several scenarios for which the model including population density information outperforms the model excluding population. However, there are also some cases for which the model excluding population performs better, and the differences in RMSE for all of these scenarios are often negligible. Moreover, in the majority of cases the RMSE is equal under both models. This suggests that for some scenarios, the particular population density threshold that was selected for classifying an area as high or low density may be useful, but a single threshold for all scenarios is inappropriate. Overall, it seems there is value in incorporating population density information into the model, however the full benefit is perhaps not being realised due to the general and coarse discretisation applied in these

analyses. Thus we leave this to future work, and excluded population density for the remainder of the analysis in this article.

| Country | Conflict type | Non-population | Population |
|---|---|---:|---:|
| Bangladesh | Battles | 0.71 | 0.71 |
| | Explosions/Remote violence | 0.30 | 0.30 |
| | Protests | 1.35 | **1.31** |
| | Riots | 2.17 | **2.15** |
| | Strategic developments | 0.28 | 0.28 |
| | Violence against civilians | **0.87** | 0.88 |
| Sri Lanka | Battles | 0.13 | 0.13 |
| | Explosions/Remote violence | 0.13 | 0.13 |
| | Protests | 0.38 | **0.37** |
| | Riots | 0.20 | 0.20 |
| | Strategic developments | 0.11 | 0.11 |
| | Violence against civilians | 0.17 | 0.17 |
| Nepal | Battles | 0.21 | 0.21 |
| | Explosions/Remote violence | 0.42 | 0.42 |
| | Protests | 2.69 | **2.65** |
| | Riots | 2.17 | **2.14** |
| | Strategic developments | 0.46 | 0.46 |
| | Violence against civilians | 0.61 | 0.61 |
| Pakistan | Battles | 1.84 | 1.84 |
| | Explosions/Remote violence | 2.57 | 2.57 |
| | Protests | **12.20** | 12.25 |
| | Riots | 1.47 | **1.46** |
| | Strategic developments | 0.41 | 0.41 |
| | Violence against civilians | 2.13 | **2.11** |

Table 3: RMSE between the estimated number of events using the maximum likelihood model and observed number of events for the models with and without population density

# D  Output from ARIMA analysis

To determine the extent of temporal autocorrelation, and thus set the maximum excitation time in our model, ARIMA models were fit for all country and conflict type combinations. The output from these analysis are below, and show that the maximum lag for any scenario is less than 3 months. The residuals were also checked for each of these models. There are some irregularities but they are generally within threshold limits.

```
[1] "Bangladesh"
[1] "Battles"
Series: tseries[[c]][[t]]
ARIMA(0,1,1) with drift

Coefficients:
          ma1    drift
       -0.8187  -0.2719
s.e.    0.0815   0.1552

sigma^2 = 37.93:  log likelihood = -190.51
AIC=387.01   AICc=387.45   BIC=393.25


Ljung-Box test

data:  Residuals from ARIMA(0,1,1) with drift
Q* = 8.2276, df = 9, p-value = 0.5114

Model df: 1.   Total lags used: 10

Saving 9.14 x 8.31 in image
[1] "Bangladesh"
[1] "Explosions/Remote violence"
Series: tseries[[c]][[t]]
ARIMA(0,1,1)

Coefficients:
           ma1
       -0.8795
s.e.    0.0671
```

```
sigma^2 = 5.08:  log likelihood = -131.9
AIC=267.8   AICc=268.02   BIC=271.96


Ljung-Box test

data:  Residuals from ARIMA(0,1,1)
Q* = 6.8066, df = 9, p-value = 0.6572

Model df: 1.   Total lags used: 10


Saving 9.14 x 8.31 in image
[1] "Bangladesh"
[1] "Protests"
Series: tseries[[c]][[t]]
ARIMA(0,1,2)

Coefficients:
          ma1      ma2
      -0.2114  -0.3613
s.e.   0.1284   0.1422

sigma^2 = 263.2:  log likelihood = -247.3
AIC=500.6   AICc=501.03   BIC=506.83


Ljung-Box test

data:  Residuals from ARIMA(0,1,2)
Q* = 11.505, df = 8, p-value = 0.1747

Model df: 2.   Total lags used: 10


Saving 9.14 x 8.31 in image
[1] "Bangladesh"
[1] "Riots"
Series: tseries[[c]][[t]]
ARIMA(1,0,0) with non-zero mean
```

```
Coefficients:
         ar1      mean
      0.6564   85.8648
s.e.  0.0961   13.5511

sigma^2 = 1431:  log likelihood = -302.38
AIC=610.76   AICc=611.18   BIC=617.04


Ljung-Box test

data:  Residuals from ARIMA(1,0,0) with non-zero mean
Q* = 12.612, df = 9, p-value = 0.181

Model df: 1.   Total lags used: 10


Saving 9.14 x 8.31 in image
[1] "Bangladesh"
[1] "Strategic developments"
Series: tseries[[c]][[t]]
ARIMA(0,1,2)

Coefficients:
         ma1      ma2
      -0.547  -0.2125
s.e.   0.135   0.1391

sigma^2 = 6.124:  log likelihood = -136.53
AIC=279.07   AICc=279.51   BIC=285.3


Ljung-Box test

data:  Residuals from ARIMA(0,1,2)
Q* = 4.9268, df = 8, p-value = 0.7654

Model df: 2.   Total lags used: 10
```

```
Saving 9.14 x 8.31 in image
[1] "Bangladesh"
[1] "Violence against civilians"
Series: tseries[[c]][[t]]
ARIMA(1,1,1)

Coefficients:
         ar1      ma1
      0.3387  -0.9037
s.e.  0.1531   0.0852

sigma^2 = 85.62:  log likelihood = -214.51
AIC=435.03   AICc=435.47   BIC=441.26


	Ljung-Box test

data:  Residuals from ARIMA(1,1,1)
Q* = 1.7261, df = 8, p-value = 0.9883

Model df: 2.   Total lags used: 10

Saving 9.14 x 8.31 in image
[1] "Nepal"
[1] "Battles"
Series: tseries[[c]][[t]]
ARIMA(0,1,1)

Coefficients:
          ma1
       -0.8991
s.e.   0.0661

sigma^2 = 0.7275:  log likelihood = -74.65
AIC=153.31   AICc=153.52   BIC=157.46


	Ljung-Box test
```

```
data:  Residuals from ARIMA(0,1,1)
Q* = 6.0984, df = 9, p-value = 0.73

Model df: 1.   Total lags used: 10


Saving 9.14 x 8.31 in image
[1] "Nepal"
[1] "Explosions/Remote violence"
Series: tseries[[c]][[t]]
ARIMA(1,0,0) with non-zero mean

Coefficients:
         ar1    mean
      0.1966  1.5415
s.e.  0.1259  0.3639

sigma^2 = 5.346:  log likelihood = -134.43
AIC=274.86   AICc=275.29   BIC=281.14

Ljung-Box test

data:  Residuals from ARIMA(1,0,0) with non-zero mean
Q* = 2.3461, df = 9, p-value = 0.9847

Model df: 1.   Total lags used: 10


Saving 9.14 x 8.31 in image
[1] "Nepal"
[1] "Protests"
Series: tseries[[c]][[t]]
ARIMA(0,0,0) with non-zero mean

Coefficients:
         mean
      18.9667
s.e.   2.2113
```

```
sigma^2 = 298.4:  log likelihood = -255.58
AIC=515.16   AICc=515.38   BIC=519.35


Ljung-Box test

data:  Residuals from ARIMA(0,0,0) with non-zero mean
Q* = 7.6303, df = 10, p-value = 0.6649

Model df: 0.   Total lags used: 10

Saving 9.14 x 8.31 in image
[1] "Nepal"
[1] "Riots"
Series: tseries[[c]][[t]]
ARIMA(0,0,1) with non-zero mean

Coefficients:
         ma1     mean
      0.2892  13.7135
s.e.  0.1245   3.0635

sigma^2 = 353.1:  log likelihood = -260.16
AIC=526.33   AICc=526.76   BIC=532.61


Ljung-Box test

data:  Residuals from ARIMA(0,0,1) with non-zero mean
Q* = 3.7622, df = 9, p-value = 0.9264

Model df: 1.   Total lags used: 10

Saving 9.14 x 8.31 in image
[1] "Nepal"
[1] "Strategic developments"
Series: tseries[[c]][[t]]
ARIMA(0,0,1) with non-zero mean
```

```
Coefficients:
         ma1    mean
      0.3103  1.3828
s.e.  0.1229  0.5853

sigma^2 = 12.48:  log likelihood = -159.9
AIC=325.79   AICc=326.22   BIC=332.08


    Ljung-Box test

data:  Residuals from ARIMA(0,0,1) with non-zero mean
Q* = 1.552, df = 9, p-value = 0.9967

Model df: 1.   Total lags used: 10


Saving 9.14 x 8.31 in image
[1] "Nepal"
[1] "Violence against civilians"
Series: tseries[[c]][[t]]
ARIMA(1,0,0) with non-zero mean

Coefficients:
         ar1    mean
      0.3058  2.9415
s.e.  0.1225  0.9023

sigma^2 = 24.71:  log likelihood = -180.38
AIC=366.76   AICc=367.19   BIC=373.05


    Ljung-Box test

data:  Residuals from ARIMA(1,0,0) with non-zero mean
Q* = 3.6536, df = 9, p-value = 0.9327

Model df: 1.   Total lags used: 10


Saving 9.14 x 8.31 in image
```

```
[1] "Pakistan"
[1] "Battles"
Series: tseries[[c]][[t]]
ARIMA(1,0,2) with non-zero mean

Coefficients:
         ar1      ma1     ma2     mean
      0.5513  -0.1805  0.4891  39.9163
s.e.  0.1690   0.1572  0.1440   5.1788

sigma^2 = 213:  log likelihood = -244.42
AIC=498.83   AICc=499.95   BIC=509.31


	Ljung-Box test

data:  Residuals from ARIMA(1,0,2) with non-zero mean
Q* = 2.4517, df = 7, p-value = 0.9307

Model df: 3.   Total lags used: 10

Saving 9.14 x 8.31 in image
[1] "Pakistan"
[1] "Explosions/Remote violence"
Series: tseries[[c]][[t]]
ARIMA(2,0,2) with non-zero mean

Coefficients:
          ar1      ar2     ma1     ma2     mean
      -0.3002  -0.5983  0.4515  0.9662  60.8505
s.e.   0.1545   0.1528  0.0888  0.1719   3.1316

sigma^2 = 398.1:  log likelihood = -263.47
AIC=538.94   AICc=540.52   BIC=551.51


	Ljung-Box test

data:  Residuals from ARIMA(2,0,2) with non-zero mean
```

```
Q* = 4.1697, df = 6, p-value = 0.6537

Model df: 4.   Total lags used: 10

Saving 9.14 x 8.31 in image
[1] "Pakistan"
[1] "Protests"
Series: tseries[[c]][[t]]
ARIMA(1,0,0) with non-zero mean

Coefficients:
         ar1       mean
      0.3387   277.0373
s.e.  0.1217    26.3073

sigma^2 = 19077:  log likelihood = -379.87
AIC=765.73   AICc=766.16   BIC=772.02

Ljung-Box test

data:  Residuals from ARIMA(1,0,0) with non-zero mean
Q* = 4.6214, df = 9, p-value = 0.866

Model df: 1.   Total lags used: 10

Saving 9.14 x 8.31 in image
[1] "Pakistan"
[1] "Riots"
Series: tseries[[c]][[t]]
ARIMA(1,0,0) with non-zero mean

Coefficients:
         ar1      mean
      0.5199   22.6802
s.e.  0.1084    4.2232

sigma^2 = 264.3:  log likelihood = -251.58
```

```
AIC=509.17   AICc=509.6   BIC=515.45
```

Ljung-Box test

```
data:  Residuals from ARIMA(1,0,0) with non-zero mean
Q* = 3.0805, df = 9, p-value = 0.961

Model df: 1.   Total lags used: 10
```

```
Saving 9.14 x 8.31 in image
[1] "Pakistan"
[1] "Strategic developments"
Series: tseries[[c]][[t]]
ARIMA(1,0,0) with non-zero mean

Coefficients:
         ar1    mean
      0.4912  3.1458
s.e.  0.1106  0.5852

sigma^2 = 5.68:  log likelihood = -136.37
AIC=278.73   AICc=279.16   BIC=285.02
```

Ljung-Box test

```
data:  Residuals from ARIMA(1,0,0) with non-zero mean
Q* = 5.161, df = 9, p-value = 0.8201

Model df: 1.   Total lags used: 10
```

```
Saving 9.14 x 8.31 in image
[1] "Pakistan"
[1] "Violence against civilians"
Series: tseries[[c]][[t]]
ARIMA(1,0,0) with non-zero mean

Coefficients:
```

```
            ar1       mean
         0.3708   34.2850
s.e.     0.1224    3.1886


sigma^2 = 253.7:  log likelihood = -250.28
AIC=506.56   AICc=506.99   BIC=512.84


	Ljung-Box test

data:  Residuals from ARIMA(1,0,0) with non-zero mean
Q* = 5.5022, df = 9, p-value = 0.7885

Model df: 1.   Total lags used: 10


Saving 9.14 x 8.31 in image
[1] "Sri Lanka"
[1] "Battles"
Series: tseries[[c]][[t]]
ARIMA(2,0,0) with non-zero mean

Coefficients:
           ar1     ar2    mean
        0.1281  0.3535  0.3310
s.e.    0.1198  0.1198  0.1496


sigma^2 = 0.3999:  log likelihood = -56.26
AIC=120.52   AICc=121.24   BIC=128.89


	Ljung-Box test

data:  Residuals from ARIMA(2,0,0) with non-zero mean
Q* = 2.2225, df = 8, p-value = 0.9734

Model df: 2.   Total lags used: 10


Saving 9.14 x 8.31 in image
[1] "Sri Lanka"
```

```
[1] "Explosions/Remote violence"
Series: tseries[[c]][[t]]
ARIMA(0,1,1)

Coefficients:
         ma1
      -0.7071
s.e.   0.1888

sigma^2 = 1.427:  log likelihood = -94.06
AIC=192.12   AICc=192.33   BIC=196.27


	Ljung-Box test

data:  Residuals from ARIMA(0,1,1)
Q* = 12.108, df = 9, p-value = 0.2073

Model df: 1.   Total lags used: 10

Saving 9.14 x 8.31 in image
[1] "Sri Lanka"
[1] "Protests"
Series: tseries[[c]][[t]]
ARIMA(0,0,0) with non-zero mean

Coefficients:
        mean
      11.1000
s.e.   0.8621

sigma^2 = 45.35:  log likelihood = -199.06
AIC=402.12   AICc=402.33   BIC=406.31


	Ljung-Box test

data:  Residuals from ARIMA(0,0,0) with non-zero mean
Q* = 15.33, df = 10, p-value = 0.1205
```

```
Model df: 0.   Total lags used: 10

Saving 9.14 x 8.31 in image
[1] "Sri Lanka"
[1] "Riots"
Series: tseries[[c]][[t]]
ARIMA(2,1,0)

Coefficients:
         ar1      ar2
      -0.5520  -0.4800
s.e.   0.1187   0.1373

sigma^2 = 25.71:  log likelihood = -178.82
AIC=363.65   AICc=364.09   BIC=369.88


    Ljung-Box test

data:  Residuals from ARIMA(2,1,0)
Q* = 4.6727, df = 8, p-value = 0.7919

Model df: 2.   Total lags used: 10

Saving 9.14 x 8.31 in image
[1] "Sri Lanka"
[1] "Strategic developments"
Series: tseries[[c]][[t]]
ARIMA(0,0,2) with non-zero mean

Coefficients:
         ma1     ma2    mean
      0.0931  0.5736  0.3547
s.e.  0.1313  0.1162  0.1519

sigma^2 = 0.5342:  log likelihood = -65.19
AIC=138.37   AICc=139.1   BIC=146.75
```

```
Ljung-Box test

data:  Residuals from ARIMA(0,0,2) with non-zero mean
Q* = 4.9934, df = 8, p-value = 0.7583

Model df: 2.   Total lags used: 10

Saving 9.14 x 8.31 in image
[1] "Sri Lanka"
[1] "Violence against civilians"
Series: tseries[[c]][[t]]
ARIMA(1,0,0) with non-zero mean

Coefficients:
         ar1    mean
      0.3122  2.8027
s.e.  0.1772  0.6369

sigma^2 = 11.53:  log likelihood = -157.51
AIC=321.01   AICc=321.44   BIC=327.3


Ljung-Box test

data:  Residuals from ARIMA(1,0,0) with non-zero mean
Q* = 4.9006, df = 9, p-value = 0.8429

Model df: 1.   Total lags used: 10

Saving 9.14 x 8.31 in image
```

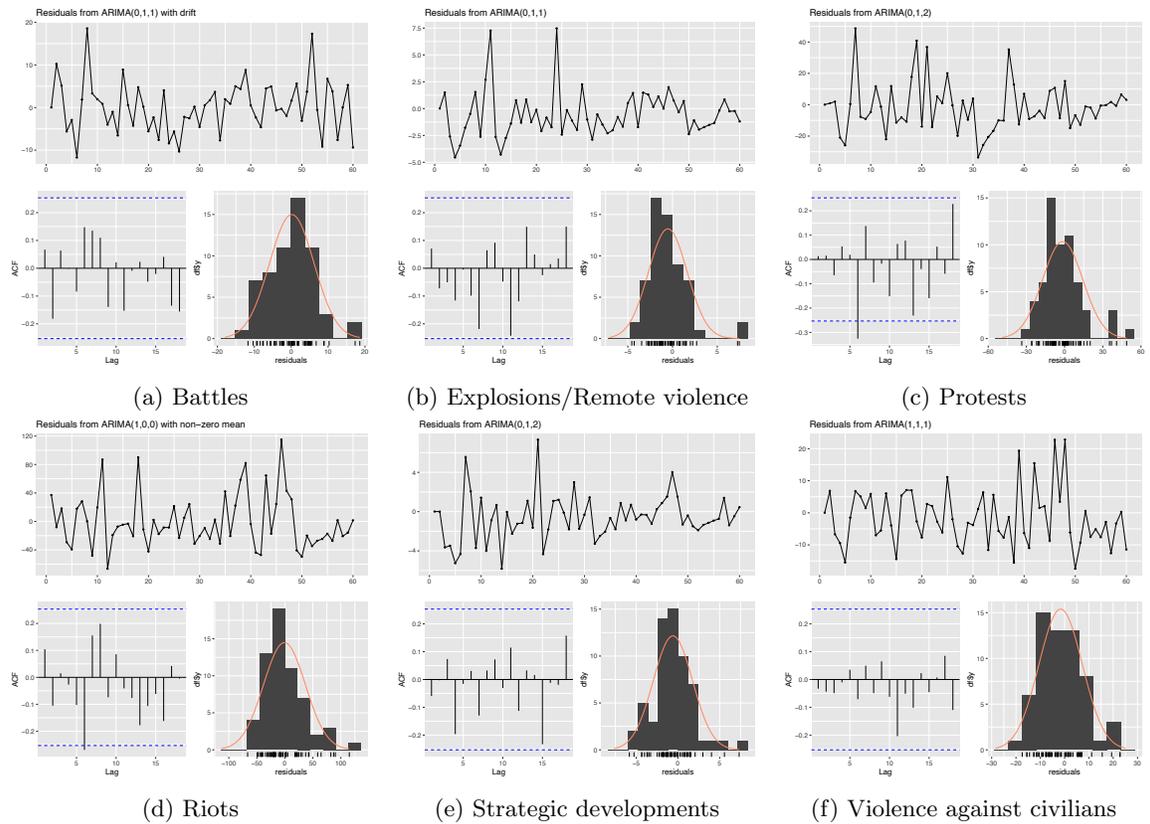

Figure 16: Bangladesh. Time plot of the residuals, the corresponding ACF, and a histogram.

54

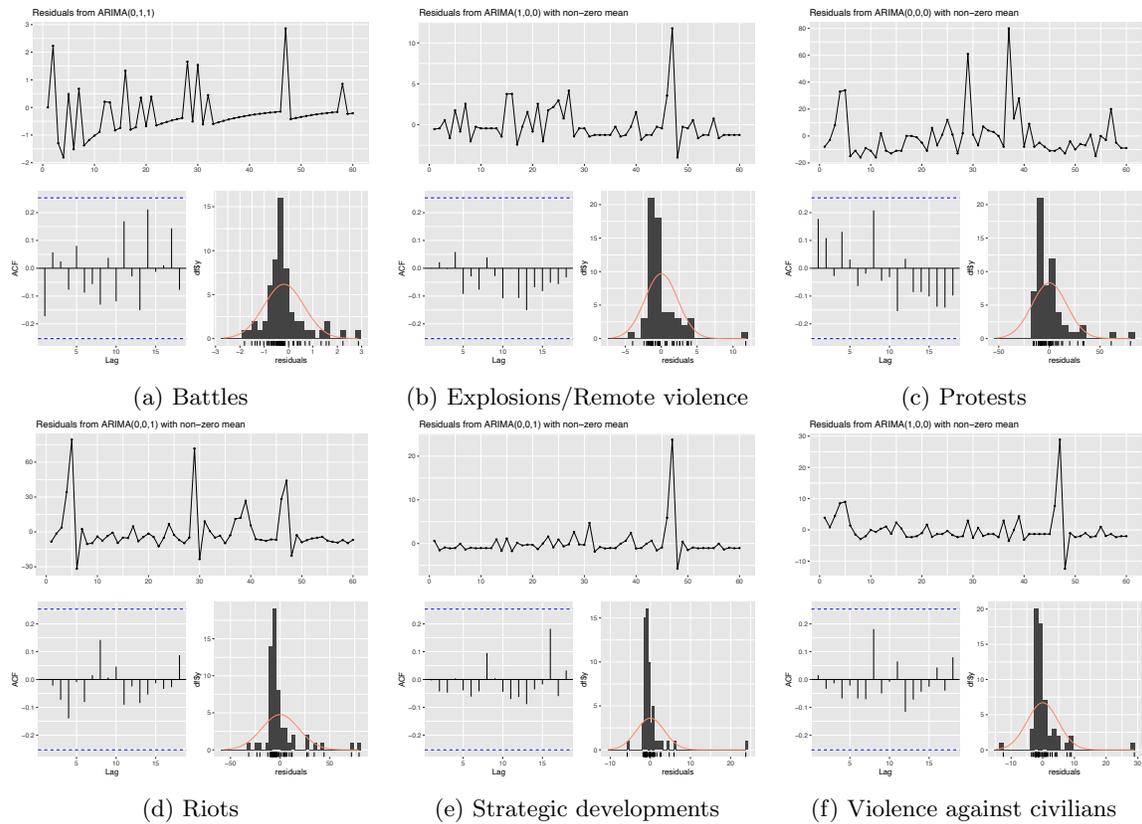

Figure 17: Nepal. Time plot of the residuals, the corresponding ACF, and a histogram.

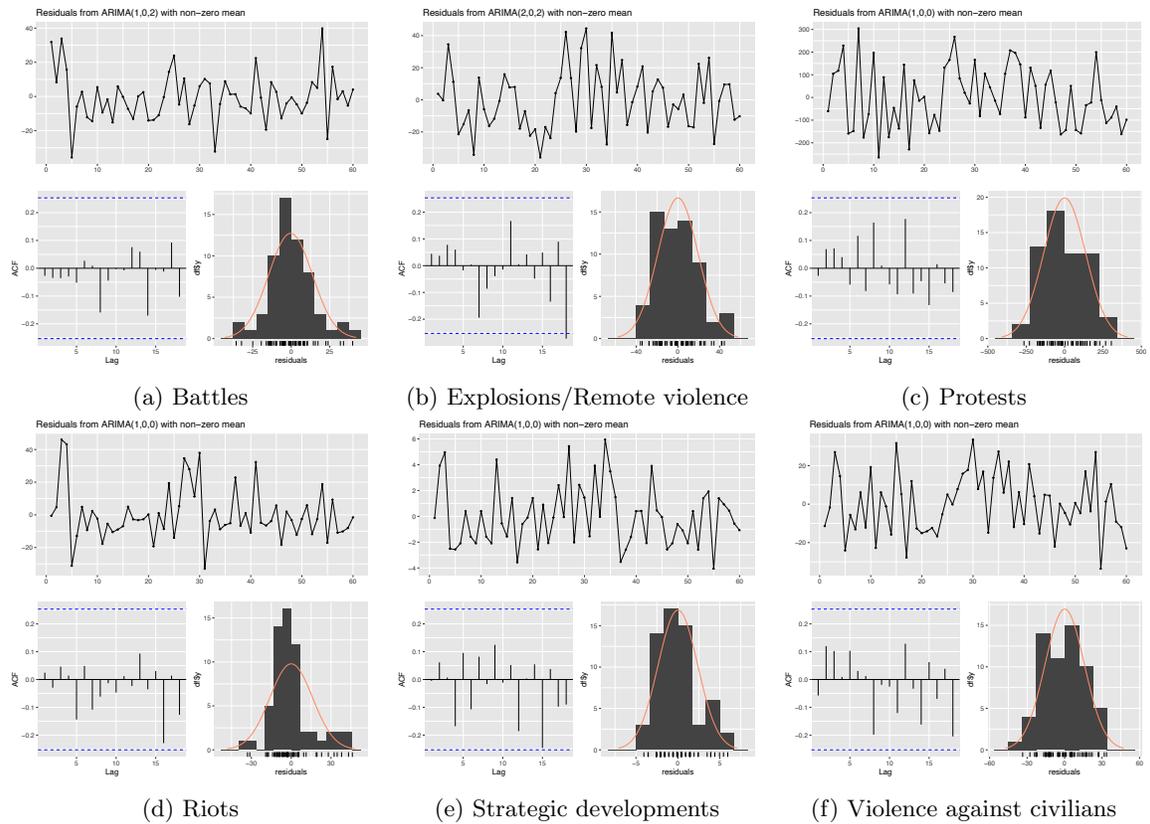

Figure 18: Pakistan. Time plot of the residuals, the corresponding ACF, and a histogram.

56

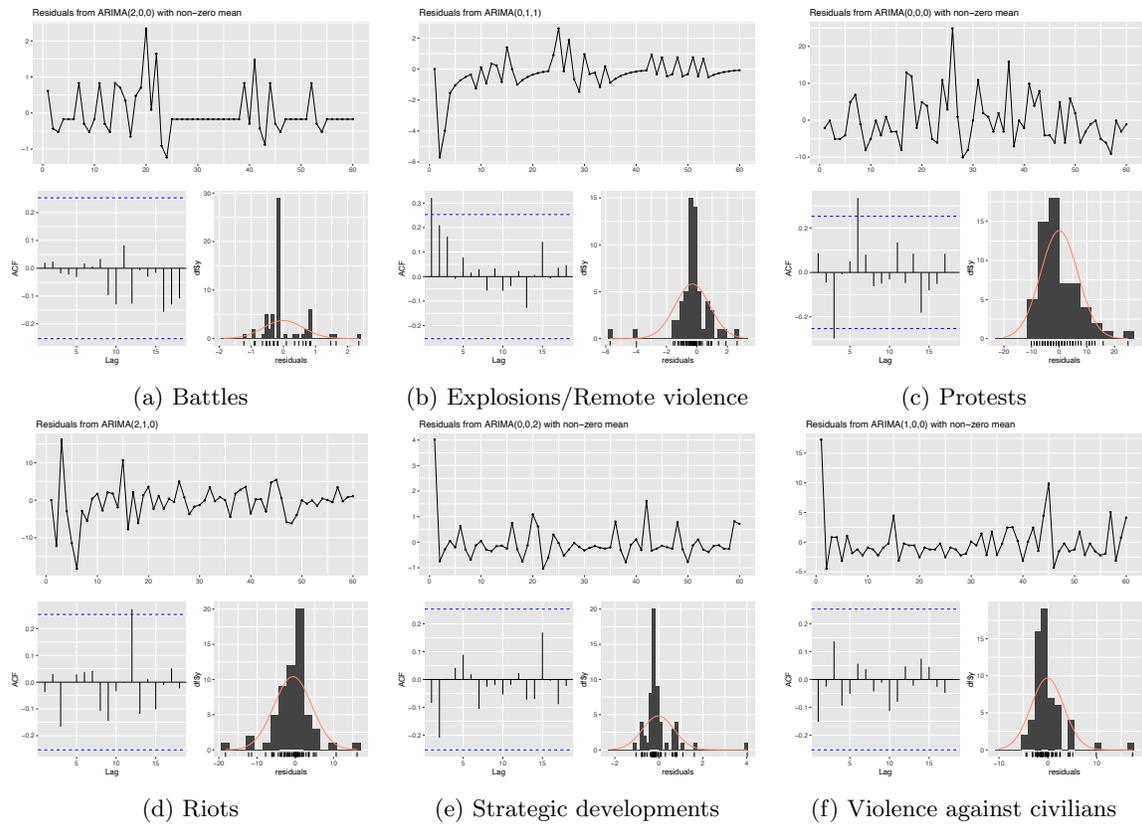

Figure 19: Sri Lanka. Time plot of the residuals, the corresponding ACF, and a histogram.

# E MCMC diagnostics

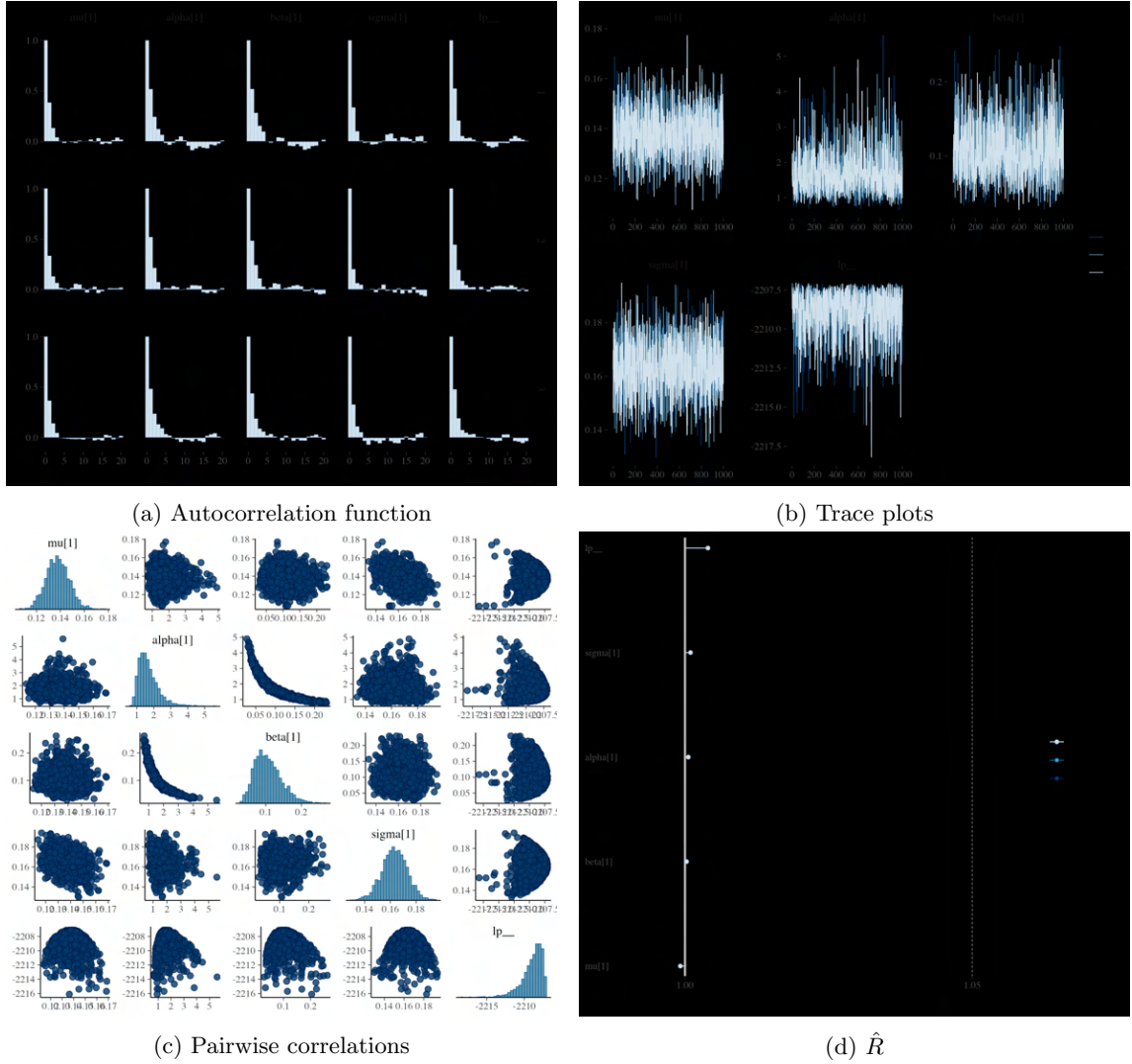

(a) Autocorrelation function

(b) Trace plots

(c) Pairwise correlations

(d) $\hat{R}$

Figure 20: MCMC diagnostic plots for Battles in Bangladesh.

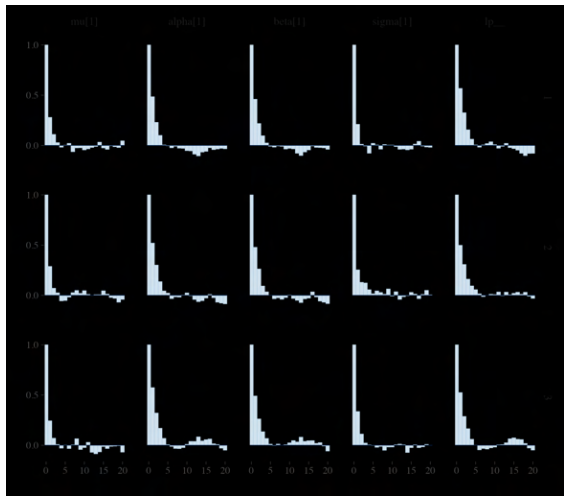
(a) Autocorrelation function

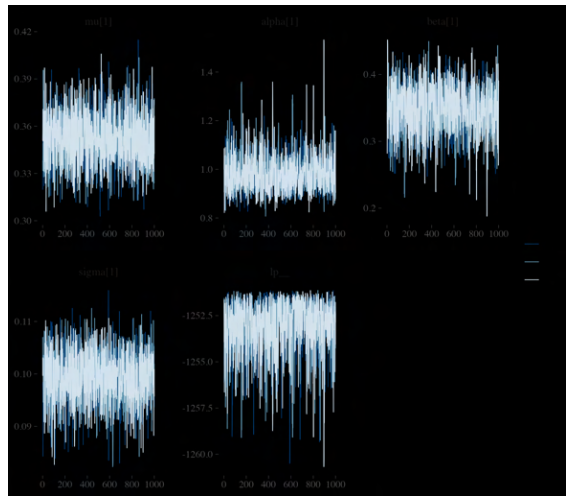
(b) Trace plots

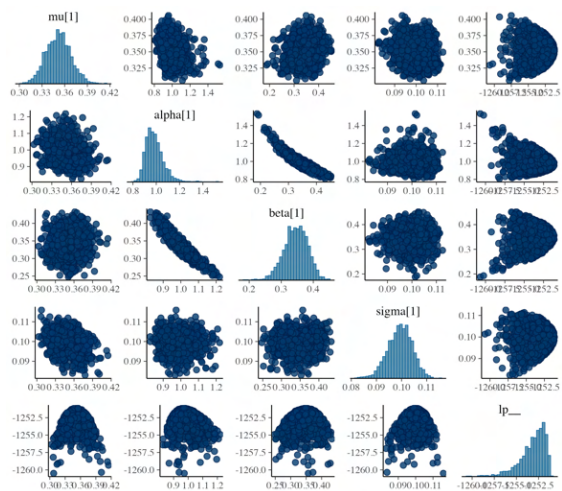
(c) Pairwise correlations

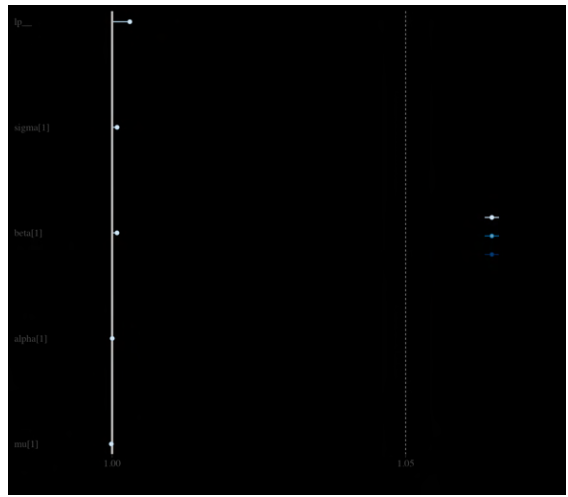
(d) $\hat{R}$

Figure 21: MCMC diagnostic plots for Riots in Bangladesh.

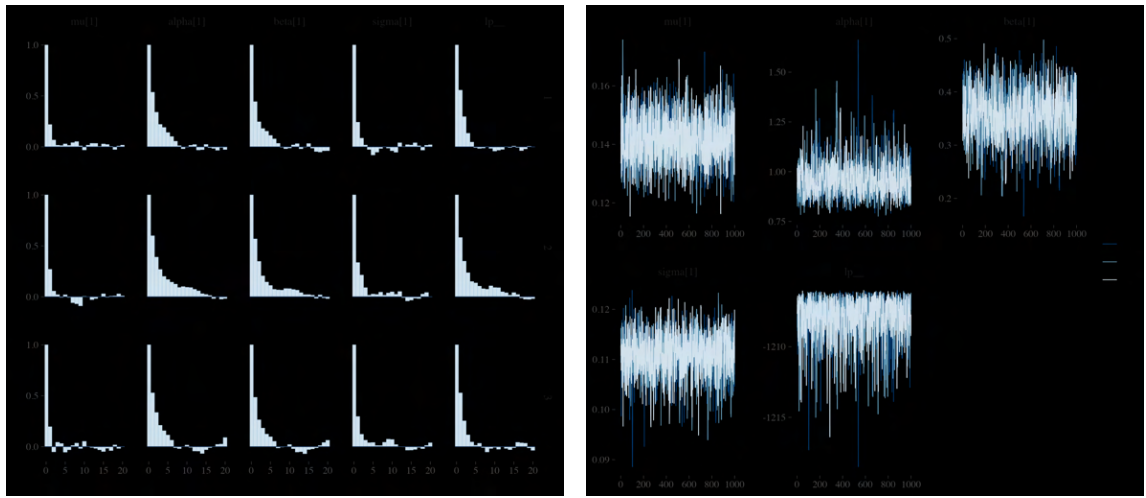

(a) Autocorrelation function

(b) Trace plots

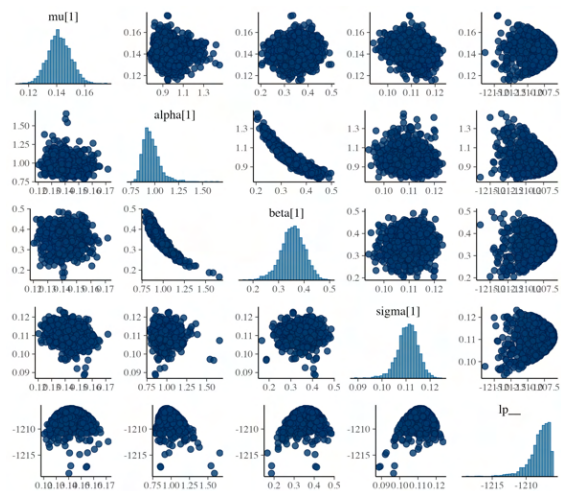

(c) Pairwise correlations

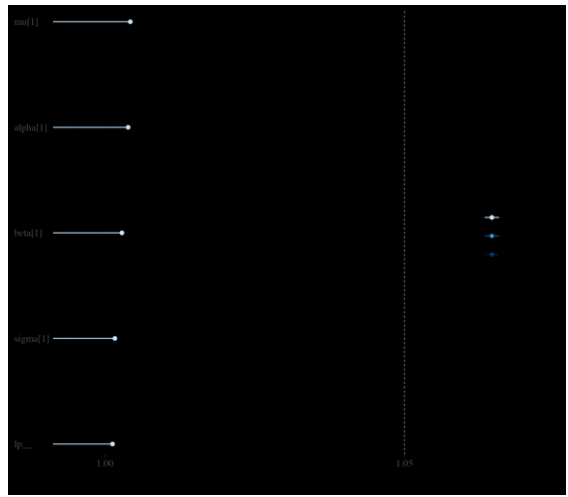

(d) $\hat{R}$

Figure 22: MCMC diagnostic plots for Protests in Bangladesh.

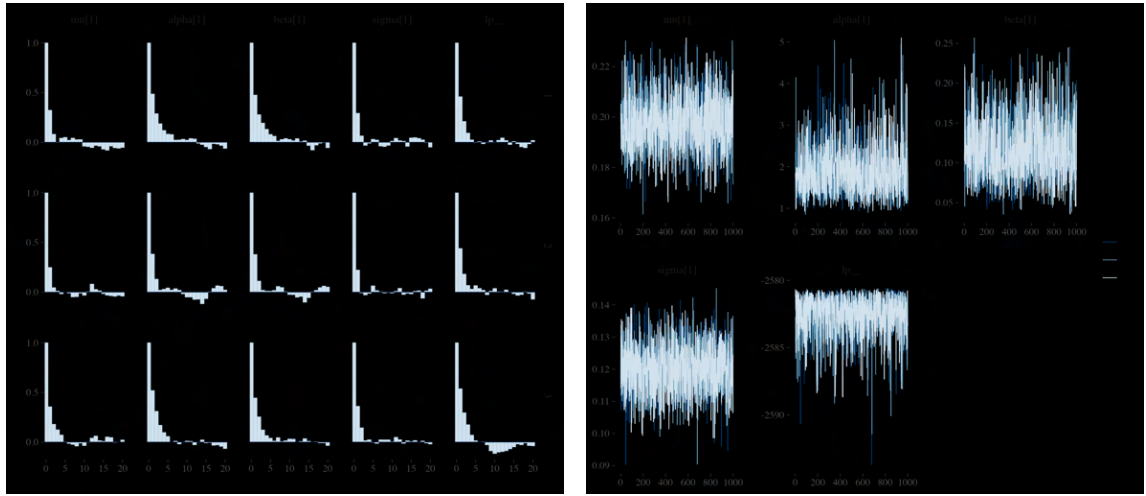

(a) Autocorrelation function

(b) Trace plots

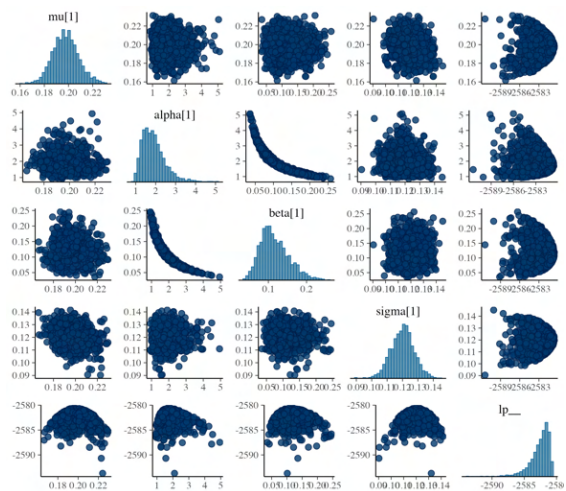

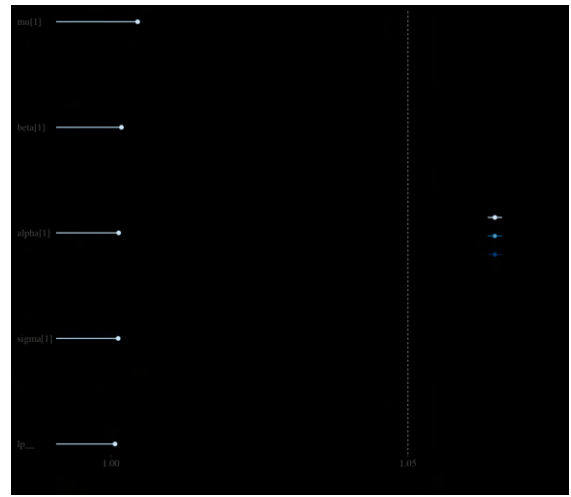

(c) Pairwise correlations

(d) $\hat{R}$

Figure 23: MCMC diagnostic plots for Violence against civilians in Bangladesh.

61

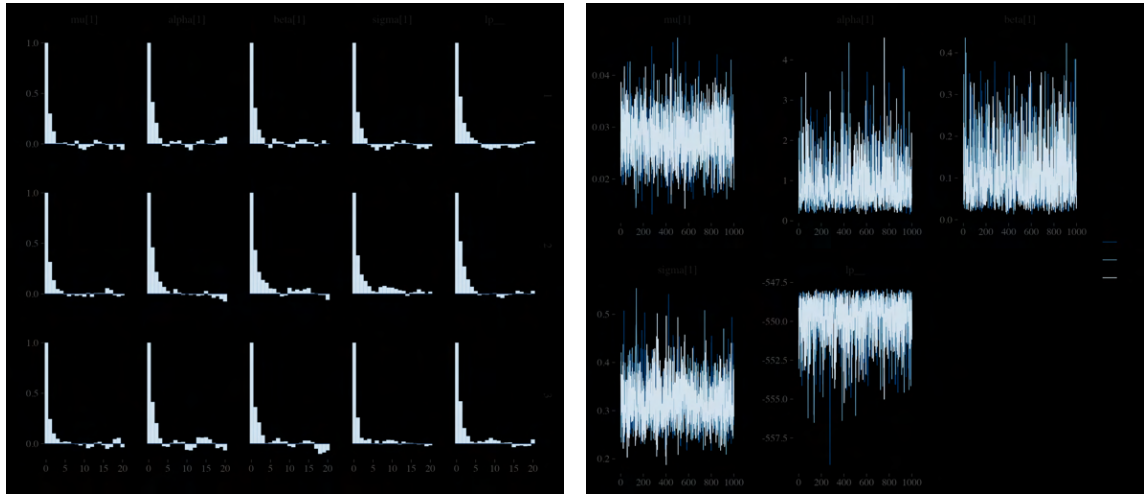

(a) Autocorrelation function

(b) Trace plots

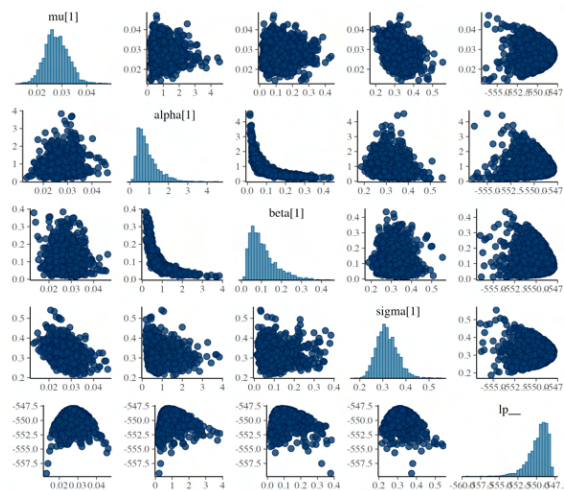

(c) Pairwise correlations

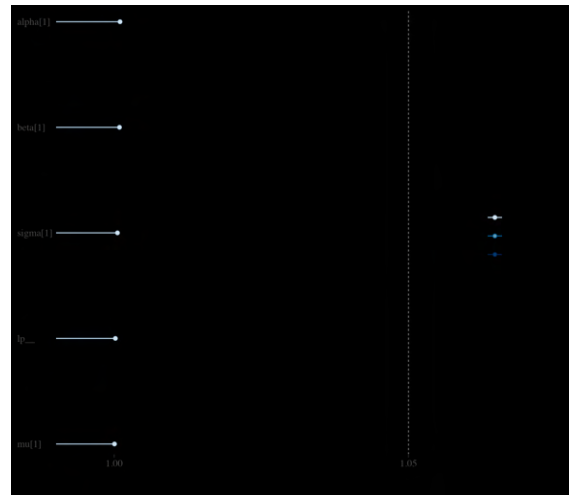

(d) $\hat{R}$

Figure 24: MCMC diagnostic plots for Strategic developments in Bangladesh.

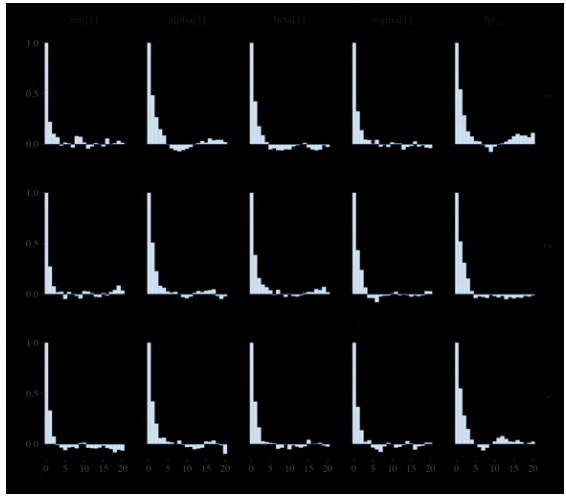

(a) Autocorrelation function

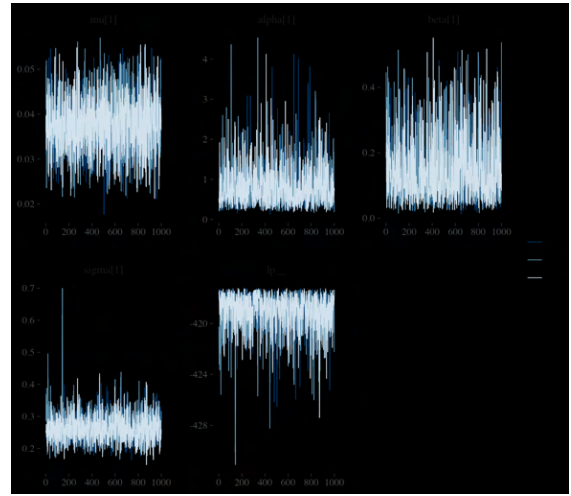

(b) Trace plots

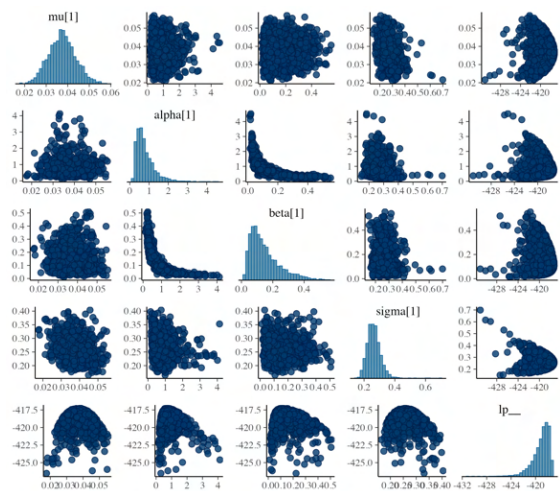

(c) Pairwise correlations

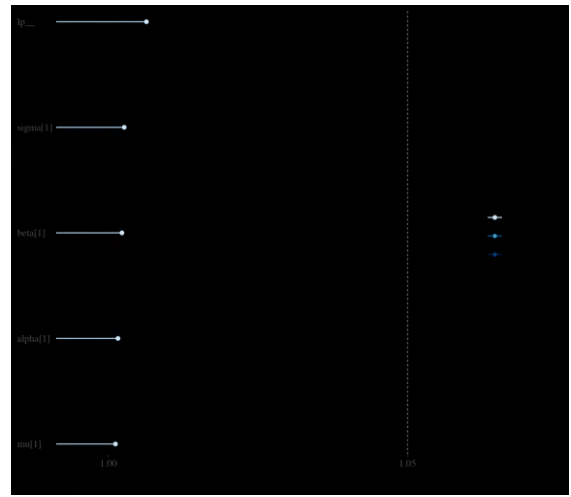

(d) $\hat{R}$

Figure 25: MCMC diagnostic plots for Explosions/remote violence in Bangladesh.

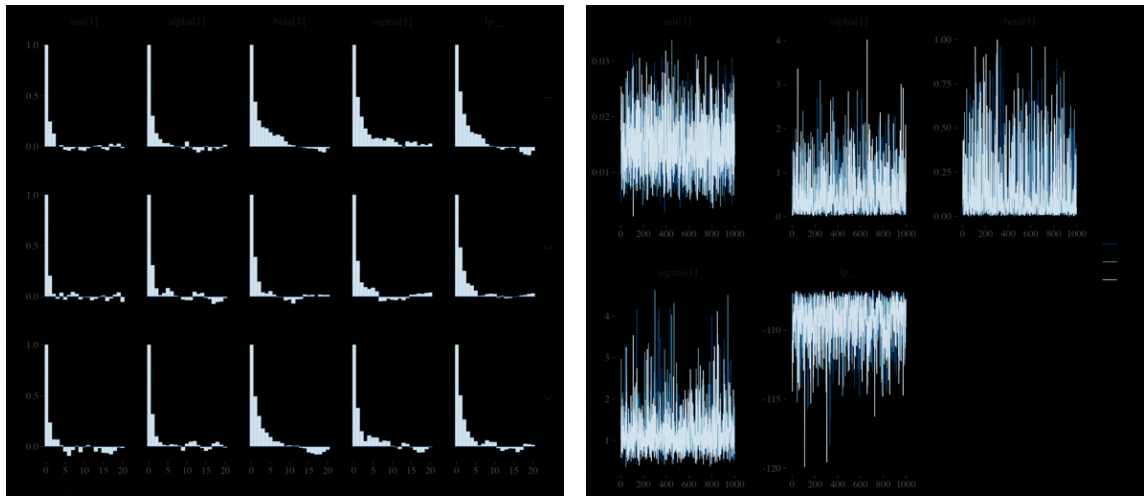

(a) Autocorrelation function

(b) Trace plots

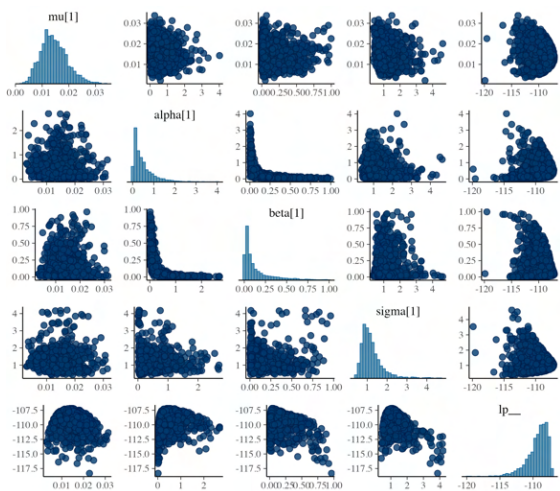

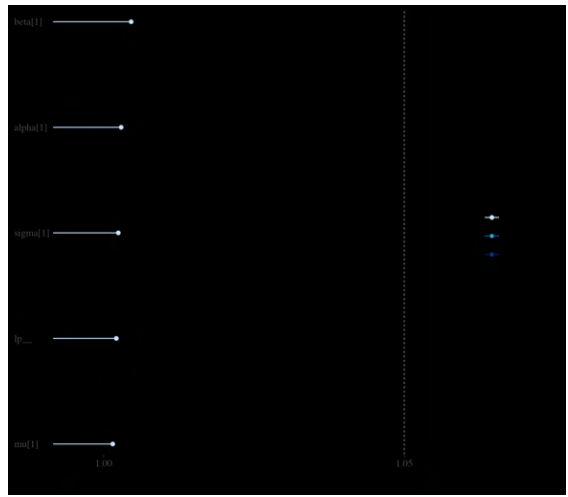

(c) Pairwise correlations

(d) $\hat{R}$

Figure 26: MCMC diagnostic plots for Battles in Sri Lanka.

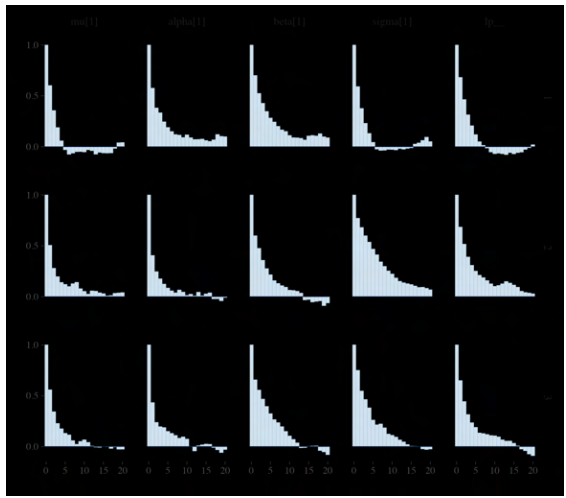
(a) Autocorrelation function

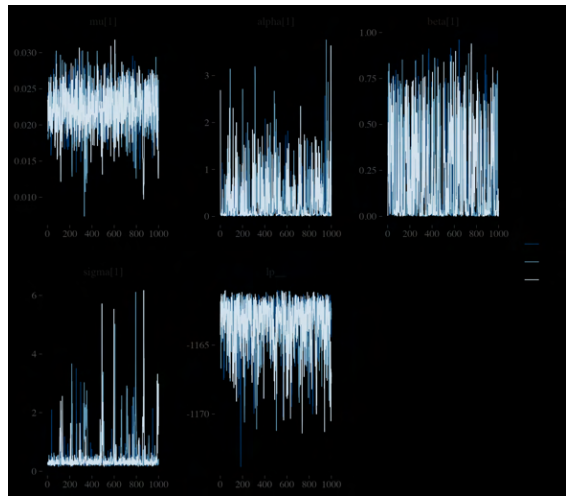
(b) Trace plots

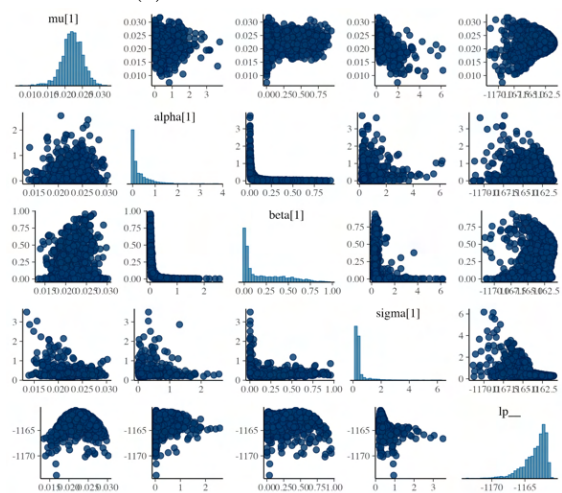
(c) Pairwise correlations

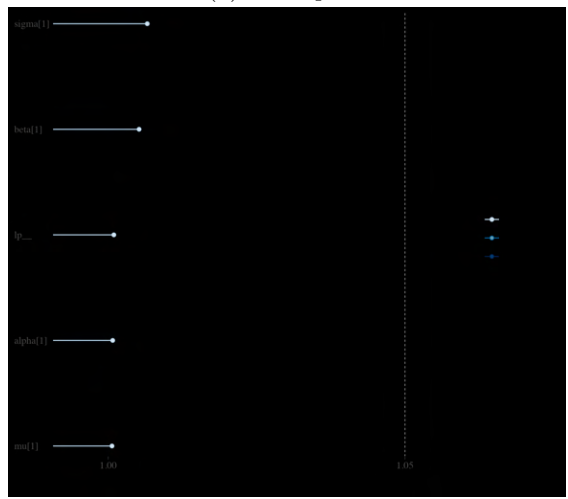
(d) $\hat{R}$

Figure 27: MCMC diagnostic plots for Riots in Sri Lanka.

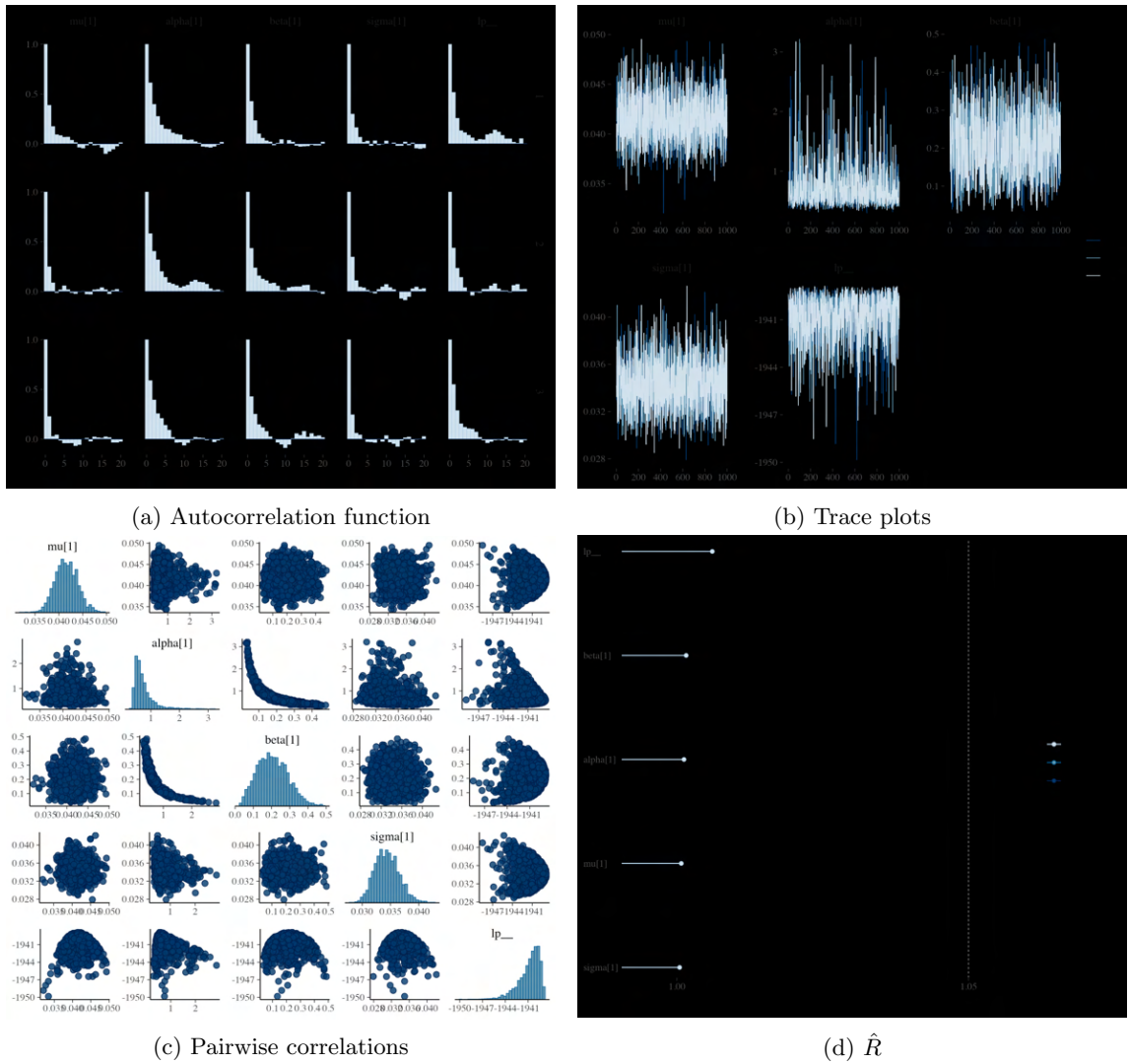

(a) Autocorrelation function
(b) Trace plots
(c) Pairwise correlations
(d) $\hat{R}$

Figure 28: MCMC diagnostic plots for Protests in Sri Lanka.

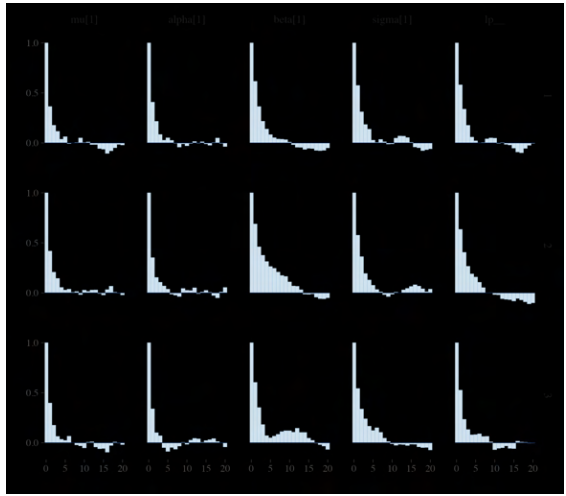
(a) Autocorrelation function

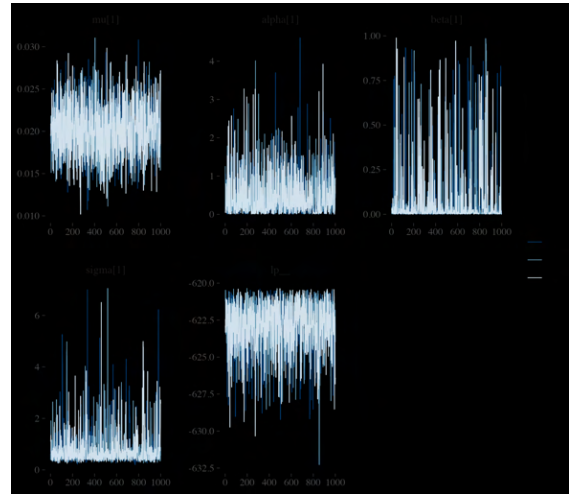
(b) Trace plots

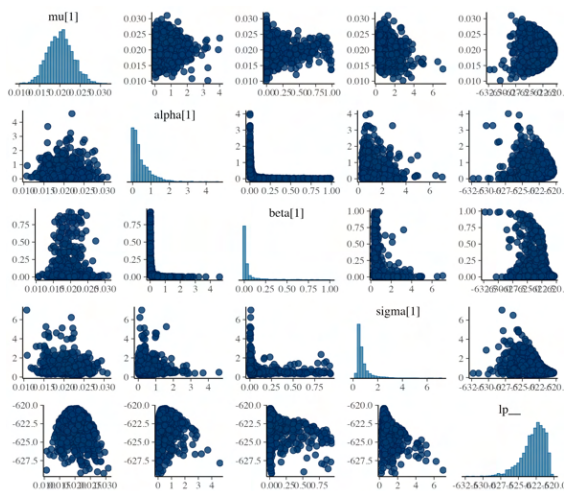
(c) Pairwise correlations

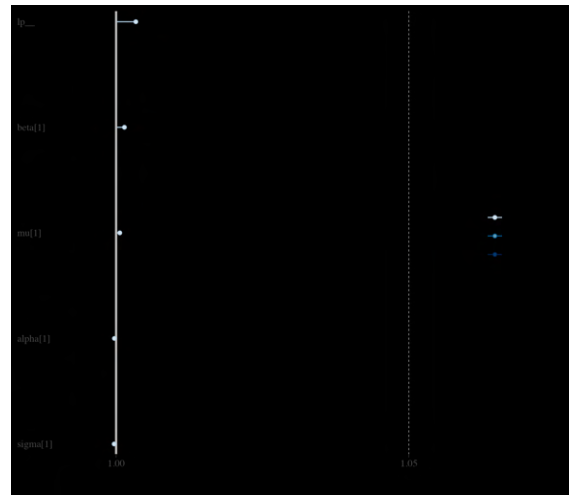
(d) $\hat{R}$

Figure 29: MCMC diagnostic plots for Violence against civilians in Sri Lanka.

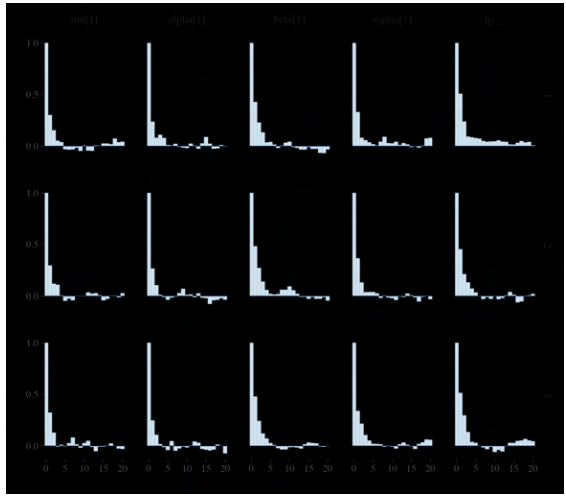
(a) Autocorrelation function

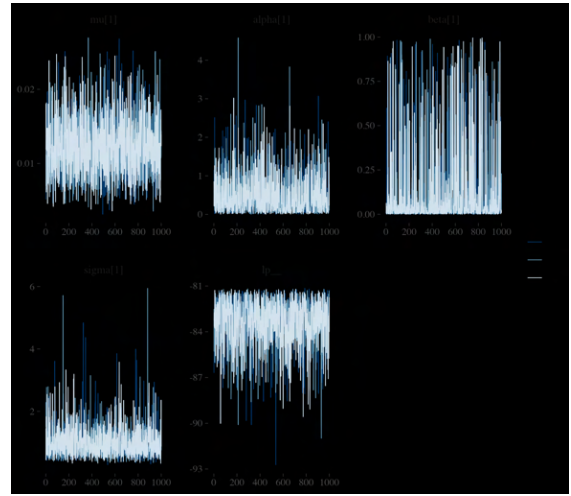
(b) Trace plots

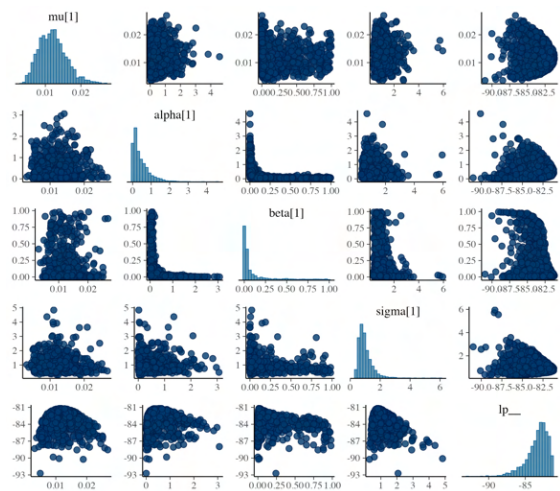
(c) Pairwise correlations

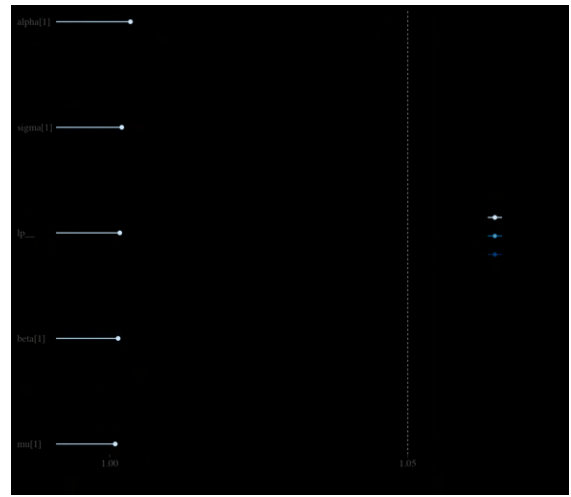
(d) $\hat{R}$

Figure 30: MCMC diagnostic plots for Strategic developments in Sri Lanka.

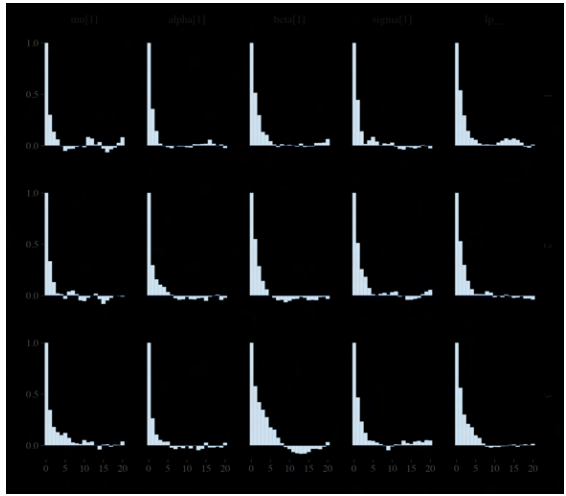
(a) Autocorrelation function

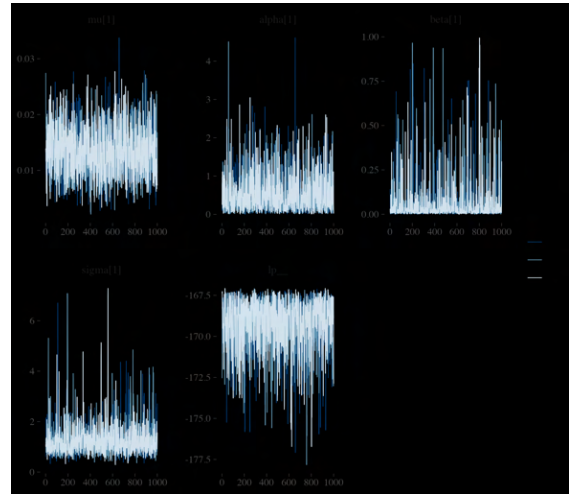
(b) Trace plots

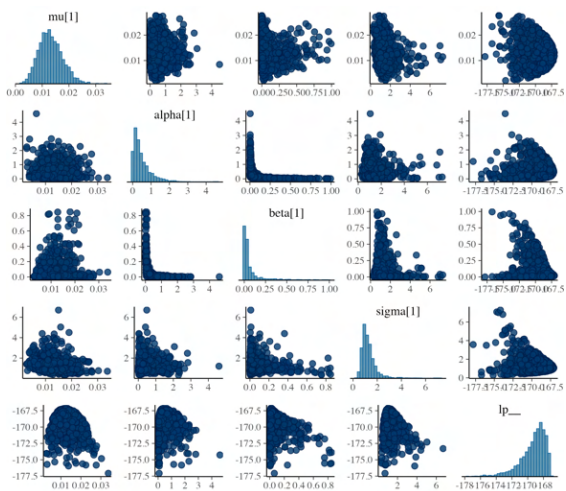
(c) Pairwise correlations

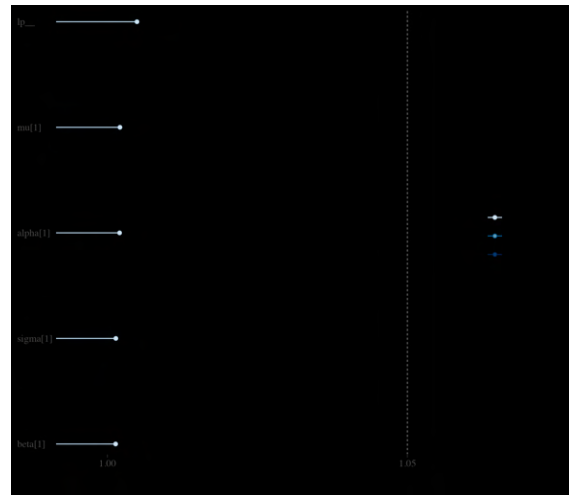
(d) $\hat{R}$

Figure 31: MCMC diagnostic plots for Explosions/remote violence in Sri Lanka.

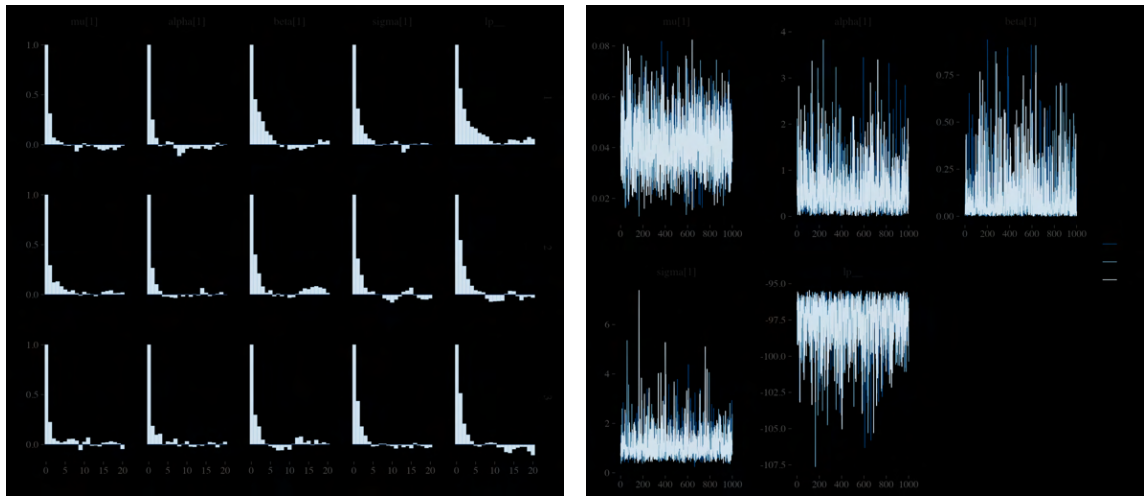

(a) Autocorrelation function  (b) Trace plots

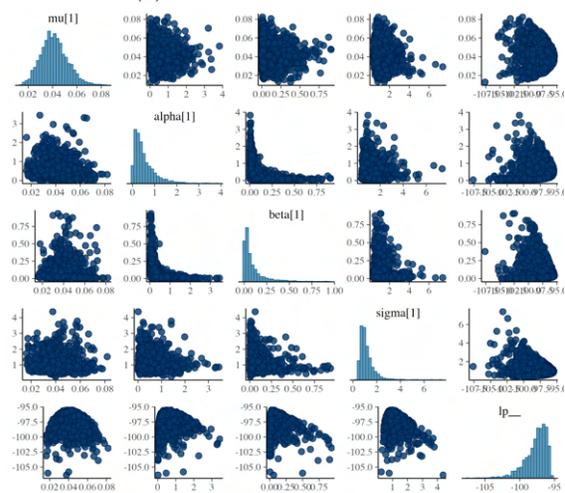
(c) Pairwise correlations

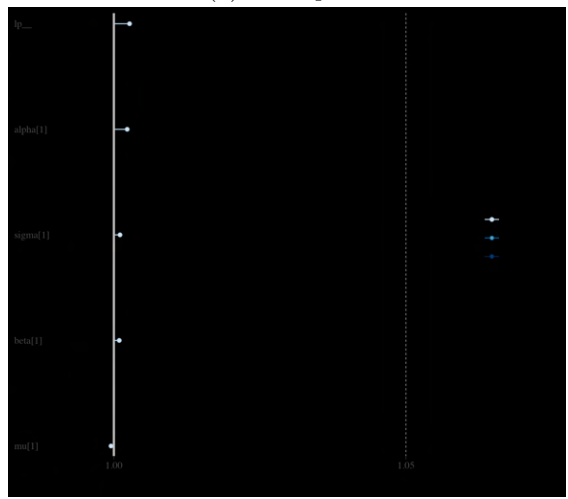
(d) $\hat{R}$

Figure 32: MCMC diagnostic plots for Battles in Nepal.

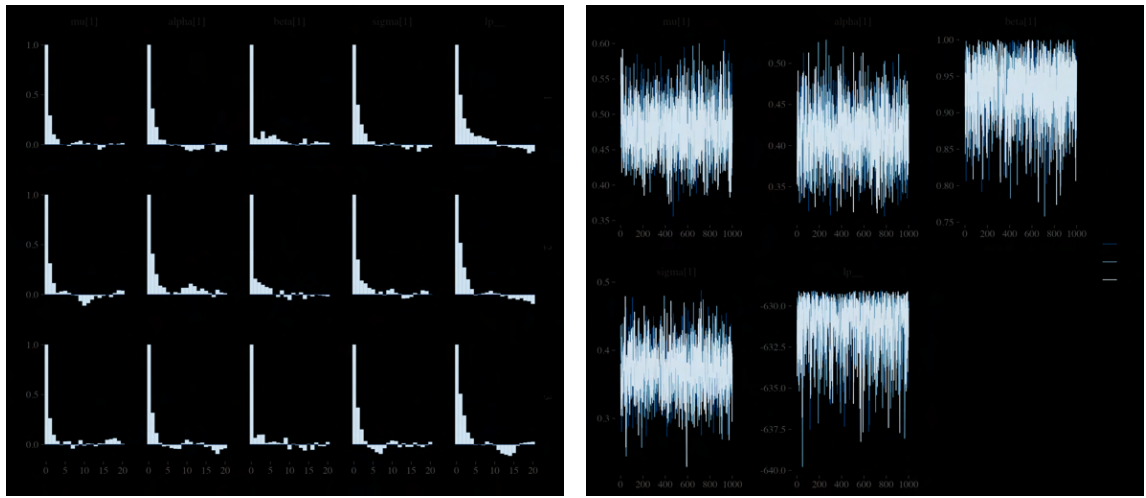

(a) Autocorrelation function  (b) Trace plots

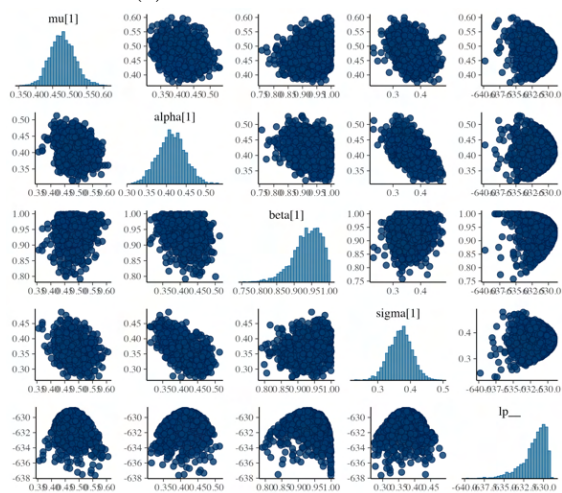

(c) Pairwise correlations

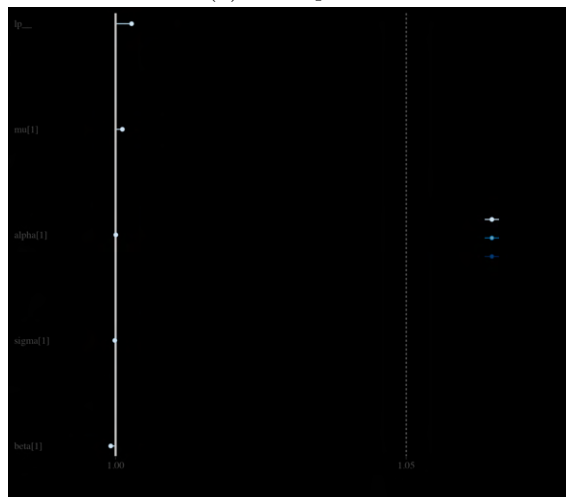

(d) $\hat{R}$

Figure 33: MCMC diagnostic plots for Riots in Nepal.

71

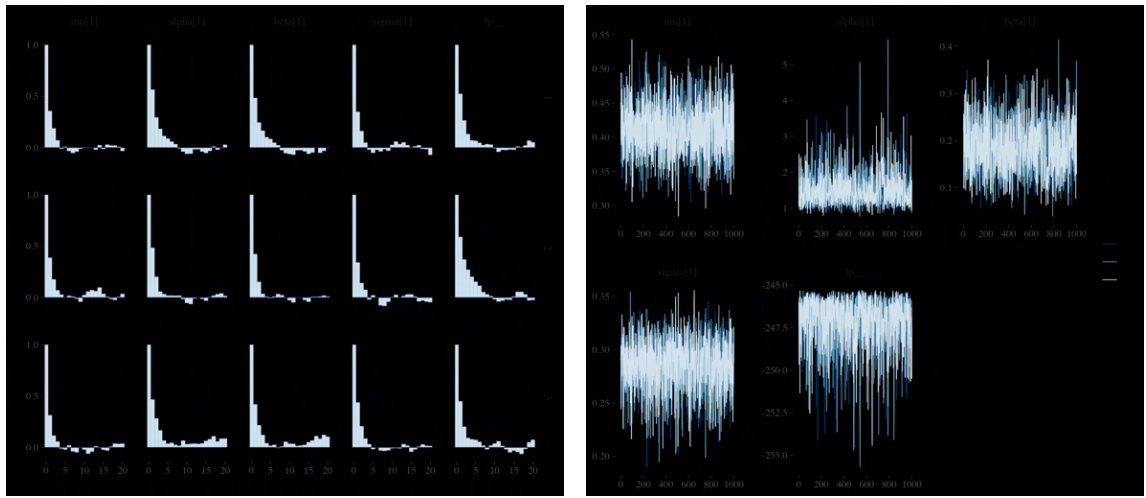

(a) Autocorrelation function
(b) Trace plots

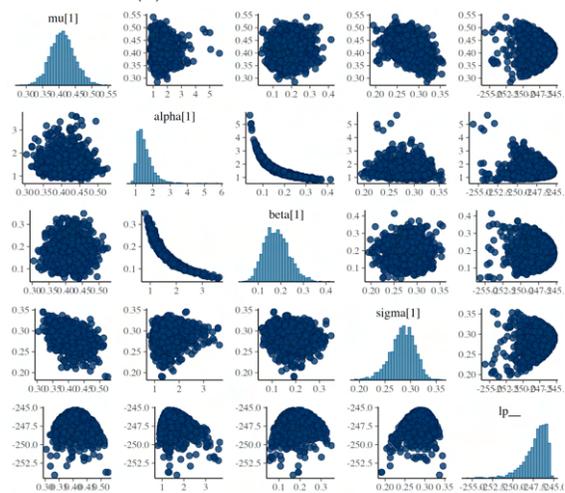
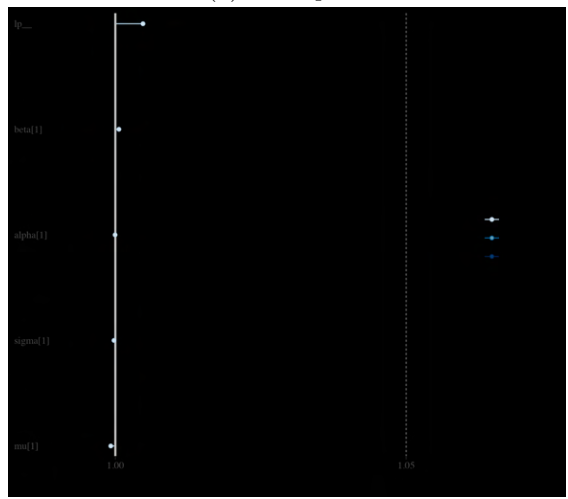

(c) Pairwise correlations
(d) $\hat{R}$

Figure 34: MCMC diagnostic plots for Protests in Nepal.

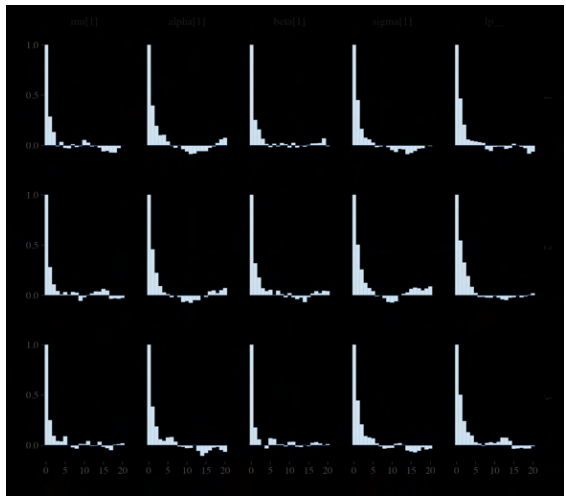
(a) Autocorrelation function

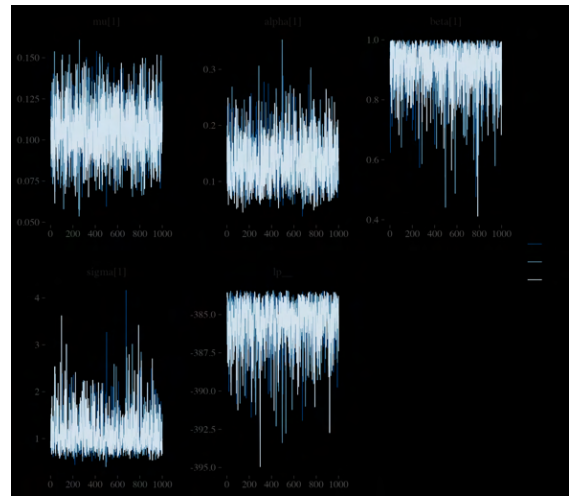
(b) Trace plots

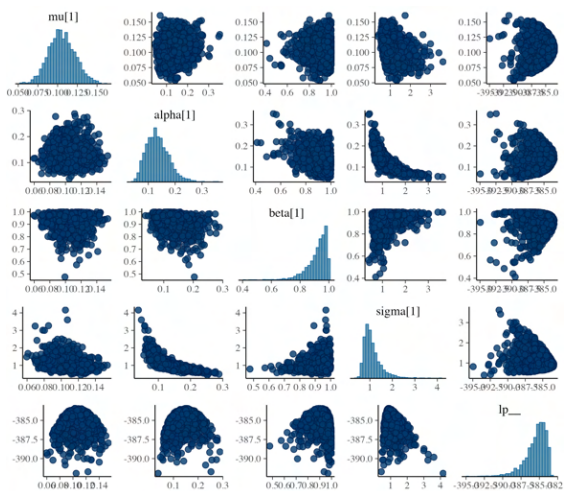
(c) Pairwise correlations

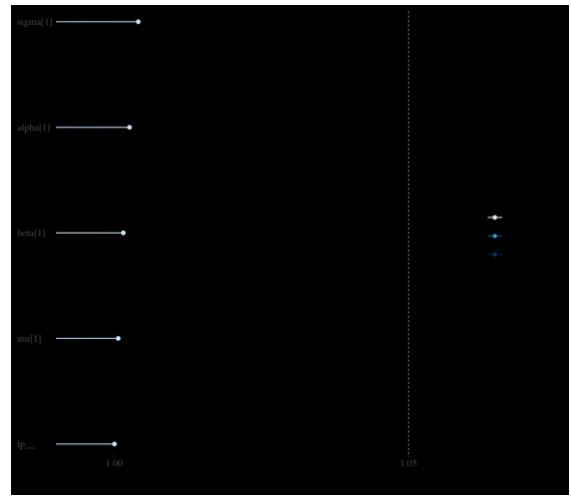
(d) $\hat{R}$

Figure 35: MCMC diagnostic plots for Violence against civilians in Nepal.

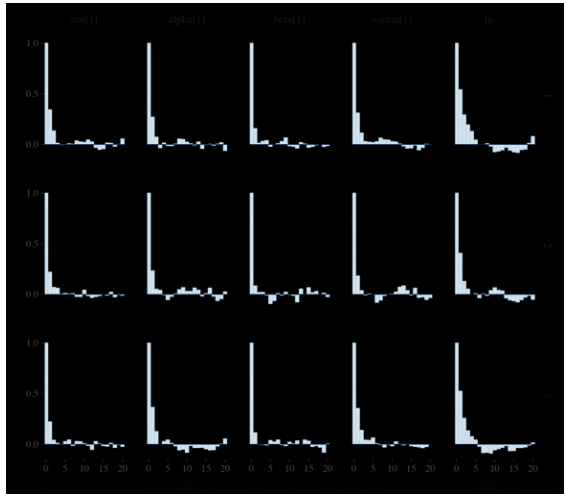

(a) Autocorrelation function

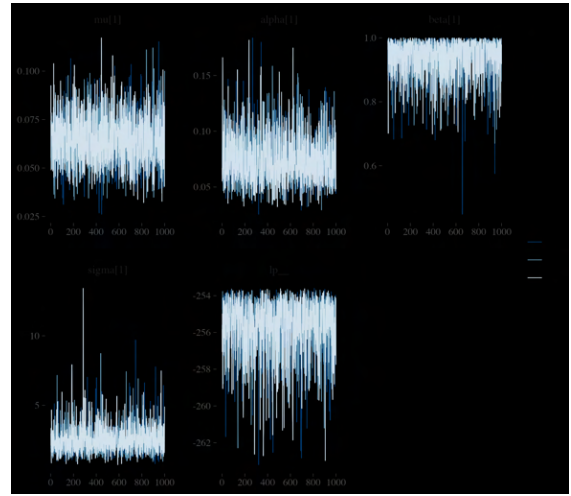

(b) Trace plots

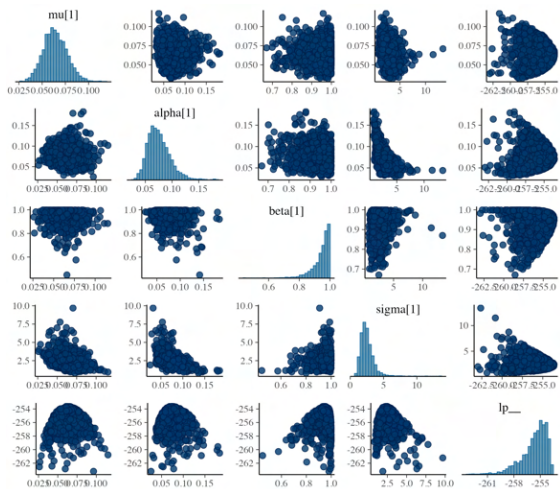

(c) Pairwise correlations

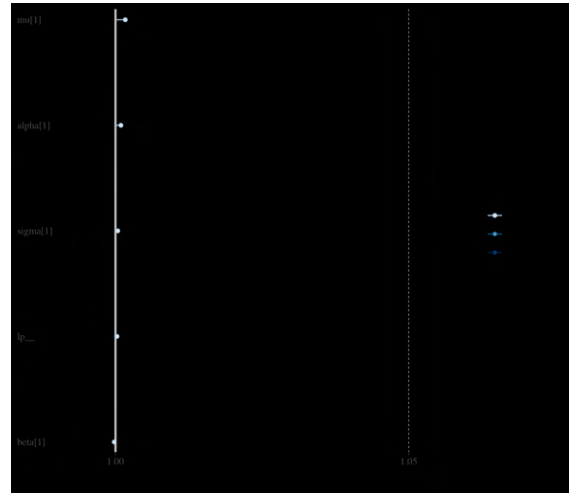

(d) $\hat{R}$

Figure 36: MCMC diagnostic plots for Strategic developments in Nepal.

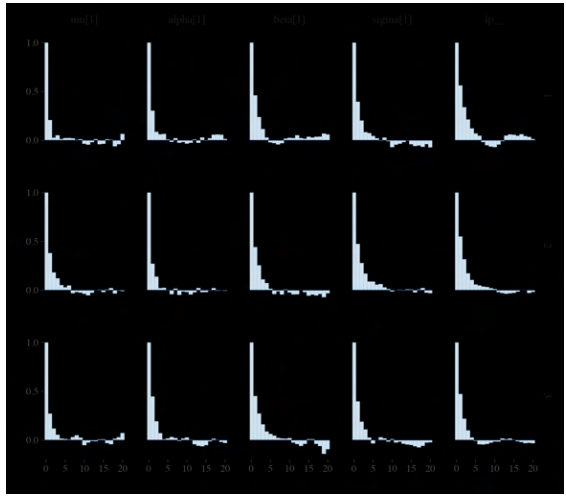
(a) Autocorrelation function

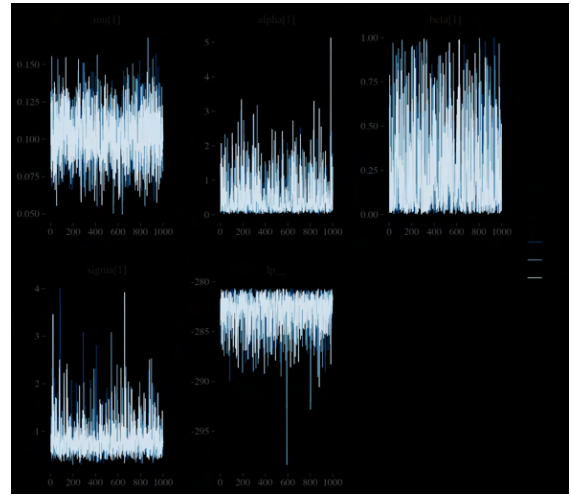
(b) Trace plots

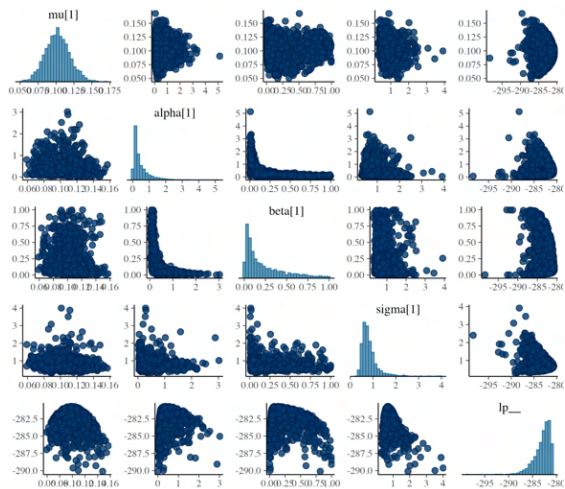
(c) Pairwise correlations

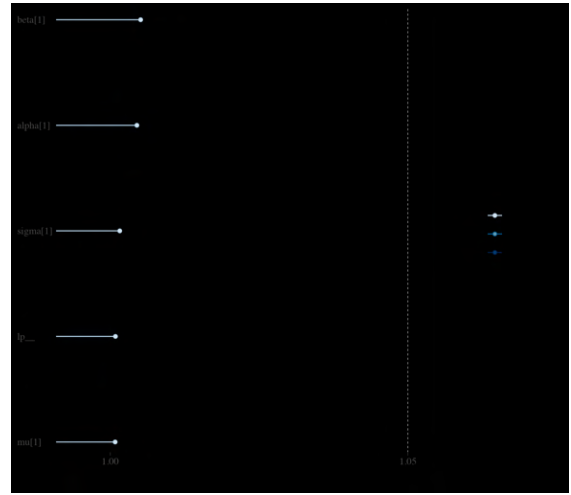
(d) $\hat{R}$

Figure 37: MCMC diagnostic plots for Explosions/remote violence in Nepal.

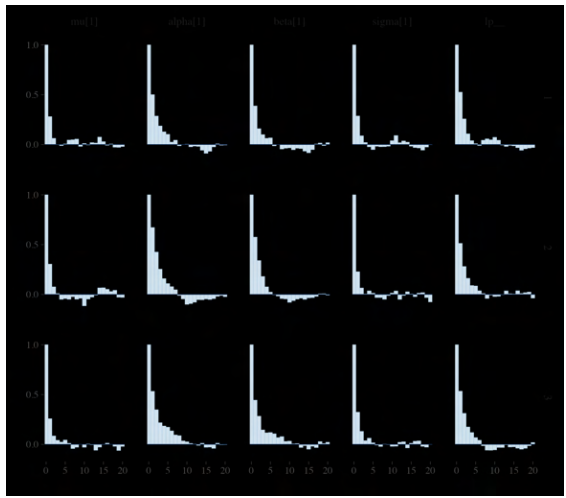
(a) Autocorrelation function

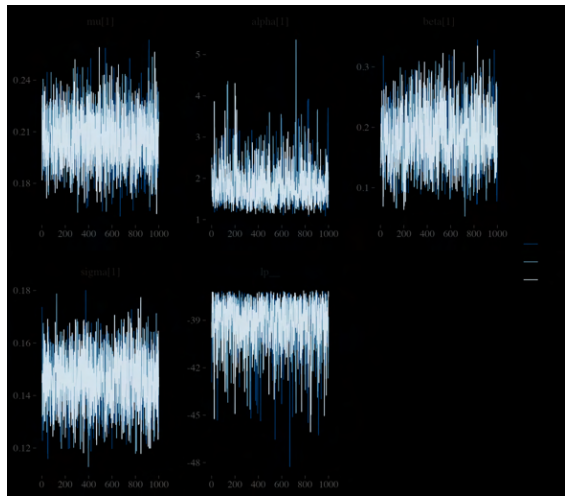
(b) Trace plots

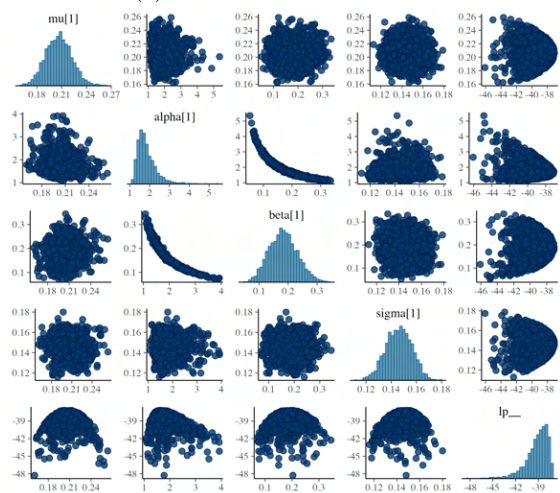
(c) Pairwise correlations

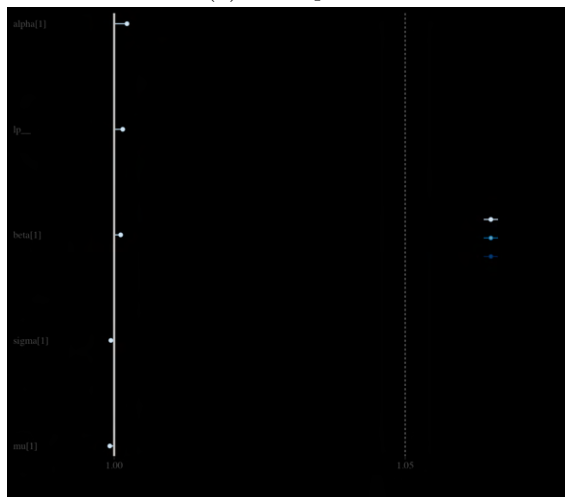
(d) $\hat{R}$

Figure 38: MCMC diagnostic plots for Battles in Pakistan.

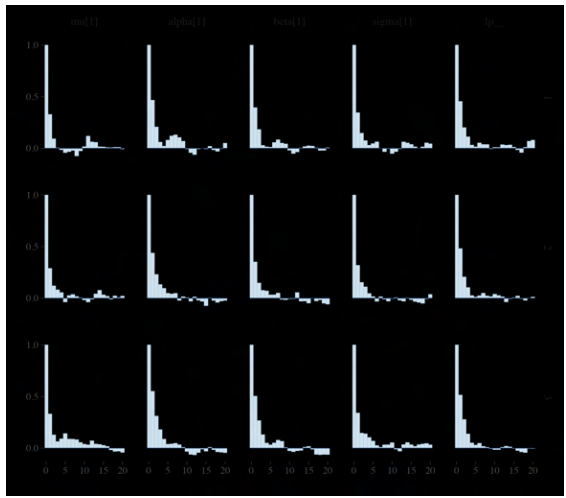
(a) Autocorrelation function

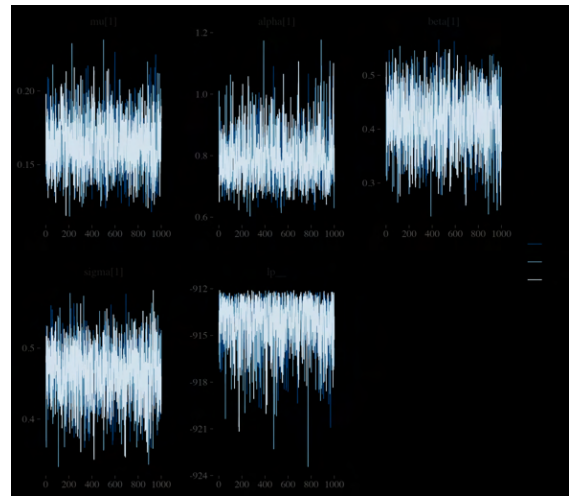
(b) Trace plots

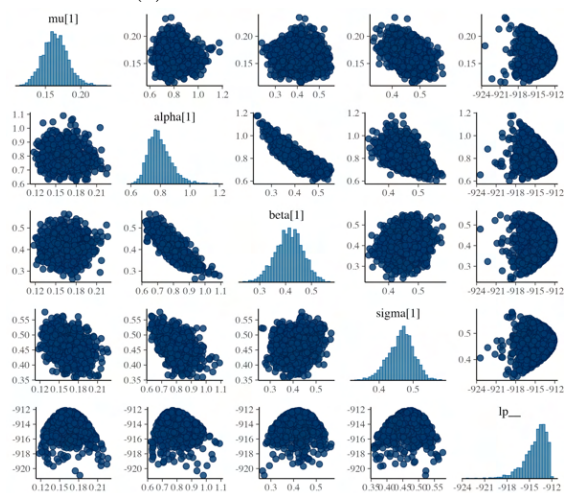
(c) Pairwise correlations

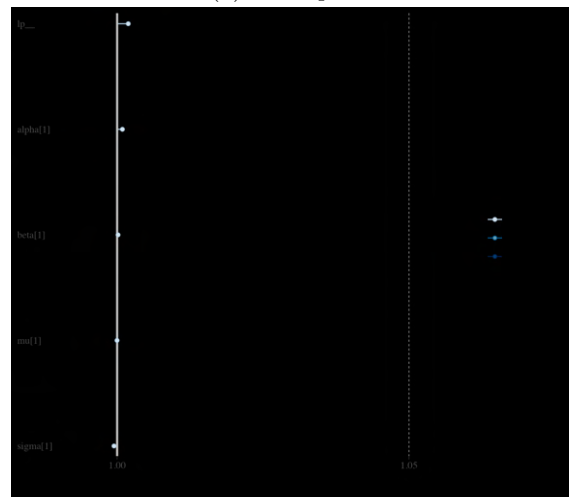
(d) $\hat{R}$

Figure 39: MCMC diagnostic plots for Riots in Pakistan.

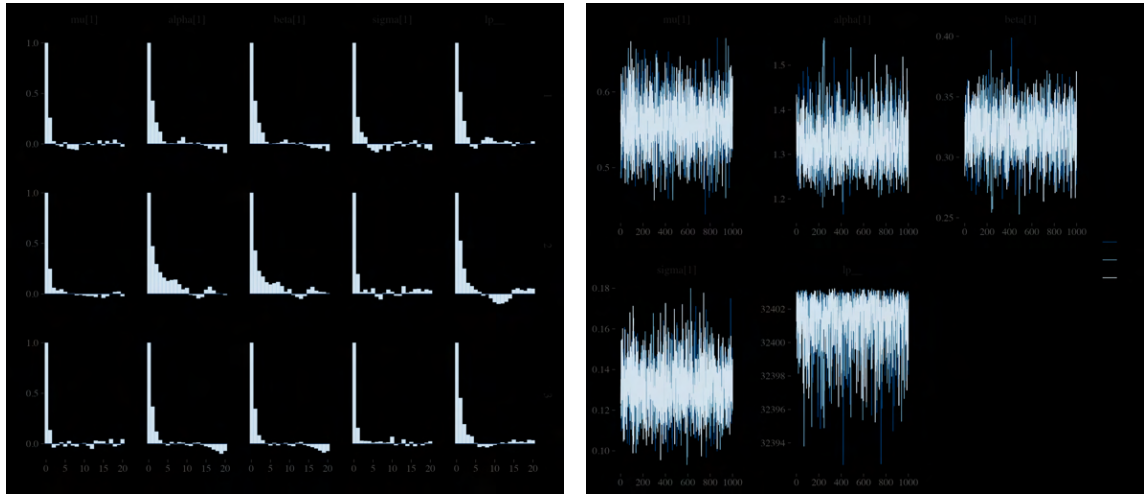

(a) Autocorrelation function
(b) Trace plots

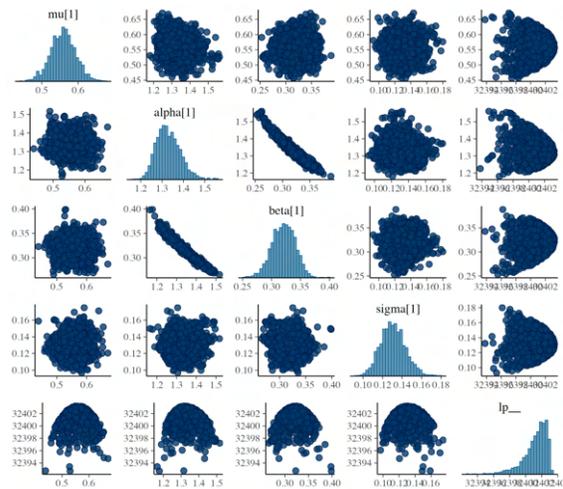
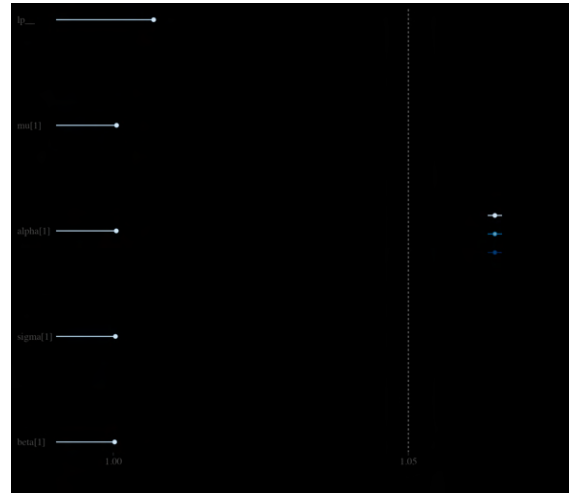

(c) Pairwise correlations
(d) $\hat{R}$

Figure 40: MCMC diagnostic plots for Protests in Pakistan.

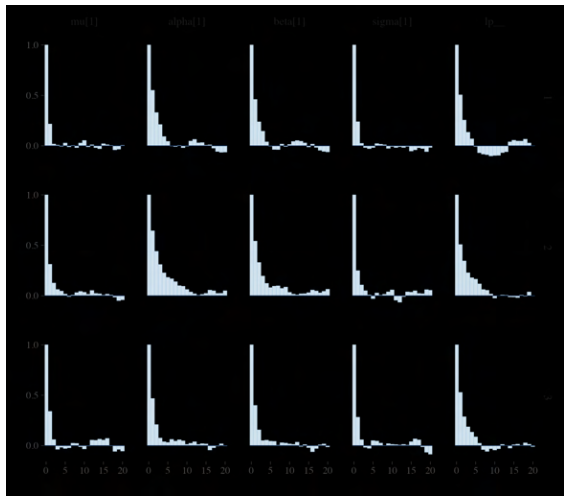
(a) Autocorrelation function

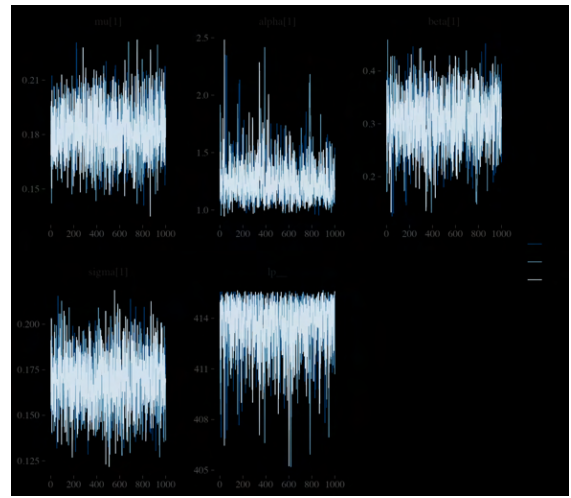
(b) Trace plots

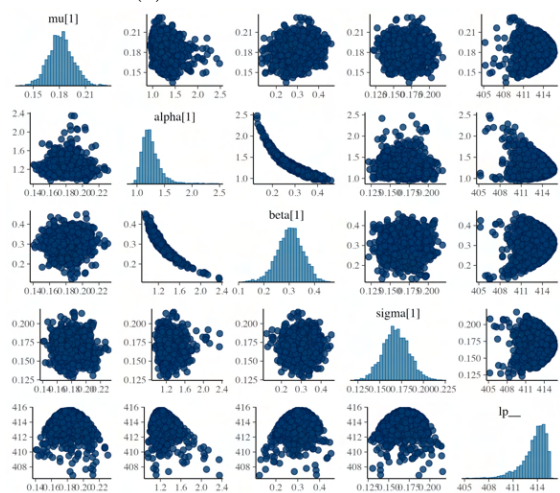
(c) Pairwise correlations

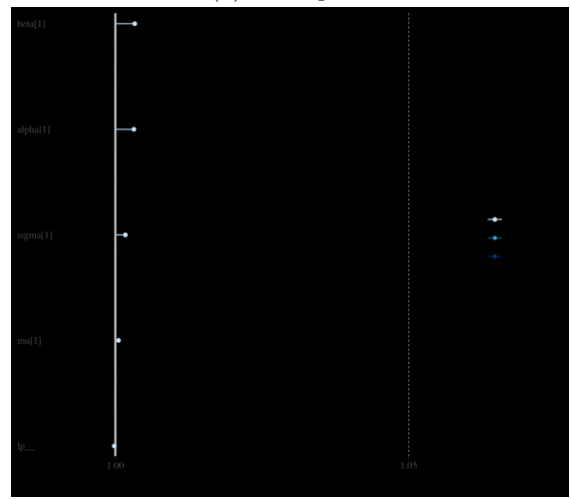
(d) $\hat{R}$

Figure 41: MCMC diagnostic plots for Violence against civilians in Pakistan.

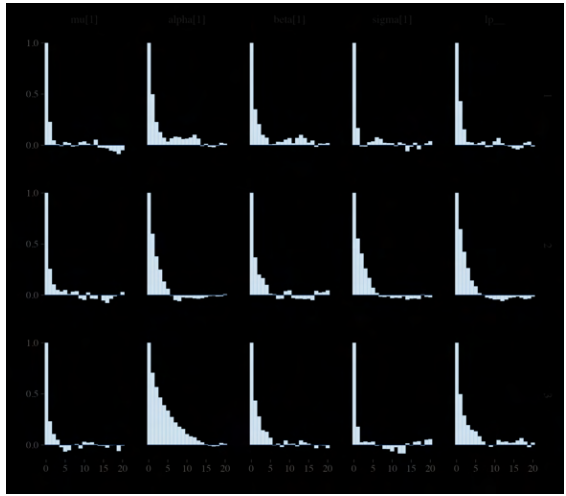
(a) Autocorrelation function

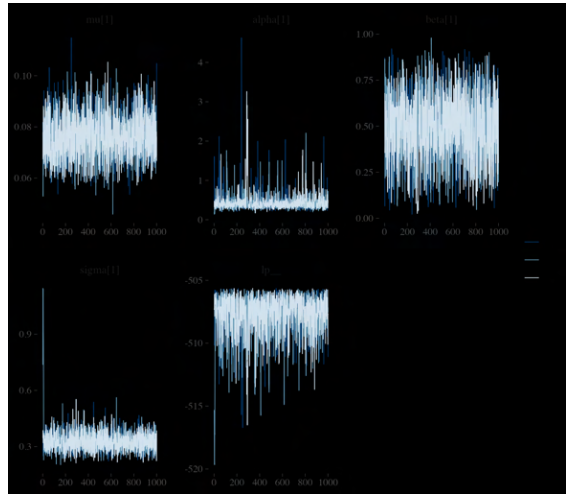
(b) Trace plots

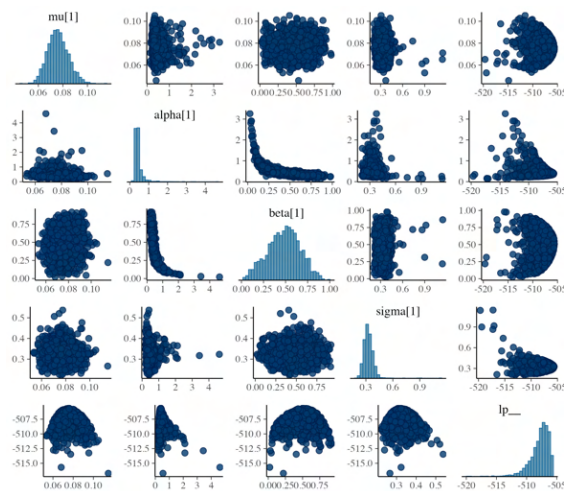
(c) Pairwise correlations

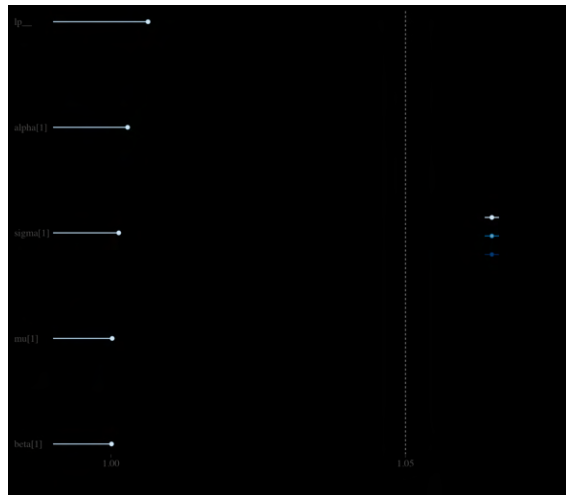
(d) $\hat{R}$

Figure 42: MCMC diagnostic plots for Strategic developments in Pakistan.

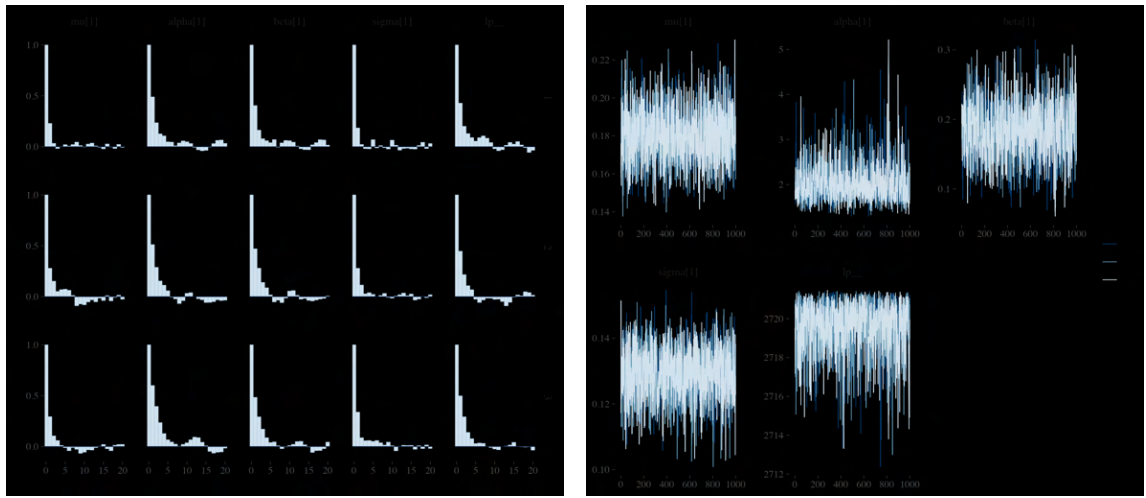

(a) Autocorrelation function

(b) Trace plots

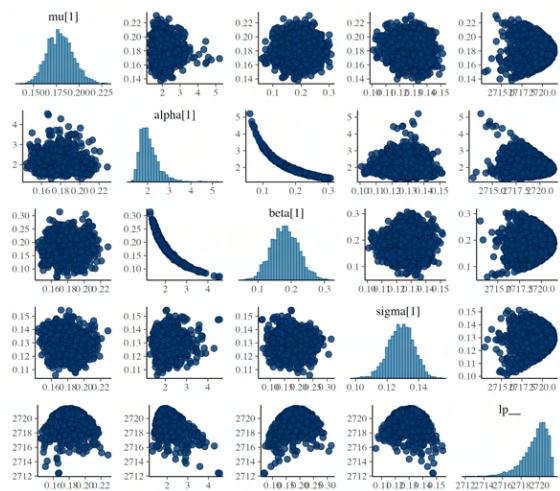

(c) Pairwise correlations

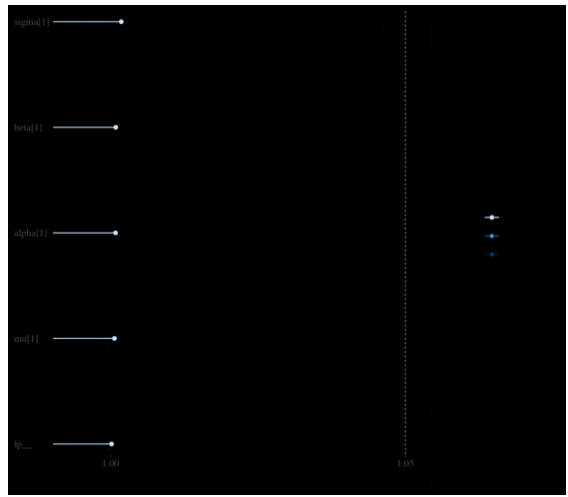

(d) $\hat{R}$

Figure 43: MCMC diagnostic plots for Explosions/remote violence in Pakistan.

81

# F  Residual maps

Maps of the absolute value of the corresponding residuals for each country can be found in the supplementary materials. The residual number of events over this 5 year period was determined by calculating the absolute value of the difference between the total number of observed events and the median of the estimated expected number of events.

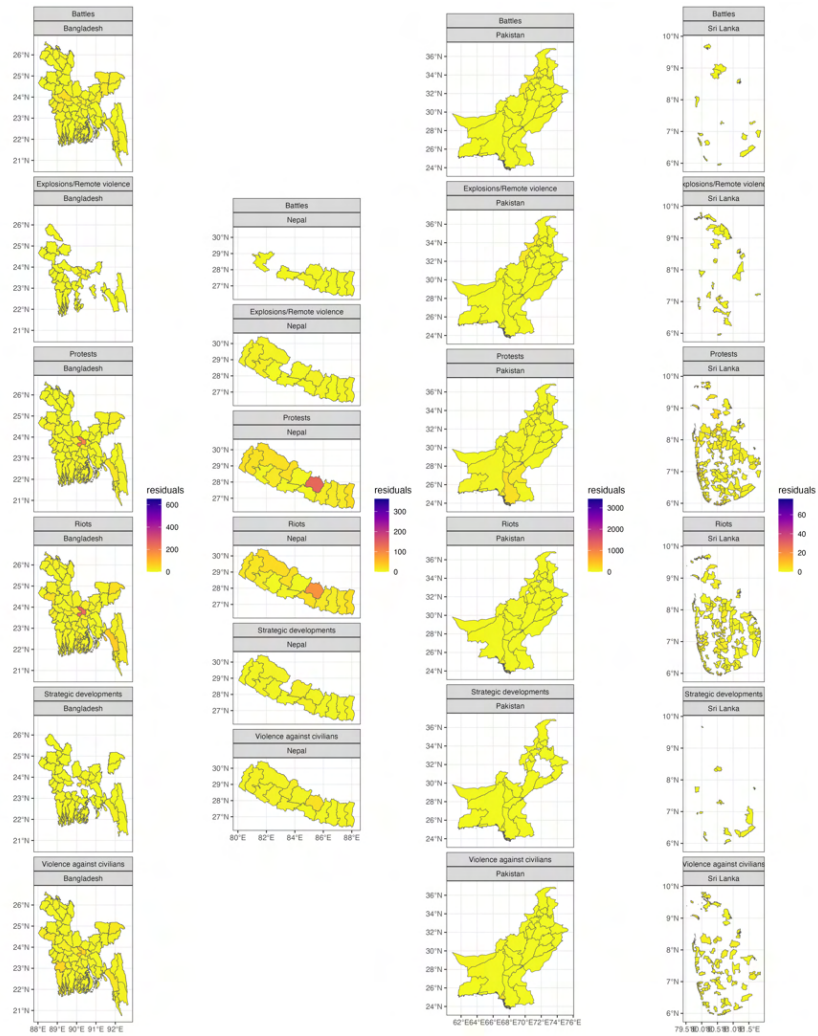

Figure 44: Residuals over 5 year observation window (absolute value of the sum of observed events minus the sum of median expected number of events)

# G  Maximum likelihood results for spatiotemporal DTHP

Table 4 presents the MLEs for the spatiotemporal DTHP model for each country and conflict type.

| Country | Conflict type | $\mu$ | $\alpha$ | $\beta$ | $\sigma$ |
|---|---|---|---|---|---|
| Bangladesh | Battles | 0.14 | 0.45 | 0.02 | 0.14 |
|  | Riots | 0.37 | 0.68 | 0.41 | 0.00 |
|  | Protests | 0.15 | 0.70 | 0.34 | 0.00 |
|  | Violence against civilians | 0.20 | 0.55 | 0.04 | 0.02 |
|  | Strategic developments | 0.03 | 0.25 | 0.02 | 0.23 |
|  | Explosions/Remote violence | 0.04 | 0.34 | 0.02 | 0.15 |
| Sri Lanka | Battles | 0.01 | 0.06 | 0.11 | 1.28 |
|  | Riots | 0.03 | 0.12 | 0.37 | 0.05 |
|  | Protests | 0.04 | 0.45 | 0.46 | 0.00 |
|  | Violence against civilians | 0.02 | 0.04 | 0.52 | 0.14 |
|  | Strategic developments | 0.01 | 0.00 | 0.00 | 0.03 |
|  | Explosions/Remote violence | 0.01 | 0.02 | 0.08 | 2.02 |
| Nepal | Battles | 0.04 | 0.04 | 0.00 | 1.18 |
|  | Riots | 0.49 | 0.42 | 0.95 | 0.35 |
|  | Protests | 0.46 | 0.65 | 0.22 | 0.00 |
|  | Violence against civilians | 0.10 | 0.08 | 1.00 | 1.90 |
|  | Strategic developments | 0.06 | 0.05 | 1.00 | 4.30 |
|  | Explosions/Remote violence | 0.10 | 0.12 | 0.33 | 0.60 |
| Pakistan | Battles | 0.21 | 0.79 | 0.15 | 0.03 |
|  | Riots | 0.16 | 0.63 | 0.42 | 0.46 |
|  | Protests | 0.55 | 0.91 | 0.39 | 0.00 |
|  | Violence against civilians | 0.18 | 0.82 | 0.30 | 0.05 |
|  | Strategic developments | 0.07 | 0.35 | 0.53 | 0.00 |
|  | Explosions/Remote violence | 0.18 | 0.89 | 0.15 | 0.01 |

Table 4: MLEs for spatiotemporal DTHP

Figure 45 shows the observed event counts for each month compared to the expected number of events $\lambda(t, x, y)$, estimated via maximum likelihood estimation. Using this approach, it is apparent that the performance is similar to the Bayesian approach, whereby the estimated mean number of events on a given month closely aligns with the observed data and the model is able to react quickly to recent events. However, while the MLE approach struggles to capture the self-excitation present for scenarios with low event counts, for example strategic developments in Sri Lanka, the Bayesian model can capture some self-excitation.

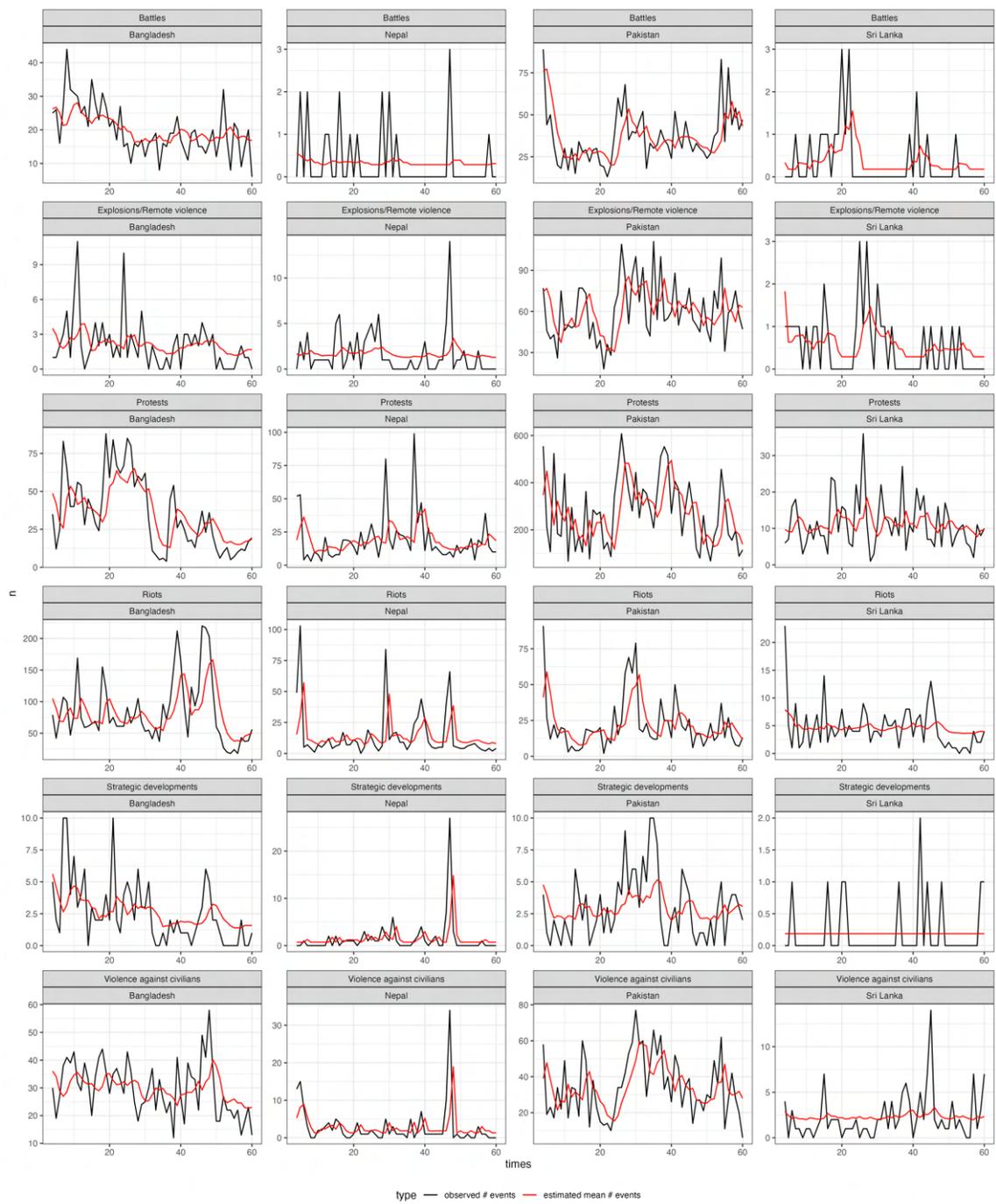

Figure 45: Observed data on month $t$ (black line) versus estimated $\lambda(t)$ (red line). Model estimated via maximum likelihood.

84

# H  Summary of counts in data

Figure 46 presents a temporal summary of the data used in this study and Table 5 summarises the total event counts.

| Country | Conflict type | Total event count | 3 month rolling average |
|---|---|---|---|
| Bangladesh | Battles | 1223.00 | 20.43 |
| | Explosions/Remote violence | 136.00 | 2.22 |
| | Protests | 2170.00 | 36.16 |
| | Riots | 5115.00 | 85.14 |
| | Strategic developments | 176.00 | 2.83 |
| | Violence against civilians | 1822.00 | 30.40 |
| Nepal | Battles | 25.00 | 0.40 |
| | Explosions/Remote violence | 93.00 | 1.59 |
| | Protests | 1138.00 | 19.23 |
| | Riots | 827.00 | 14.09 |
| | Strategic developments | 83.00 | 1.41 |
| | Violence against civilians | 176.00 | 2.93 |
| Pakistan | Battles | 2348.00 | 38.33 |
| | Explosions/Remote violence | 3661.00 | 61.16 |
| | Protests | 16738.00 | 282.24 |
| | Riots | 1372.00 | 23.06 |
| | Strategic developments | 190.00 | 3.16 |
| | Violence against civilians | 2081.00 | 35.28 |
| Sri Lanka | Battles | 20.00 | 0.33 |
| | Explosions/Remote violence | 39.00 | 0.57 |
| | Protests | 666.00 | 11.16 |
| | Riots | 327.00 | 5.27 |
| | Strategic developments | 18.00 | 0.24 |
| | Violence against civilians | 158.00 | 2.36 |

Table 5: Counts for each country and conflict type

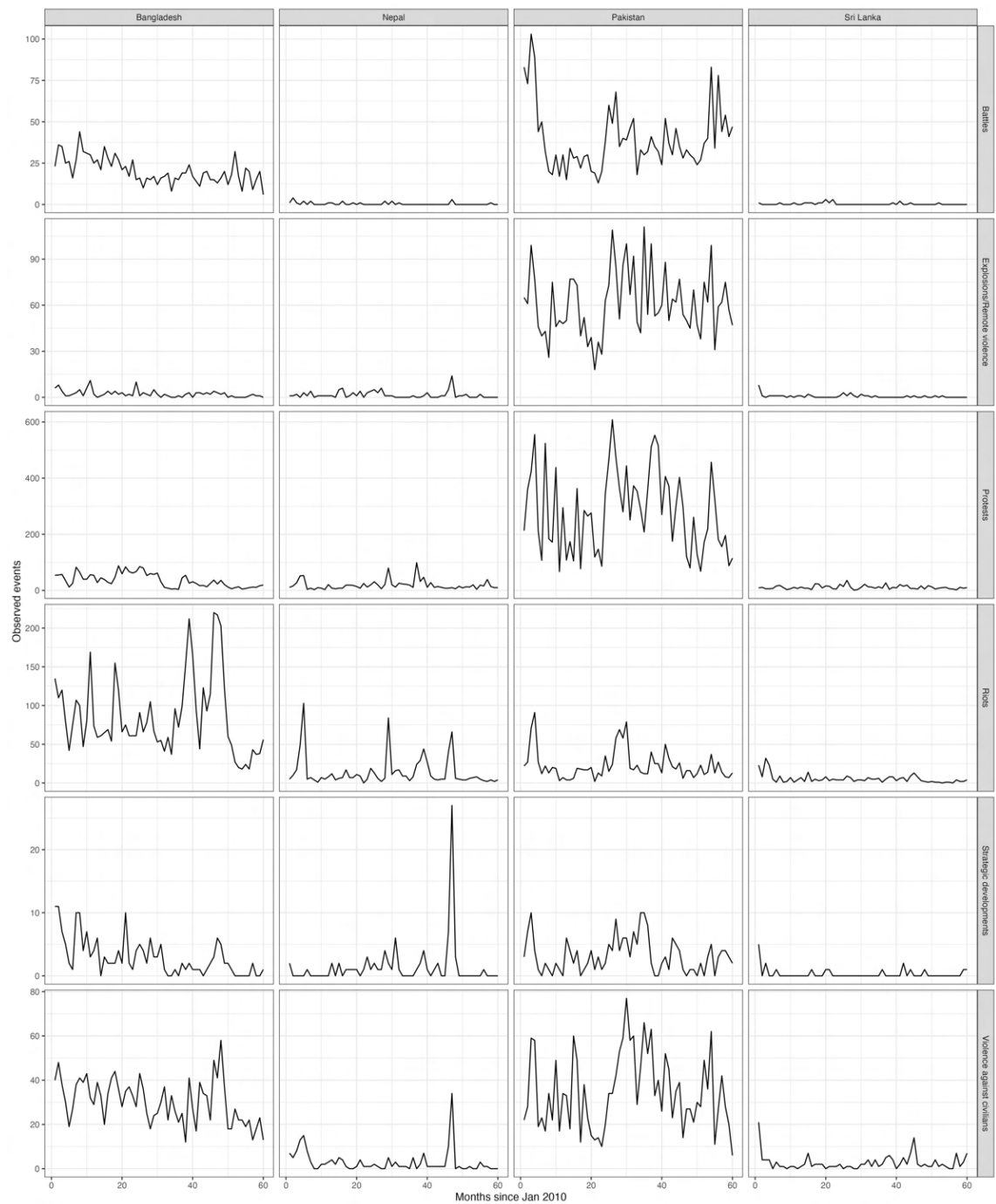

Figure 46: Observed conflict events from 2010 – 2014 by country and conflict type. Horizonal panels: conflict type. Vertical panels: country.

86



\end{document}